\documentclass[twocolumn,trackchanges]{aastex631}

\usepackage{amsmath}
\usepackage{relsize}
\usepackage[T1]{fontenc}

\shorttitle{Hot Stellar Populations in NGC\,2298}
\shortauthors{Rani et al.}
\graphicspath{{./}{figures/}}

\begin{document}

\title{ \textit{AstroSat} study of the globular cluster NGC\,2298: \\
probable evolutionary scenarios of hot HB stars}

\correspondingauthor{Sharmila Rani}
\email{sharmila.rani@iiap.res.in}

\author[0000-0003-4233-3180]{Sharmila Rani}
\affiliation{Indian Institute of Astrophysics, Bangalore, 560034,  India}
\affiliation{Pondicherry University, R.V. Nagar, Kalapet, 605014, Puducherry, India}
\author{Gajendra Pandey}
\affiliation{Indian Institute of Astrophysics, Bangalore, 560034,  India}
\author{Annapurni Subramaniam}
\affiliation{Indian Institute of Astrophysics, Bangalore, 560034,  India}
\author[0000-0001-6812-4542]{Chul Chung}
\affiliation{Department of Astronomy $\&$ Center for Galaxy Evolution Research, Yonsei University, Seoul 03722, Republic of Korea}
\author[0000-0002-0801-8745]{Snehalata Sahu}
\affiliation{Indian Institute of Astrophysics, Bangalore, 560034,  India}
\author[0000-0002-8414-8541]{N. Kameswara Rao}
\affiliation{Indian Institute of Astrophysics, Bangalore, 560034, India}





\begin{abstract}
We present the far-UV (FUV) photometry of images acquired with UVIT on \textit{AstroSat} to probe the horizontal branch (HB) population of the Galactic globular cluster NGC\,2298. UV-optical color-magnitude diagrams (CMDs) are constructed for member stars in combination with \textit{HST} UV Globular Cluster Survey (HUGS) data for the central region and \textit{Gaia} and ground-based photometric data for the outer region. 
Blue HB (BHB) sequence with a spread and four hot HB stars are detected in all FUV-optical CMDs and are compared with theoretical updated BaSTI isochrones and synthetic HB models with a range in helium abundance, suggesting that the hot HB stars are helium enhanced when compared to the BHB. The estimated effective temperature, radius, and luminosity of HB stars, using best SED fits, were compared with various HB models. BHB stars span a temperature range from 7,500-12,250 K. 
The three hot HB stars have 35,000-40,000 K, whereas one star has around $\sim 100,000 K$. We suggest the following evolutionary scenarios: two stars are likely to be the progeny of extreme HB (EHB) stars formed through an early hot-flasher scenario; one is likely to be an EHB star with probable helium enrichment, the hottest HB star is about to enter the WD cooling phase, could have evolved from BHB phase. 
Nevertheless, these are interesting spectroscopic targets to understand the late stages of evolution.

\end{abstract}


\section{Introduction} \label{sec:intro}
Galactic globular clusters (GGCs) are one of the oldest (age $\sim$10-13 Gyr) stellar systems known to exist in our galaxy. It has been well established from last two decades that globular clusters (GCs) host multiple stellar populations (MSPs) instead of simple stellar populations (SSPs). 
Complex populations with different chemical compositions are detected in all phases of stellar evolution, such as, Main Sequence (MS, \citealp{2007ApJ...661L..53P}), Sub-Giant Branch (SGB, \citealp{2008ApJ...673..241M, 2009ApJ...697L..58A, 2012ApJ...760...39P}), Red Giant Branch (RGB, \citealp{2008A&A...490..625M, 2008ApJ...672L..29Y, 2009Natur.462..480L, 2015AJ....149...91P, 2017MNRAS.464.3636M, 2018MNRAS.481.5098M}) and Asymptotic Giant Branch (AGB, \citealp{2021ApJ...910....6L}). Helium abundance variations among the distinct sequences along MS, SGB, and RGB have been found from various photometric and spectroscopic studies within a few clusters \citep{2005ApJ...631..868D, 2005ApJ...621..777P, 2007ApJ...661L..53P, 2020ApJ...897...32H}. Recently, \cite{2021ApJ...906...76D} identified and characterized, for the first time, multiple stellar populations along the red horizontal branch (RHB) in 14 GCs based on the distribution of RHB stars in UV-optical two-color diagrams. UV observations are an essential tool to detect and analyze the exotic populations that reside in GCs, that tend not to follow standard stellar evolution, such as blue straggler stars (BSSs), cataclysmic variables, extreme HB (EHB), and blue-hook (BHk) stars. Identifying these hot populations in optical images can be extremely difficult mainly because of two reasons 1) the severe crowding of optical images (especially in the cores of GCs) which are dominated by MS and RGB stars, 2) most exotic stars are optically faint and present in the core of the GCs. However, all of these are hotter than other cluster members and emit much of their radiation in the UV. Crowding is generally not a problem in the UV images as normal cluster stars (MS and RGB) are cooler than late A type stars and considerably fainter at wavelengths less than 2000 {\AA}. 
Thus, a combination of optical and far-UV (FUV) magnitudes is the most powerful tool to analyze the hot stars in GCs \citep{1998ApJ...500..311F, 2010ApJ...710..332D, 2011MNRAS.410..694D, 2013MNRAS.430..459D, 2017AJ....154..233S, 2019MNRAS.482.1080S, 2020ApJ...905...44S, 2020JApA...41...35R, 2021MNRAS.501.2140R, 2021ApJ...908...66P}.

HB stars represent the late stages in the evolution of low mass stars where helium is burning in the core of mass $\sim0.5M_{\odot}$ surrounded by hydrogen burning shell \citep{1955ApJ...121..776H}. 
The HB is of particular interest since the morphology of HB varies from cluster to cluster. The change in the HB morphology is actually a well known problem in GCs known as second parameter problem. It has been first suggested that metallicity is a principal parameter governing the shape of HBs in GGCs. In general, metal rich clusters have red HBs whereas metal poor ones have stars distributed at higher effective temperatures (bluer colors) along HB. Moreover, there are several clusters sharing the same metal content, but the different HB morphology, e.g. the GC pairs M3-M13, NGC\,288-NGC\,362. Even some metal rich clusters, namely NGC\,6388 and NGC\,6441, have blue HBs \citep{1997ApJ...484L..25R, 2007A&A...474..105B, 2008ApJ...677.1069D}. These exceptions have indicated the need of a second and possibly a third parameter to explain the HB distributions in GCs. The parameters other than metallicity playing a role in shaping the HB are suggested to be age, helium abundance, mass loss along RGB etc., but the answer is not obvious as some of these parameters are not well constrained from theory. For more information, see \cite{2010A&A...517A..81G} and references therein. \cite{2014ApJ...785...21M} analyzed the HBs of 74 GGCs using \textit{HST} data in optical filters, and concluded that age and metallicity are the main global parameters, while the range of helium abundance within a GC is the main non-global parameter defining the HB morphology. Recently, \cite{2020MNRAS.498.5745T} studied the HBs of 46 GCs by comparing the \textit{HST} dataset with theoretical stellar evolutionary models, and concluded that helium enhancement and mass loss both contribute to the HB morphology.\\

The hot sub-populations, namely EHB and BHk, located at the blue and hot end of the BHB, follow a vertical sequence in the optical color-magnitude diagrams (CMDs) as their optical colors become degenerate at high temperatures because of the large increase of the bolometric corrections. In general, EHB and BHk stars are defined as the HB stars that have effective temperatures ({\it $T_{eff}$}) higher than $\sim$20,000 and $\sim$32,000 K, respectively. BHk stars, also known as sub-luminous stars, are fainter than canonical EHB stars in optical as well as in FUV CMDs. They form a hook like structure at the hot end of the EHB in (FUV$-$NUV, FUV) CMDs and hence known as ‘blue-hook stars’ (See FUV CMDs in \cite{2000ApJ...530..352D, 2001ApJ...562..368B, 2010ApJ...718.1332B}). The hottest EHB and BHk stars have extremely thin envelope masses ($M_{env} < 0.01M_{\odot}$) due to severe mass loss on the RGB. Therefore, these stars, when helium is exhausted in their core, evolve into the AGB-manqu\'e stars or post-early AGB (peAGB) stars, but do not ascend to the AGB \citep{1990ApJ...364...35G, 1993ApJ...419..596D}. 
It is extremely difficult to detect them in the optical images due to their faintness and location near the crowded core of the GCs. 
Here UV CMDs play an important role as these stars are bright compared to the cooler stars in UV images, and also they follow a particular sequence in UV CMDs separating them from BHB and hot BSSs. It has been well demonstrated from spectroscopy and spatially resolved imaging that EHB stars are the dominant contributors to the “UV upturn” in the spectra of elliptical galaxies \citep{1995ApJ...442..105D, 1997ApJ...482..685B, 2000ApJ...532..308B, 2008ApJ...682..319B, 2011ApJ...740L..45C, 2012ApJ...747...78B}. The dominating mechanism for producing these hot objects in GCs are still unclear. Until now, many theories have been proposed to explain the formation of both type of stars in GCs, although population of these stars mainly depend on the cluster mass and density \citep{2004ApJ...603..135R, 2004A&A...415..313M}. Very massive and dense clusters such as NGC\,2808 and $\omega$ Cen, have a large samples of EHB and BHk stars. The possible formation scenarios suggested for EHB and BHk stars in clusters are dynamical interactions among binary stars, helium-mixing, early hot-flasher, late hot-flasher and helium enhancement \citep{1976ApJ...204..488M, 1997ApJ...474L..23S, 2001ApJ...562..368B, 2010ApJ...718.1332B, 2015MNRAS.449.2741L, 2016PASP..128h2001H}.


      The southern GGC NGC\,2298 is located in the constellation Puppis at a distance of 10.6 kpc and has metallicity \big[Fe/H\big] = -1.92 dex \citep{2009A&A...508..695C, 2010arXiv1012.3224H, 2018ApJ...865..160M}. The adopted reddening value and age for the cluster in this work is ($0.2\pm0.01$) mag and ($13.2\pm0.4$) Gyr, respectively \citep{2018ApJ...865..160M}. NGC\,2298 is also known for hosting multiple stellar populations along the MS and the RGB \citep{2015AJ....149...91P, 2017MNRAS.464.3636M}.
       This cluster is widely studied in optical, but the UV studies of this cluster are sparse. 
As this cluster is metal-poor, its HB mainly comprises BHB and a few EHB stars and hence is ideal for studying the UV properties of the very hot HB population. In our quest to understand the formation and evolution of hot HB stars, here we present, for the first time, an FUV photometric study of hot HB stars in NGC\,2298. This work aims to identify and shed light on the properties of hot HB stars in this cluster and find their formation mechanisms in comparison with theory. 

The structure of this paper is as follows. Section~\ref{sec:data} describes the observational data and their reduction. In Section~\ref{sec:members}, the proper motion membership determination using \textit{Gaia} EDR3 data is presented. Section~\ref{sec:CMDs} presents the selection of HB and BSS samples as well as details of the observed UV and Optical CMDs. In Section~\ref{sec:models}, a comparison of observed HB with theoretical models is discussed. In Sections~\ref{sec:SEDs} and ~\ref{sec:status}, we describe the properties of HB stars derived from the UVIT photometry along with \textit{HST}, \textit{Gaia} and ground-based photometry and their evolutionary status. A detailed discussion of all results is provided in Section~\ref{sec:dis}. Finally, we summarize and conclude our results in Section~\ref{sec:summary}.

\section{Data used and Analysis} 
\label{sec:data}
\subsection{UVIT Data}
  \label{subsec:uvitdata}
  To investigate the stellar populations along the HB in GGC NGC\,2298, we have used images taken with the UVIT instrument on board \textit{AstroSat} satellite. 
 The observations of NGC\,2298 were carried out on 13 December 2018 in three FUV filters: F148W, F154W and F169M. 
 UVIT consists of twin 38 cm telescopes, one for the FUV region (130-180 nm) and the other for the NUV (200-300 nm) and visible (VIS) regions (320-550 nm). UVIT is primarily an imaging instrument, simultaneously generating images in the FUV, NUV and VIS channels over a circular field of diameter 28$'$. Full details on the telescope and instruments, including initial calibration results, can be found in \cite{2017JApA...38...28T}. The magnitude system adopted for UVIT filters is similar to that used for \textit{GALEX} filters, and hence the estimated magnitudes will be in the AB magnitude system.\\
 
\begin{figure}[!htb]
    \centering
	\includegraphics[scale=0.55]{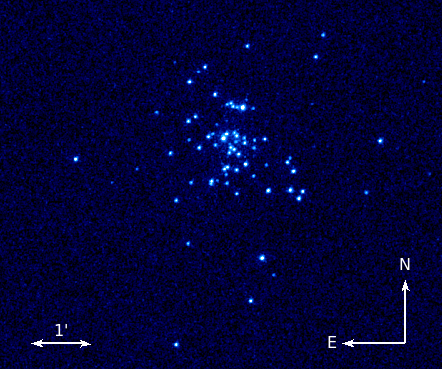}
    \caption{The color image of NGC\,2298 in UVIT FUV band F148W. The hot stellar populations displayed with blue color are well resolved at the center of the cluster.}
    \label{fuvimage}
    \end{figure}

 In order to complete the required exposure times in given filters, UVIT takes data over multiple orbits. A customised software package CCDLAB (\citealp{2017PASP..129k5002P}) was utilized to correct for the geometric distortion, flat field illumination, spacecraft drift, and create images for each orbit. Then, the orbit-wise images were co-aligned and combined to generate a science ready images in order to improve the signal-to-noise ratio. The photometry was performed on these final science ready images to derive the magnitudes of the stars detected with UVIT. A detailed log of UVIT observations of NGC\,2298 is reported in Table~\ref{tab1}. The color image of NGC\,2298 in F148W band is shown in Figure~\ref{fuvimage}.  In this figure, the blue color indicates UVIT FUV detections in the F148W filter. Crowding is not a problem here, as we are able to resolve stars well into the core of the cluster in FUV images.
      
	 \begin{table}[!htb]
    \centering
	\caption{Observational details of NGC\,2298 in UVIT FUV filters. The last column lists the extinction value calculated in each FUV filter using \cite{1999PASP..111...63F} law of extinction.}
	\label{tab1}
	\begin{tabular}{cccccc} 
		\hline
		\hline
		 Filter & $\lambda_{mean}$ & $\Delta\lambda$  & ZP  & Exposure &  $A_{\lambda}$\\
		  & ({\AA}) & ({\AA}) & (AB mag) & Time (sec) & (mag)\\
		\hline
		 F148W & 1481 & 500 & 18.016 & 2300 & 1.63  \\
		 F154W & 1541 & 380 & 17.78 & 558 & 1.59\\
		 F169M & 1608 & 290 & 17.45 & 1263 & 1.55 \\
		 \hline
	\end{tabular}
  \end{table}
  
  \subsection{Photometry}
  \label{sec:phot}
  
\begin{figure}[!htb]
  \hspace*{-0.4cm} 
	\includegraphics[height=8.5cm, width=9cm]{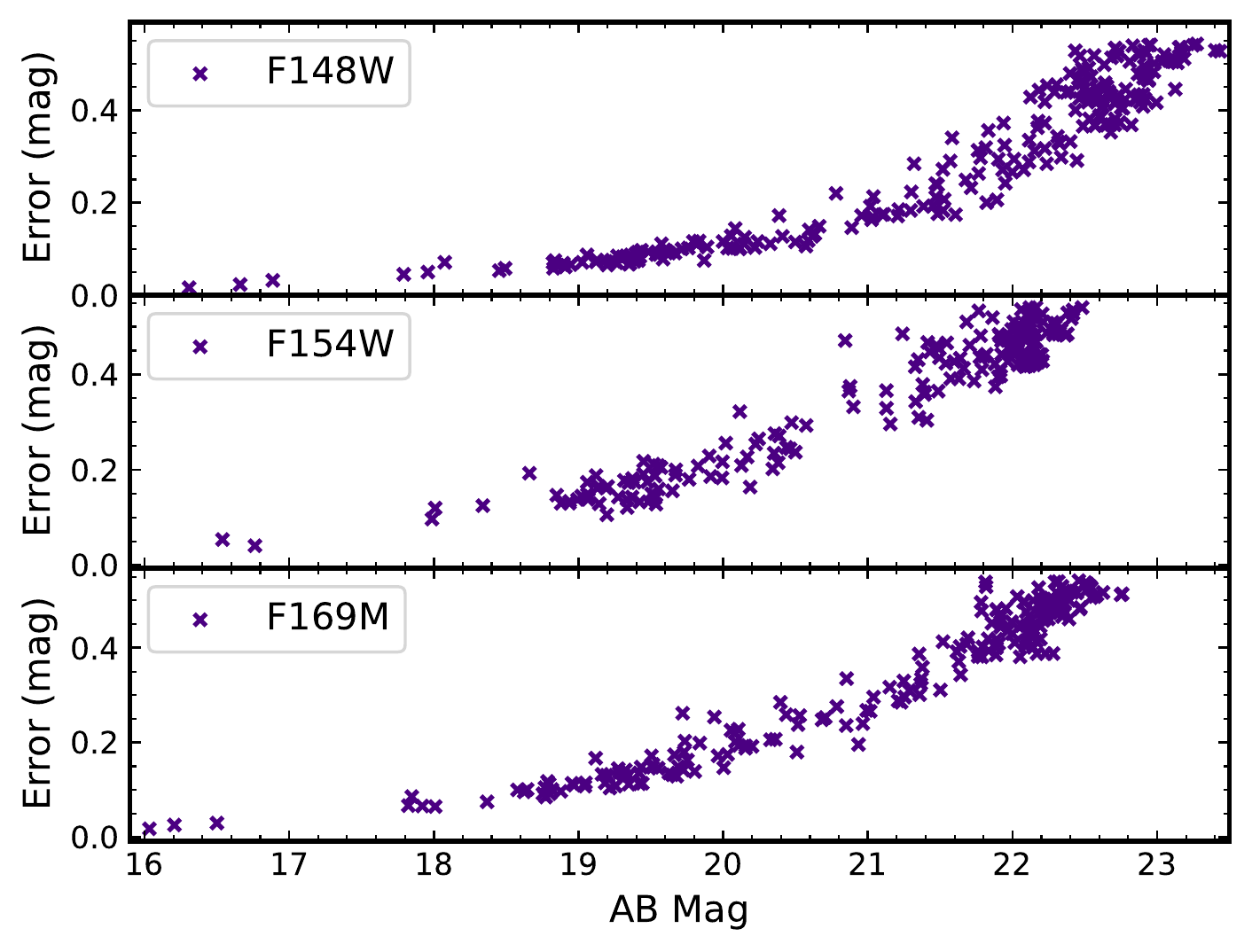}
     \caption{PSF fit errors as a function of magnitude for our UVIT observations of NGC\,2298 in FUV bandpasses. From top to bottom, the panels show results for the F148W, F154W, and F169M bandpasses, respectively.}
    \label{psferr}
\end{figure}

In order to derive the magnitudes of stellar sources, crowded field photometry was carried out on all the FUV images using DAOPHOT package in IRAF/NOAO \citep{1987PASP...99..191S}. The following  steps are taken to extract point spread function (PSF) photometry. At first, the stars were located in the images using DAOFIND task, and then the aperture photometric magnitudes were computed using PHOT task in DAOPHOT. The model PSF was constructed by selecting isolated and bright stars, and then it was applied to all the detected stars using ALLSTAR task to obtain PSF fitted magnitudes. A curve-of-growth analysis was carried out to estimate aperture correction values in each filter, which were then applied to the estimated PSF magnitudes. As the detector works in the photon counting mode, saturation correction,  (to account for more than one photon per frame), was done to the PSF generated magnitudes to get the instrumental magnitudes. The saturation correction mainly affects stars brighter than 17 magnitude. The details of the saturation correction are described in \citep{2017AJ....154..128T}. We have calibrated the instrumental magnitudes into AB system by using zero-point magnitudes provided in the calibration paper \citep{2017AJ....154..128T}. We show our PSF-fit errors for all filters as a function of magnitude in figure~\ref{psferr}. Stars detected with UVIT are considered up to $\sim22$ mag in FUV F148W filter, and $\sim21$ mag in F154W and F169M filters for further analysis, which effectively puts an upper limit of $\sim0.3$ mag on our photometric errors.
 The observed UVIT stellar magnitudes are corrected for extinction and reddening. To compute the extinction value in visual band ($A_V$), We have adopted reddening E(B$-$V) = 0.20 mag from \cite{2018ApJ...865..160M} and the ratio of total-to-selective extinction as $R_{V}$ = 3.1 from \cite{1958AJ.....63..201W} for the Milky Way. The Fitzpatrick reddening law \citep{1999PASP..111...63F} was used to calculate extinction co-efficients $A_\lambda$ for all bandpasses as tabulated in Table~\ref{tab1}. We adopted the following relation to correct for extinction and reddening in observed magnitudes:

\begin{equation}
\centering
    m_{f,corr} = m_{f,obs} - c_{f}R_{V}E(B-V)
\end{equation}
where $m_{f,corr}$ is the extinction corrected magnitude for a particular bandpass \textit{f}; $m_{f,obs}$ is the observed psf magnitude; $c_{f} = \frac{A_{\lambda}}{A_{V}}$ is the extinction law.

\subsection{Other catalogs}
 \label{sec:otherdata}
 In this study, we have combined UVIT data with UV-optical photometric data from {\it HST} UV legacy survey catalog of GCs in five filters, namely, F275W, F336W, F438W, F606W, and F814W provided by \cite{2018MNRAS.481.3382N} in the inner region (central region $\sim2\farcm7\times2\farcm7$ covered by {\it HST} WFC3/UVIS) and for the external region (not observed with {\it HST}), we used optical catalog provided by \cite{2019MNRAS.485.3042S} obtained by analyzing ground-based observations. To select the proper motion (PM) members of the cluster in the outer region, {\it Gaia} EDR3 PM catalog is utilized \citep{2020arXiv201201533G}.

\begin{figure*}[!htb]
 \centering
	\includegraphics[width=0.724\columnwidth]{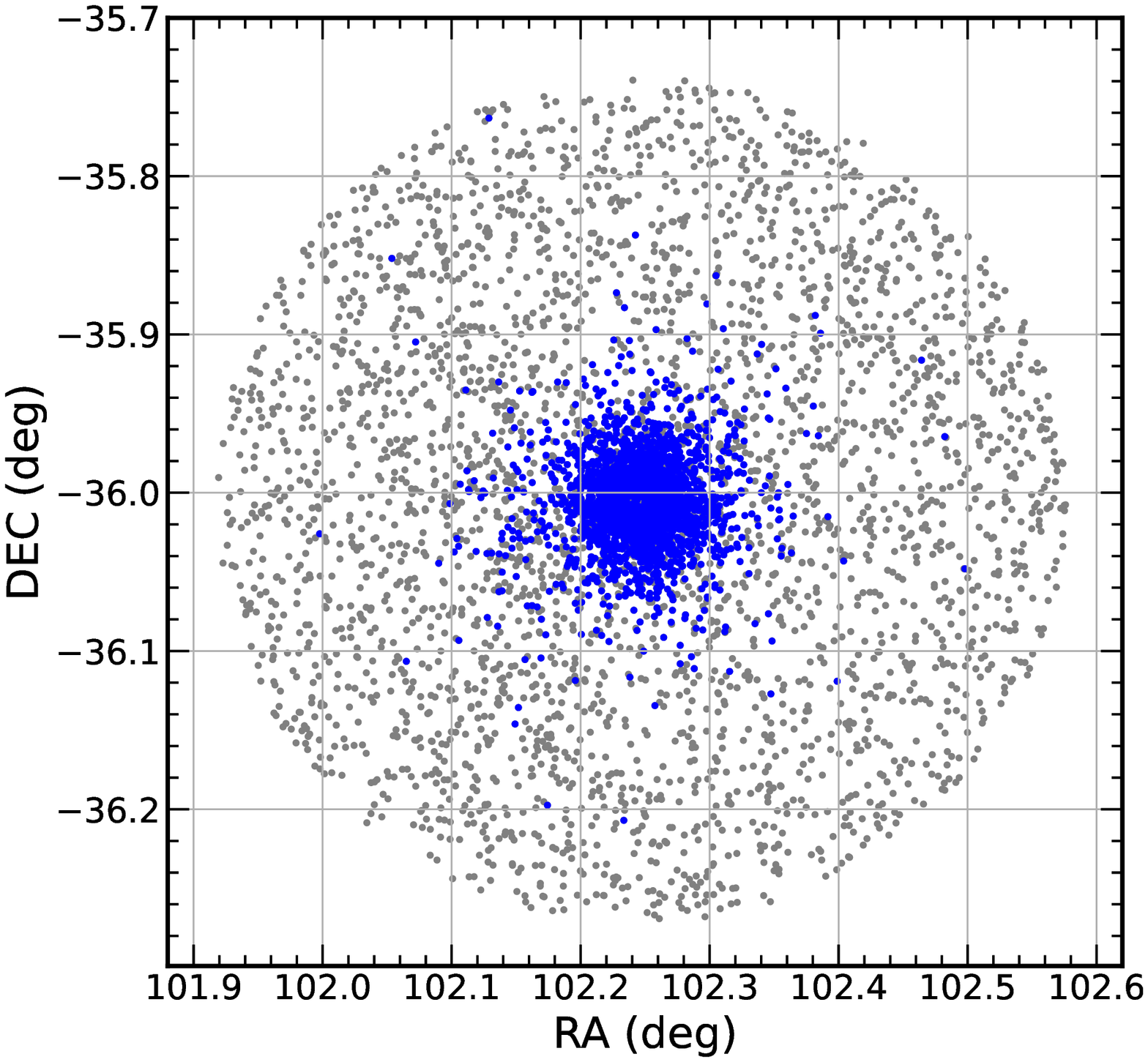}
	\includegraphics[width=0.69\columnwidth]{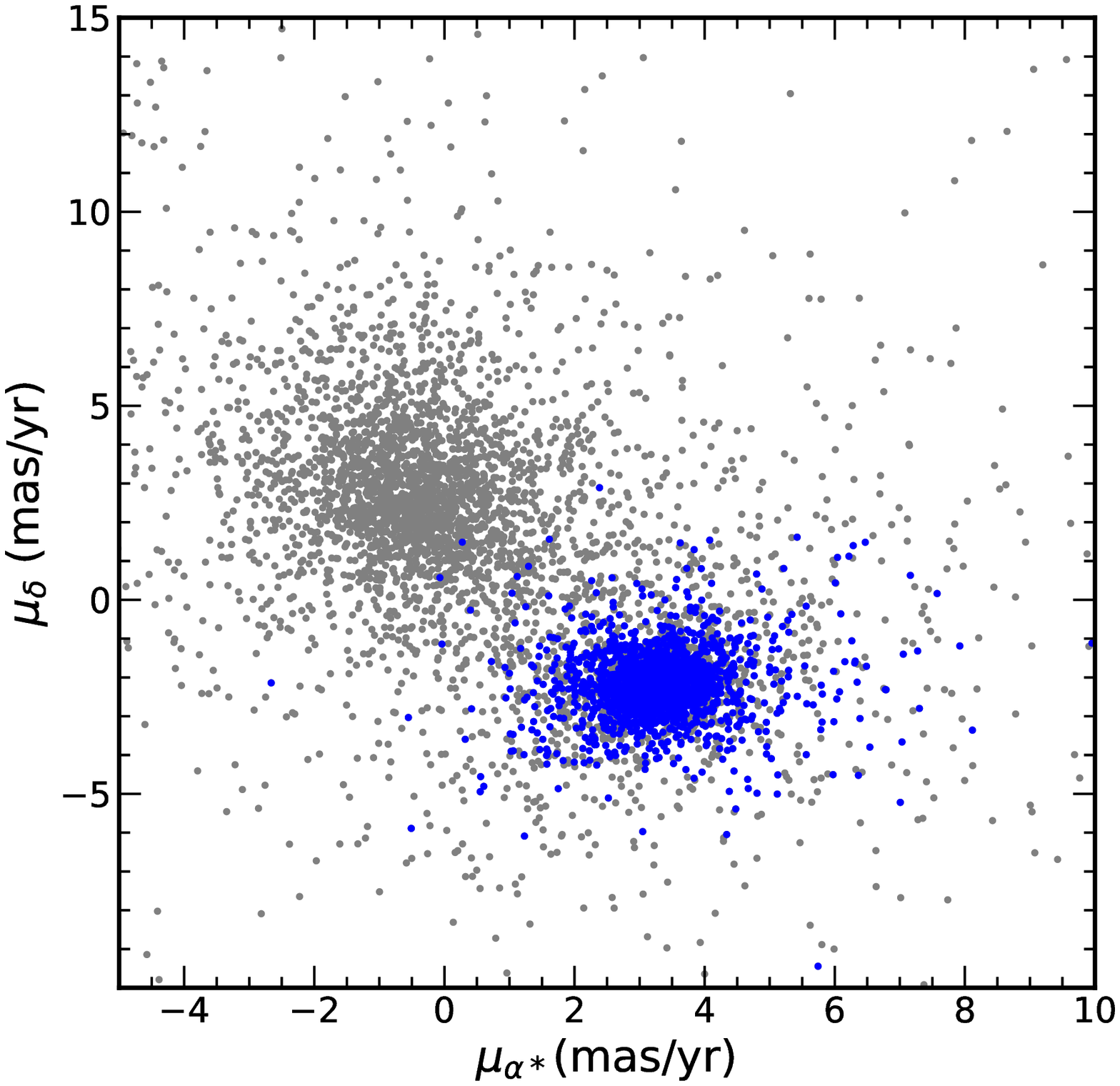}
	\includegraphics[width=0.6808\columnwidth]{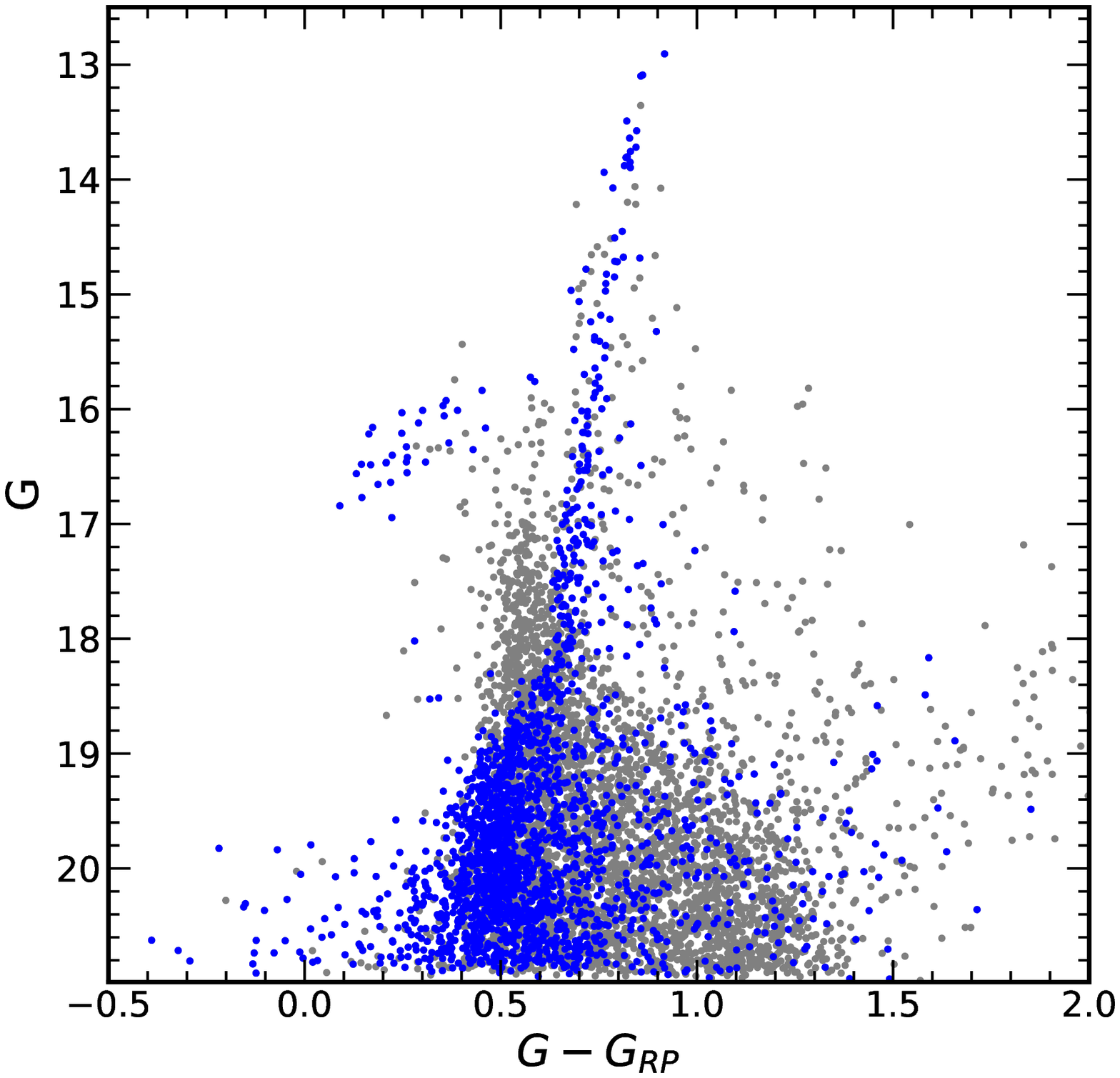}
     \caption{In three panels from left to right, PM members of the cluster are shown with blue dots and remaining \textit{Gaia} EDR3 sample marked with gray dots represent field stars. Left Panel: position in the sky; Middle Panel: Vector Point Diagram (VPD); Right Panel: \textit{Gaia} Optical CMD.}
    \label{gaiapm}
\end{figure*}

\section{Proper Motion Membership} 
\label{sec:members}
To select the confirmed members of the cluster in the inner region covered by \textit{HST}, we have used membership probability information given in astro-photometric catalog of \cite{2018MNRAS.481.3382N}. 
To obtain the membership probability of stars in the outer region of the cluster, we have utilized Gaia EDR3 PM data released on December 3, 2020 \citep{2020arXiv201201533G}. The \textit{Gaia} EDR3 catalog provides the photometric and astrometric information for all the stars with Gmag down to 21 mag. We have employed a probabilistic Gaussian mixture model (GMM) method to select member stars in the cluster and infer the intrinsic parameters of the distributions of both member and non-member stars. The distribution of stellar sources in PM space ($\mu_{\alpha}, \mu_{\delta}$) consists of well-defined clump corresponding to the cluster members and wide distribution of field stars. In most clusters, both distributions overlap each other, and we can not distinguish between the two by eye. Each of the distributions follows a Gaussian distribution, and hence these two distributions are assumed to overlap two Gaussian distributions. The Gaussian probability distribution corresponding to the sum of two distributions is
\begin{equation}
\hspace{0cm}
    f(\mu|\overline{\mu_{i}}, \mathsmaller{\sum_{i}}) = \sum_{i=1}^{2} w_{i}\frac{exp\big[-1/2(\mu - \overline{\mu_{i}})^{T}\sum_{i}^{-1}(\mu - \overline{\mu_{i}})\big]}{2\pi\sqrt{det\sum_{i}}}
\end{equation}

\begin{equation}
\hspace{0cm}
    w_{i}\geq0, \indent \sum_{i=1}^{2} w_{i}= 1
\end{equation}
where $\mu$ is individual PM vector; $\mu_{i}$ are field and cluster mean PMs; $\scriptstyle\sum$ is the symmetric covariance matrix and $w_{i}$ are weights for the two Gaussian distributions.
Full details of this method for n-dimensional case are described in \citep{2019MNRAS.484.2832V}.\\

We selected \textit{Gaia} EDR3 stars with complete astrometric data within twice the tidal radius from the cluster center \citep{2007A&A...467..107D}. In order to choose the stars with good astrometric solution, we removed those with parallaxes that deviate by more than 3$\sigma$ from the expected parallax calculated using the previously known distance to the cluster, where $\sigma$ is the error in parallax given in \textit{Gaia} EDR3 catalog. We also removed the sources with renormalized unit weight error (RUWE) exceeding 1.2 as larger values of this parameter might lead to an unreliable astrometric solution \citep{2018A&A...616A...2L, 2021A&A...649A...3R}. The PM in RA and DEC of the cluster members are supposed to follow two  Gaussian distributions. So, GMM is created for these two distributions, and at first, it is assumed that cluster members and field stars follow the isotropic Gaussian distributions. Initial guess for cluster PM $\mu_{\alpha}$ and $\mu_{\delta}$ values and internal velocity dispersion are taken from \citep{2019MNRAS.484.2832V}. We utilized GaiaTools\footnote{\url{https://github.com/GalacticDynamics-Oxford/GaiaTools}} to maximize the total log-likelihood of GMM and measure the mean PM and standard deviation of both the Gaussian distributions. The membership probabilities (MPs) of all the selected stars are calculated using the same technique simultaneously. The equations used to maximize the log-likelihood of GMM and estimate the MP of $i^{th}$ star belonging to the $k^{th}$ component are given in appendix A in \cite{2019MNRAS.484.2832V}.

We determined the PM mean and standard deviations of the cluster distribution to be $\mu_{\alpha}$ = 3.31 mas/yr and
$\mu_{\delta}$ = -2.176 mas/yr, with $\sigma_{c}$ = 0.055 mas/yr. Figure~\ref{gaiapm} shows the position of stars in the sky, in the PM space known as vector point diagram (VPD), and in an optical CMD created using \textit{Gaia} filters. In this figure, blue dots indicate the member stars belonging to the cluster, and gray dots represent the field stars. The stars with a membership probability of more than 90$\%$ are selected as confirmed cluster members. The number of PM members in the outer region of the cluster are found to be $\sim$1240 and considered for further analysis. 

\section{color Magnitude Diagrams}
\label{sec:CMDs}
\subsection{Selection of HB and BS stars}
We have data in three FUV filters of UVIT, and only very hot and bright stars are expected to be detected. It
is not easy to classify the stars belonging to different evolutionary sequences from the FUV CMDs. Therefore,
optical CMDs are needed to identify different evolutionary sequences. Since UVIT has a large field of view, it
covers the outer parts of the cluster. As the GCs are very dense and massive objects, \textit{HST} is an ideal
telescope to resolve and study the central region of the
clusters in all available bandpasses. 

\begin{figure}[!htb]
\hspace{-0.7cm}
\includegraphics[width=1.1\columnwidth]{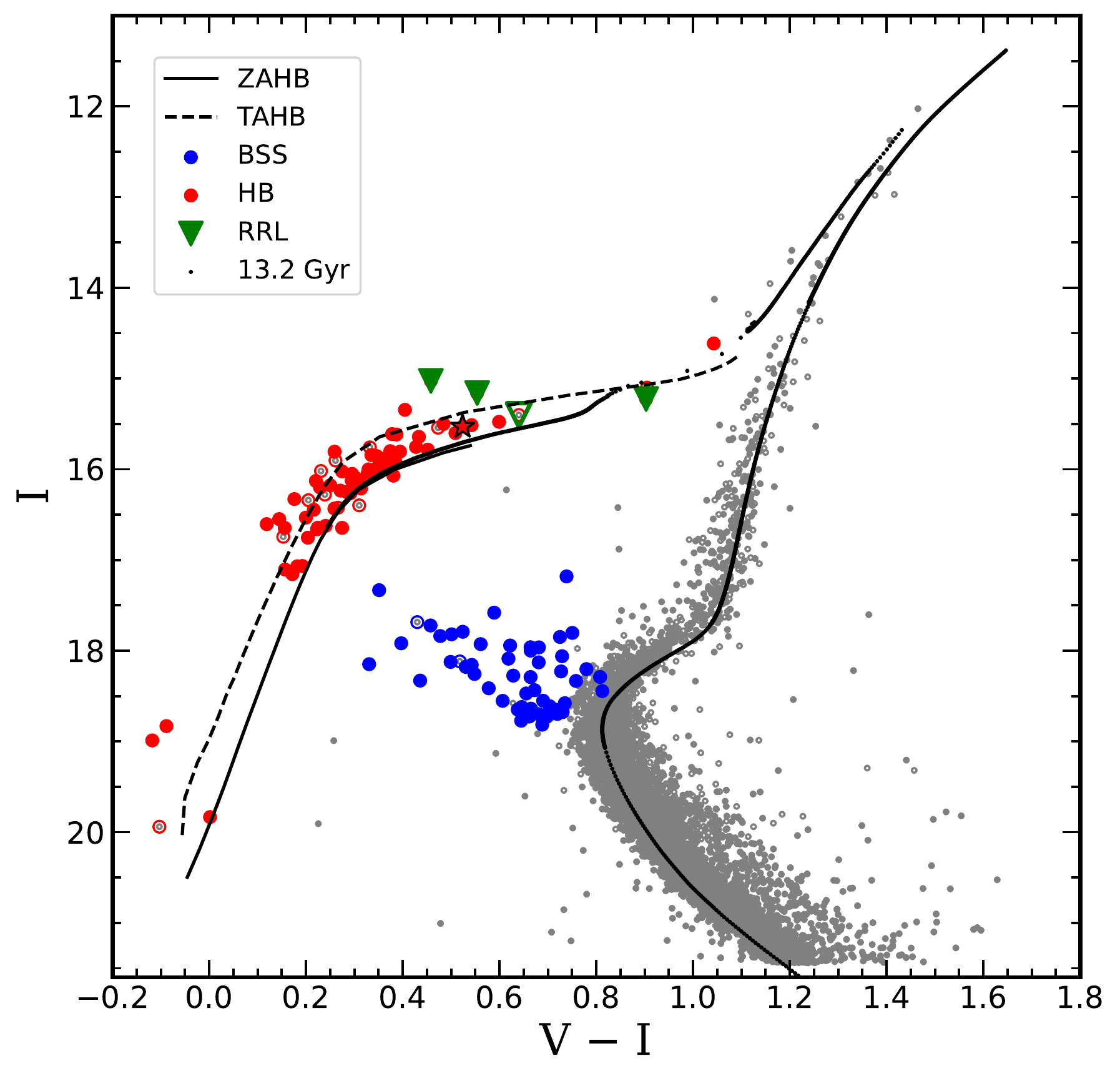}
 \caption{ Optical CMD of NGC\,2298. Gray and colored filled symbols in the optical CMD represent the \textit{HST} detected stars whereas stars shown with gray and colored open symbols are cross-identified using ground based data and \textit{Gaia} EDR3 data. 
 All colored symbols represent the selected HB and BS stars used for further cross-match with UVIT data. All the stars shown in above CMD are confirmed PM members of the cluster.
 The known RR Lyrae stars are shown with green inverted triangles. For comparison with theoretical models, we overlaid the updated BaSTI-IAC model isochrone with an age 13.2 Gyr and metallicity \big[Fe/H\big] = $-1.92$ dex shown with black dots. Along the HB locus, solid and dashed black lines indicate zero-age HB (ZAHB) and terminal-age HB (TAHB), where the star has completed 99$\%$ of its core He-burning lifetime, respectively.}
  \label{optcmds}
\end{figure}

In order to identify and classify UVIT detected stars into various evolutionary phases, we used the \textit{HST} catalog \citep{2018MNRAS.481.3382N} to cross-match with our stars inside a central region and ground-based photometric data for the region outside the \textit{HST} coverage. \cite{2018MNRAS.481.3382N} estimated the PM membership probabilities of stars detected in the inner region of the cluster using \textit{HST} data, and suggested that most likely members have a PM probability of more than $90\%$.
 Therefore,  we have selected stars with a membership probability of more than $90\%$ in the inner as well as outer region using the \textit{HST} and \textit{Gaia} EDR3 catalogs, respectively. However, there are four stars in the inner region with a membership probability of more than $80\%$, which are bright in UV images and found to lie along the HB locus. We have also included these stars in our study. To identify member stars in the outer region of the cluster, we first cross-matched ground-based photometric data \citep{2019MNRAS.485.3042S} with \textit{Gaia} EDR3 PM membership data. In order to plot stars detected in the inner and outer region in the same optical (V-I, I) CMD, we have transformed the \textit{HST} ACS/WFC photometric system into the standard Johnson-Cousins photometric system using the transformation equations given in \cite{2005PASP..117.1049S}. The optical CMD, created using member stars detected with the \textit{HST} and the ground-based observations, is shown in Figure~\ref{optcmds}, where filled and open symbols indicate the stars detected in the inner and outer regions, respectively. The magnitude system adopted for the Johnson-Cousin filters is Vega, hence, V-I color and I magnitude shown in Figure~\ref{optcmds} is in the Vega magnitude system. 

The HB stars were selected by giving a specific color and magnitude cut-off ($-$0.2<V$-$I<1.0, 14.5<I<20.1) in an optical CMD, and displayed with the filled and open red-colored symbols in Figure~\ref{optcmds}. Since this is a metal-poor cluster, the HB is populated mainly with BHB and hot HB stars. However, there are also one or two RHB stars present along HB. Because of its low metallicity, very few variable stars such as RR Lyrae were detected in the cluster. Until now, only four RR Lyrae are known in this cluster \citep{2001AJ....122.2587C}. We have cross-matched the \textit{HST} and ground-based photometric data with the available variable star catalog to identify the cluster's previously known RR Lyrae stars. The green inverted triangles correspond to the variable stars in Figure~\ref{optcmds}. In order to select the BSSs in the inner region, we have employed the same approach used by \cite{2017ApJ...839...64R}. Their study highlighted the importance of using UV CMDs over optical CMDs for proper selection of BSSs in GCs. The BSSs are selected from the UV CMD created using \textit{HST} F275W and F336W passbands and are shown with filled blue symbols in Figure~\ref{optcmds}. BSSs are selected from an optical CMD in the outer region and are displayed with open blue symbols.

\begin{figure*}[!htb]
\begin{minipage}[c][8cm][t]{0.5\textwidth}
        \centering
        \includegraphics[height=8cm]{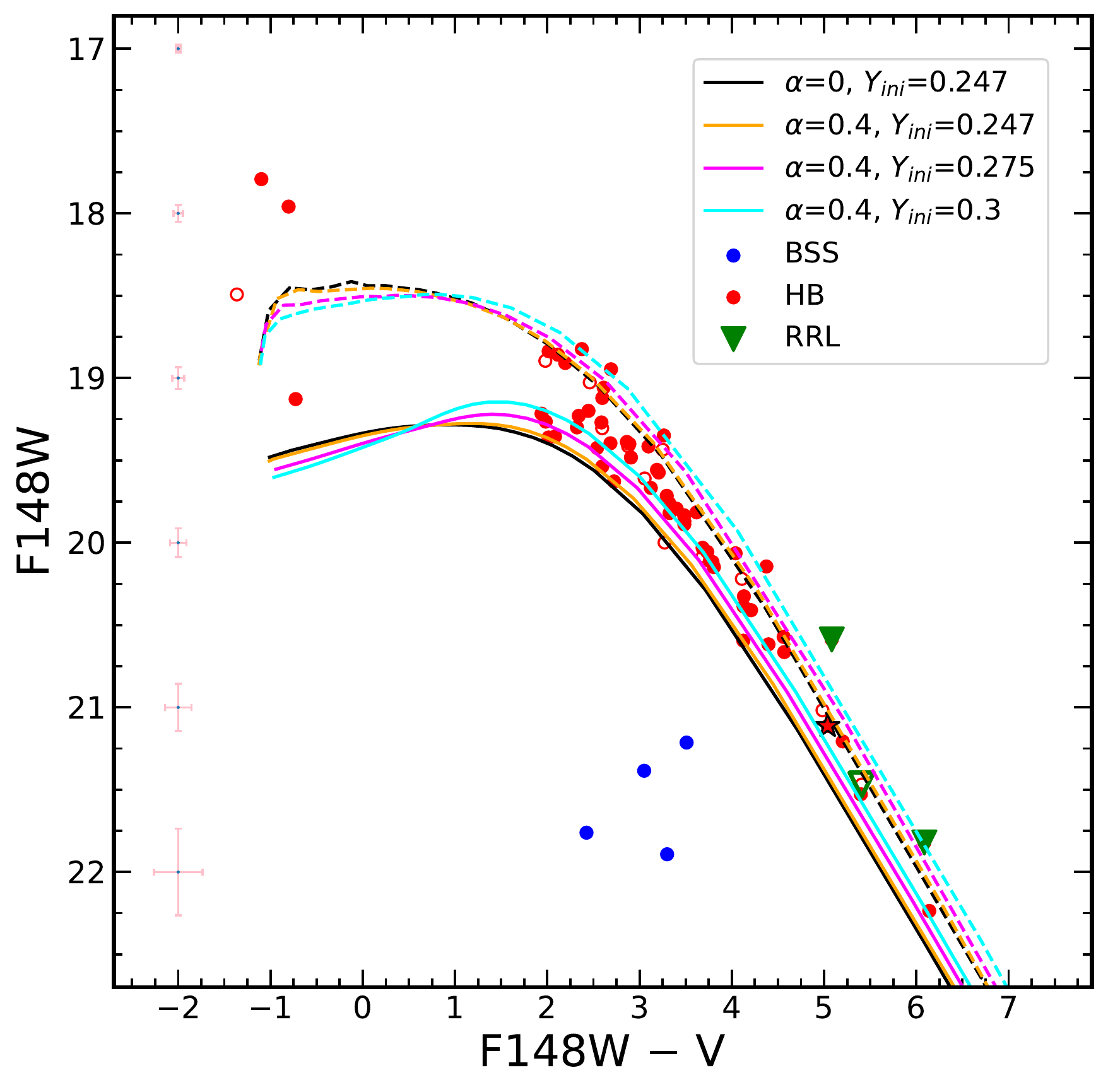}
    \end{minipage}%
    \hspace{-0.7cm}
    \begin{minipage}[c][8cm][c]{0.5\textwidth}
        \centering
        \includegraphics[height=4.2cm]{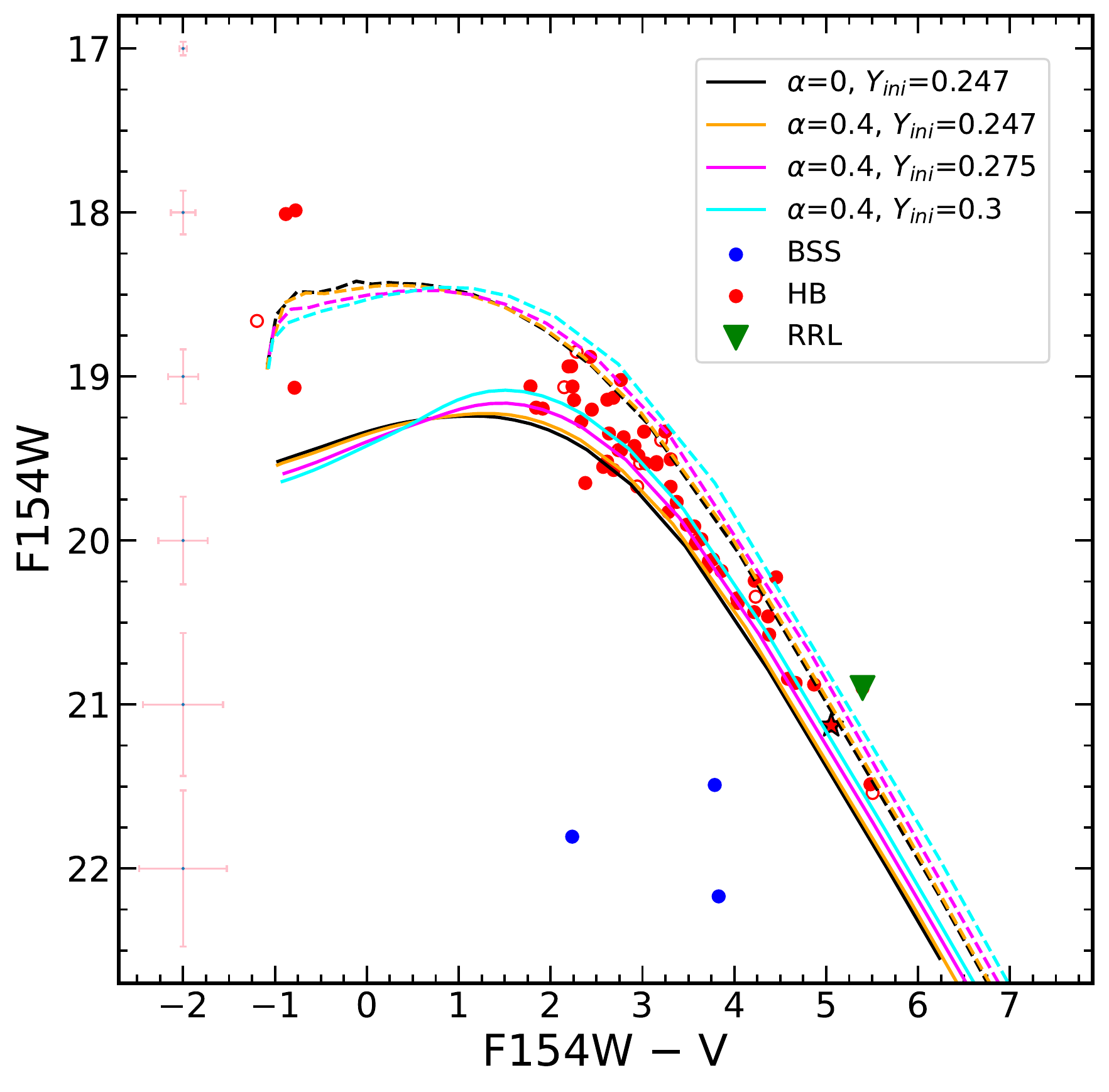}
        \includegraphics[height=4.2cm]{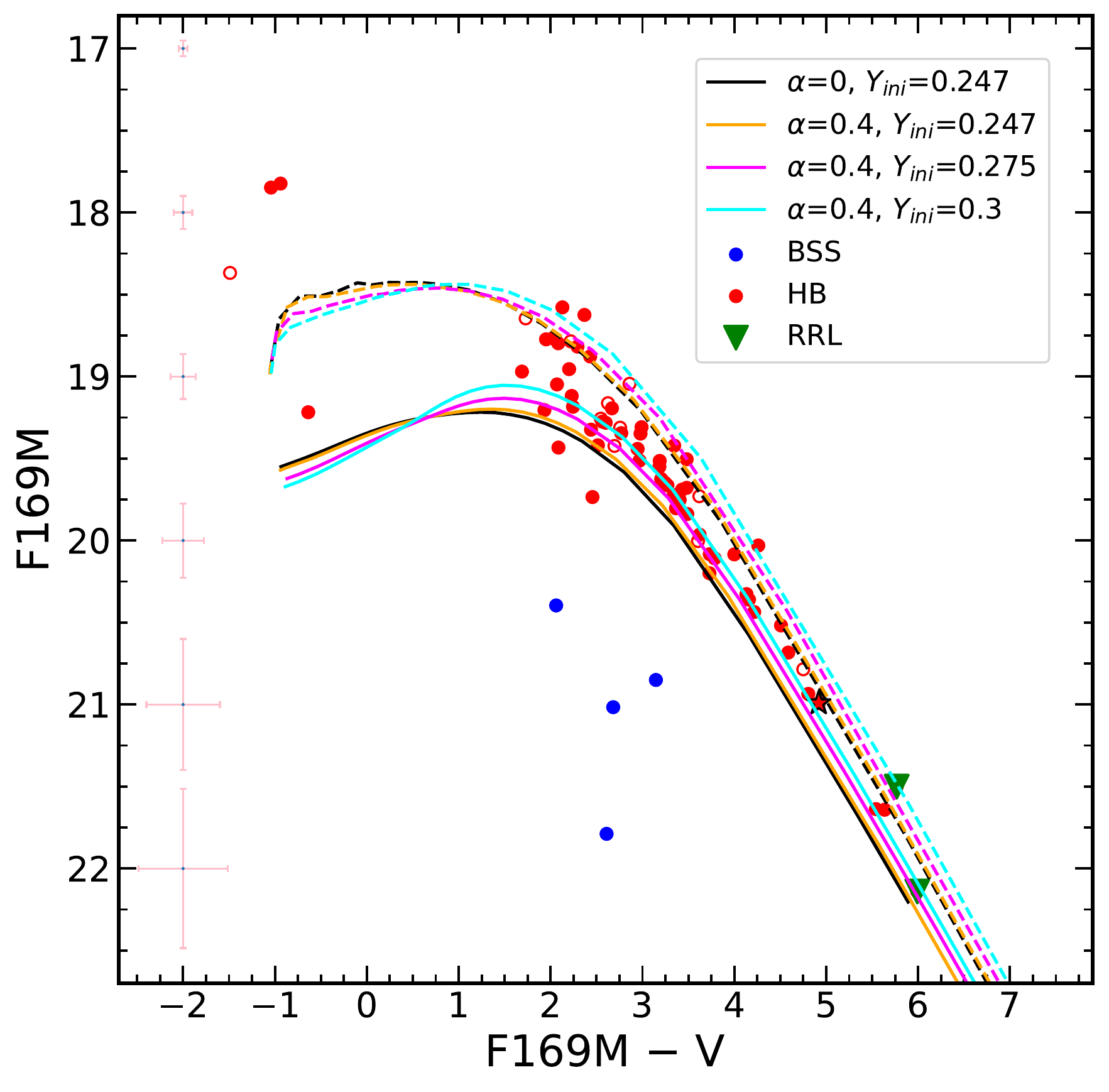} 
    \end{minipage}
	\caption{FUV-optical CMDs of NGC\,2298. The red filled and open symbols indicate the HB stars detected with UVIT in the inner and outer region of the cluster, respectively. The meaning of different colors and symbols is displayed in the panels. The photometric errors in magnitude and color are also shown in each panel. The solid black line on the HB locus is the ZAHB and the dashed one represents the TAHB for \big[$\alpha$/Fe\big] = 0 and normal helium abundance. The solid and dashed lines, shown with orange, magenta and cyan color, correspond to alpha-enhanced and helium-enhanced ZAHB and TAHB tracks.}
    \label{uvcmds}
\end{figure*}

\subsection{UV and UV-optical CMDs}
\label{sec:fuvcmds}
This section presents the FUV-optical CMDs generated by cross-identifying common stars between optical and our FUV detections. As only hot and bright stars have detectable emission in the FUV, we have cross-matched FUV detected sources with the above-selected sample of HB and BSSs in the cluster's inner and outer regions using optical passbands. The Vega magnitude system used in the standard Johnson-Cousins photometric system is converted into the AB magnitude system to make it similar to the UVIT magnitude system. 

 \begin{table}[!htb]
	\caption{The number of detected HB and BS stars in different UVIT filters are listed in this table. Here $N_{HB}$ and $N_{BSS}$ indicates number of detected HB and BS stars, respectively. Column 2 displays the sub-populations of HB where  $N_{BHB}$, $N_{Hot-HB}$ and $N_{RRL}$ denote the number of BHB, hot HB and RRL identified with UVIT. The total number of selected HB and RRL stars from an optical CMD are shown in parentheses.
	}
	\label{tab2}
	\hspace{-1.0cm}
\begin{tabular}{c|ccc|c} 
		\hline
		\hline
		 Filter &  & $N_{HB}$ & & $N_{BSS}$\\
		& $N_{BHB}$ & $N_{Hot-HB}$ & $N_{RRL}$ &\\
		\hline
		 F148W & 63(68) & 4(4) & 3(4) & 4\\ 
		 F154W & 60(68) & 4(4) & 1(4) & 3\\ 
		 F169M & 63(68) & 4(4) & 2(4) & 4\\
		\hline
	\end{tabular}
\end{table} 

In Figure~\ref{uvcmds}, three panels display the FUV-optical CMDs created using all the three FUV filters, where filled and open markers represent the detections in the inner and outer regions, respectively. The photometric error bars shown in all panels of Figure~\ref{uvcmds} are the median of the photometric errors of stars at a selected magnitude. 
 We have also recovered already known RR Lyrae variables displayed with filled and open green inverted triangles in all FUV-optical CMDs ( See Figure ~\ref{uvcmds}). In our FUV images, the variable stars are basically sampled at random phase. Out of four previously known RR Lyrae stars, we found 3 in F148W, 1 in F154W, and 2 in F169M filter. Their positions in the optical and FUV-optical CMDs are shown in Figures~\ref{optcmds} and ~\ref{uvcmds}, respectively.
 We notice that fewer HB stars are detected in the outer region of the cluster when compared to the inner region. 
 
\begin{figure*}[htb]
\centering
	\includegraphics[width=0.4\columnwidth]{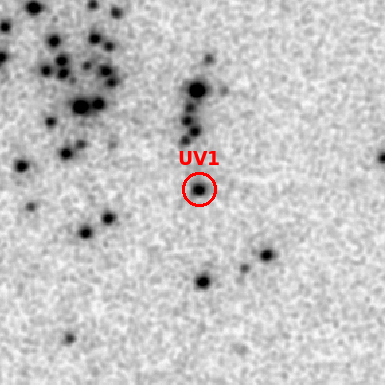}
	\includegraphics[width=0.4\columnwidth]{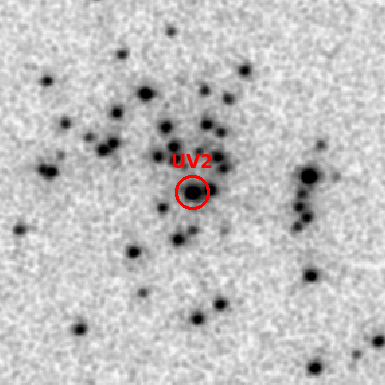}
	\includegraphics[width=0.4\columnwidth]{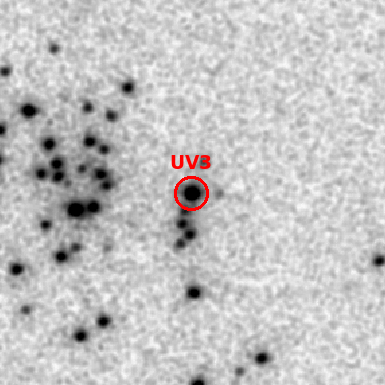}
	\includegraphics[width=0.4\columnwidth]{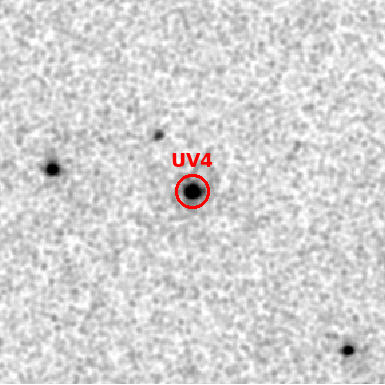}
	\includegraphics[width=0.4\columnwidth]{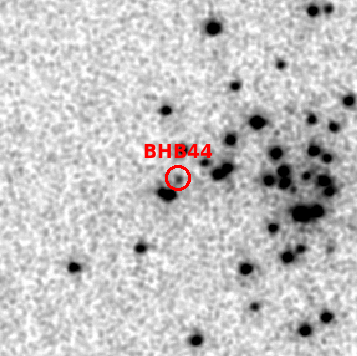}
	\caption{Location of hot HB and BHB44 stars on FUV F148W image of UVIT. The field of view of each image is 2$'$ x 2$'$.}
    \label{locuv}
\end{figure*}

  In all CMDs, we detect four stars at the extreme blue end of the HB. These stars are quite separated from the observed HB sequence and are likely to be very hot, as suggested by their FUV-optical color. Out of the four stars, three are found to be brighter and bluer than normal BHB and EHB stars. The three bluer stars might be post-HB (pHB) stars, as evident from their FUV magnitudes in all FUV bands. 
 These hot HB stars are confirmed PM members of the cluster and are well resolved in all FUV images, as shown in Figure~\ref{locuv}. The total number of detected BHB, hot HB, BSSs, and variable stars in all the UVIT filters are tabulated in Table~\ref{tab2}.  In the FUV-optical CMDs, we observe that the HB stars no longer lie along the horizontal sequence as found in the optical CMDs; instead, they follow a diagonal sequence with a significant spread.  
 In all FUV CMDs, we also detect a few (up to 4) FUV bright BSSs.\\
 
 The optical and FUV-optical CMDs overlaid with updated BaSTI-IAC isochrones and HB tracks are shown in Figures~\ref{optcmds} and ~\ref{uvcmds}, respectively \citep{2018ApJ...856..125H}. The updated BaSTI-IAC\footnote{\url{http://basti-iac.oa-abruzzo.inaf.it/}} isochrones are considered for an age 13.2 Gyr \citep{2018ApJ...865..160M}, a distance modulus of 15.75 mag \citep{2018ApJ...865..160M}, and a metallicity \big[Fe/H\big] = $-$1.92 dex \citep{2009A&A...508..695C} with helium abundance \textit{$Y_{ini}$} = 0.247, \big[$\alpha$/Fe\big] = 0, encompassing overshooting, diffusion, and mass loss efficiency parameter $\eta$ = 0.3. The BaSTI-IAC model also provides the HB model, which  comprises zero age HB (ZAHB), post-ZAHB tracks, and end of the core-helium-burning phase known as terminal age HB (TAHB) with or without diffusion for a particular mass range. We have generated the ZAHB and TAHB tracks for a metallicity \big[Fe/H\big] = $-$1.92 dex, including diffusion happening in the sub-atmospheric regions of these stars.\\

\begin{figure*}[htb]
	\includegraphics[width=0.7\columnwidth]{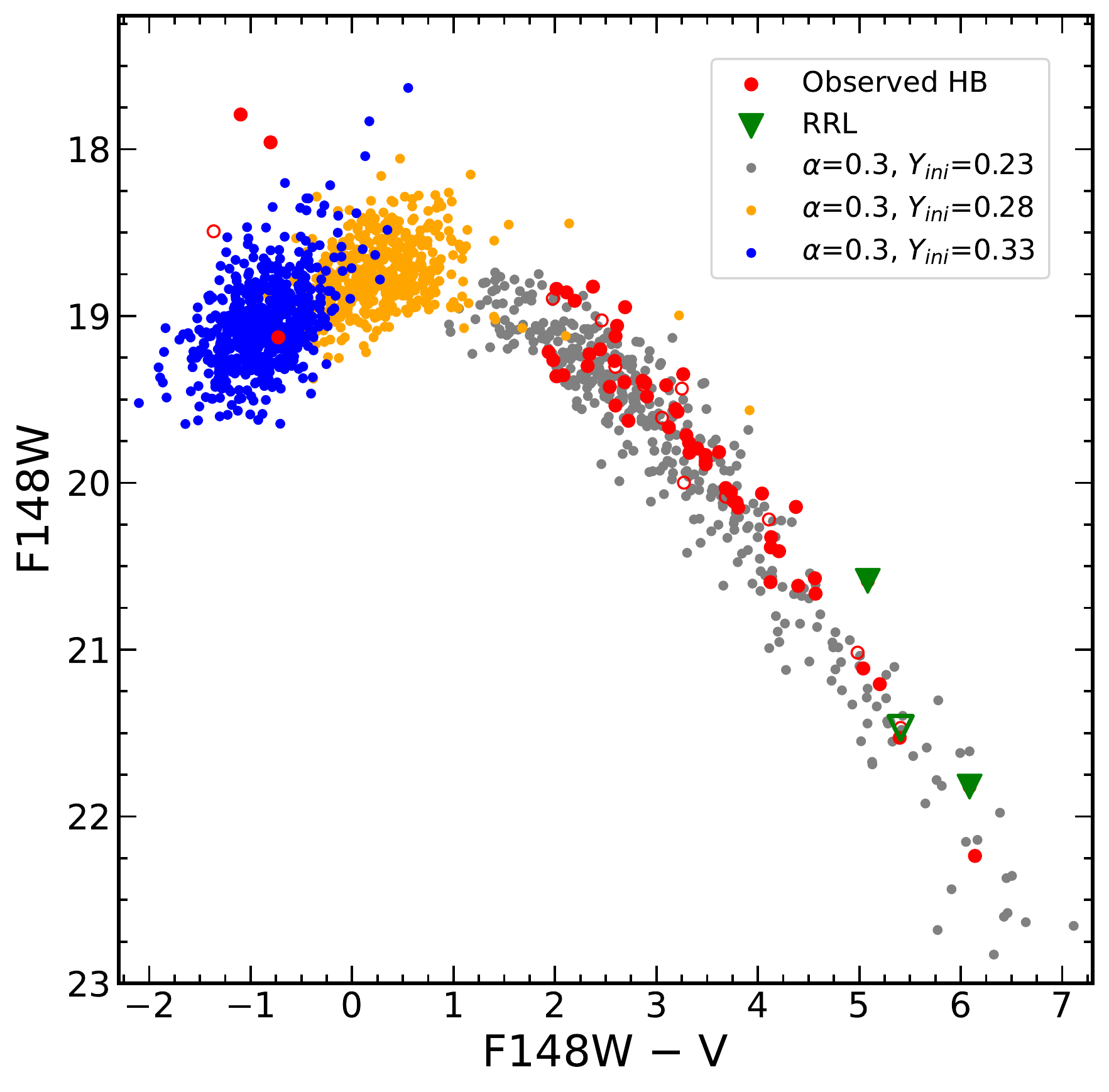}
	\includegraphics[width=0.7\columnwidth]{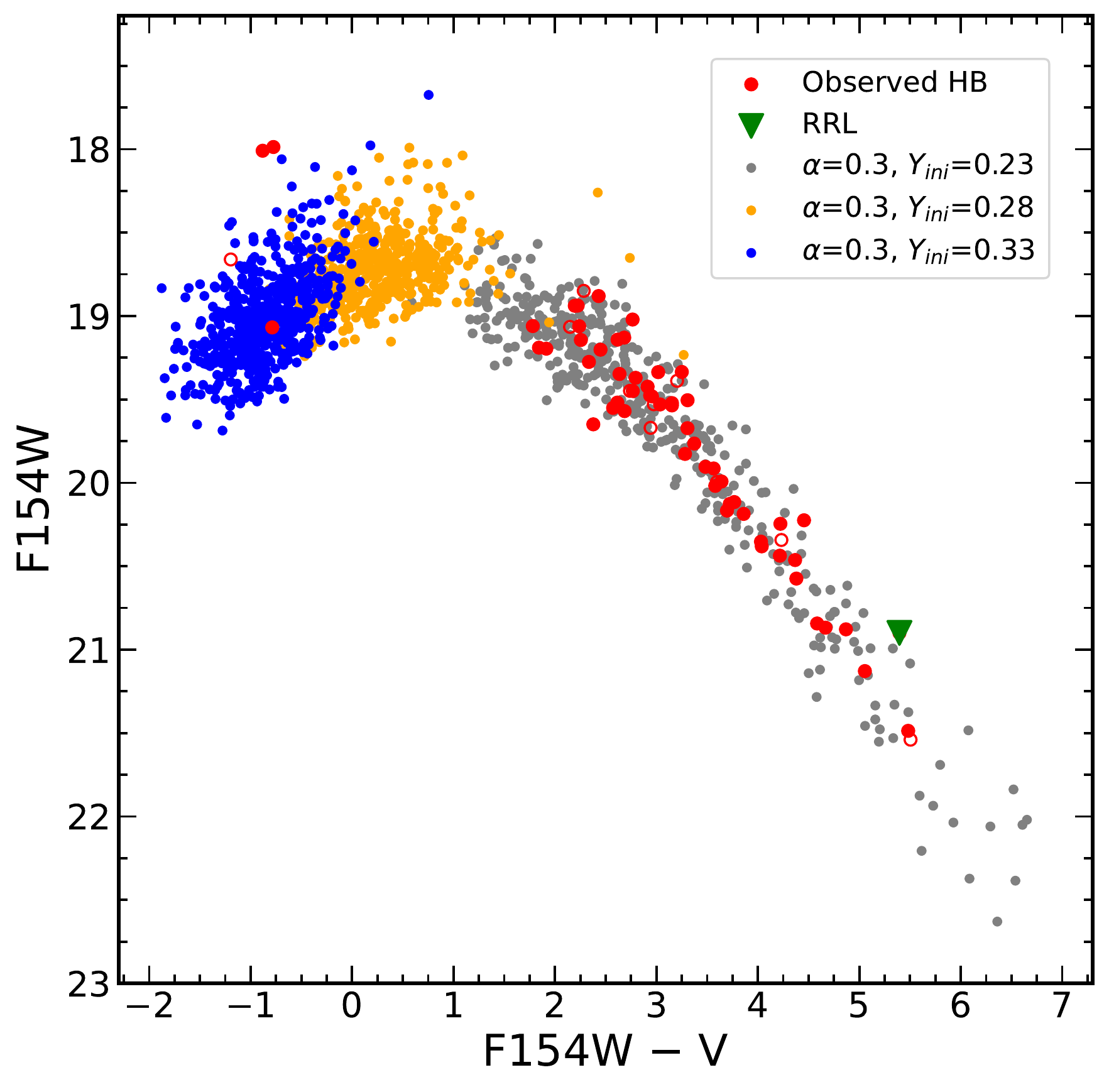}
	\includegraphics[width=0.7\columnwidth]{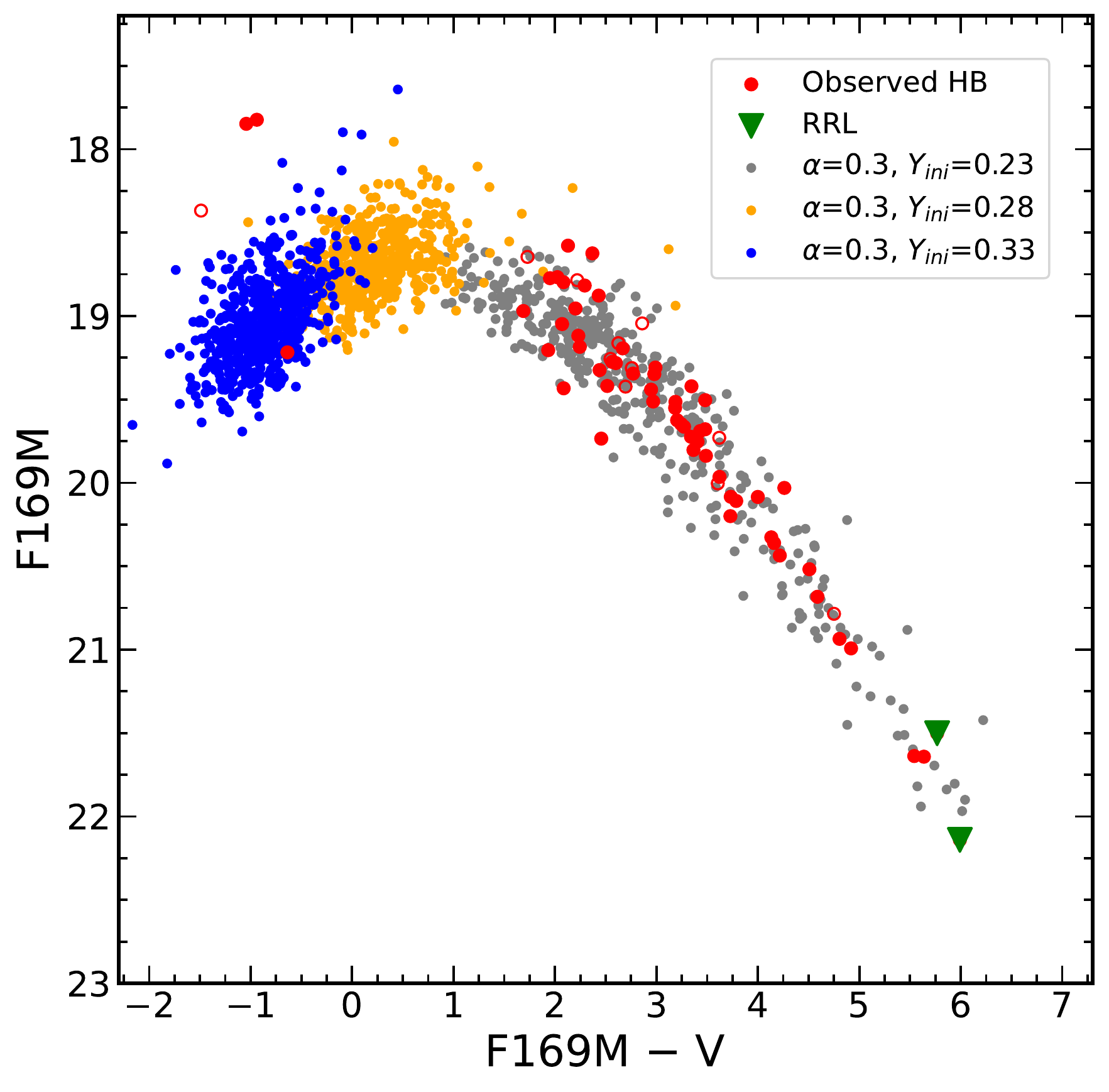}
	 \caption{FUV-optical CMDs showing the comparison of the observed HB with synthetic HB models. The meaning of red and green symbols is same as in Figure~\ref{uvcmds}. The simulated HB populations for initial helium abundance \textit{$Y_{ini}$} = 0.23, 0.28 and 0.33 are marked with gray, orange and blue dots, respectively. }
    \label{simhb}
\end{figure*}

     The overlaid HB tracks in FUV CMDs helps in defining the location of HB stars, which span an extensive range in color and magnitude, when compared to the optical CMDs. The HB sequence is not well fitted with the theoretical HB models at the brighter end (above 20 mag) with normal helium abundance (\textit{$Y_{ini}$} = 0.247 dex) shown with black color solid and dashed lines. 
     The color/magnitude spread at brighter FUV magnitudes might be related to chemical composition differences among BHB stars or the evolutionary effects from ZAHB or photometric errors. In order to probe the cause of this spread, we have compared observations with theoretical HB models generated for enhanced \big[$\alpha$/Fe\big]=0.4 dex and different initial helium abundances available in the updated BaSTI-IAC database, i.e. \textit{$Y_{ini}$} = 0.247, 0.275, and 0.3. The ZAHB and TAHB tracks corresponding to these initial helium abundances are indicated with different color solid and dashed lines in Figure~\ref{uvcmds}, respectively. We find that ZAHB tracks with different initial helium abundances are not producing the color/magnitude spread observed along the BHB sequence in all FUV-optical CMDs. We also notice that HB stars are located between the ZAHB and TAHB tracks, suggesting that some may be evolving from the HB. 
     Out of four hotter HB stars, three stars are found to be brighter than TAHB tracks, indicating that they are in the pHB phase, whereas one is lying within the ZAHB and TAHB tracks implying that it is an EHB star. In order to confirm the nature of these stars further, effective temperatures and bolometric luminosities are measured using the spectral energy distribution (SED) fitting technique described in Section~\ref{sec:SEDs}.
     
\begin{figure*}[htb]
	\includegraphics[width=0.71\columnwidth]{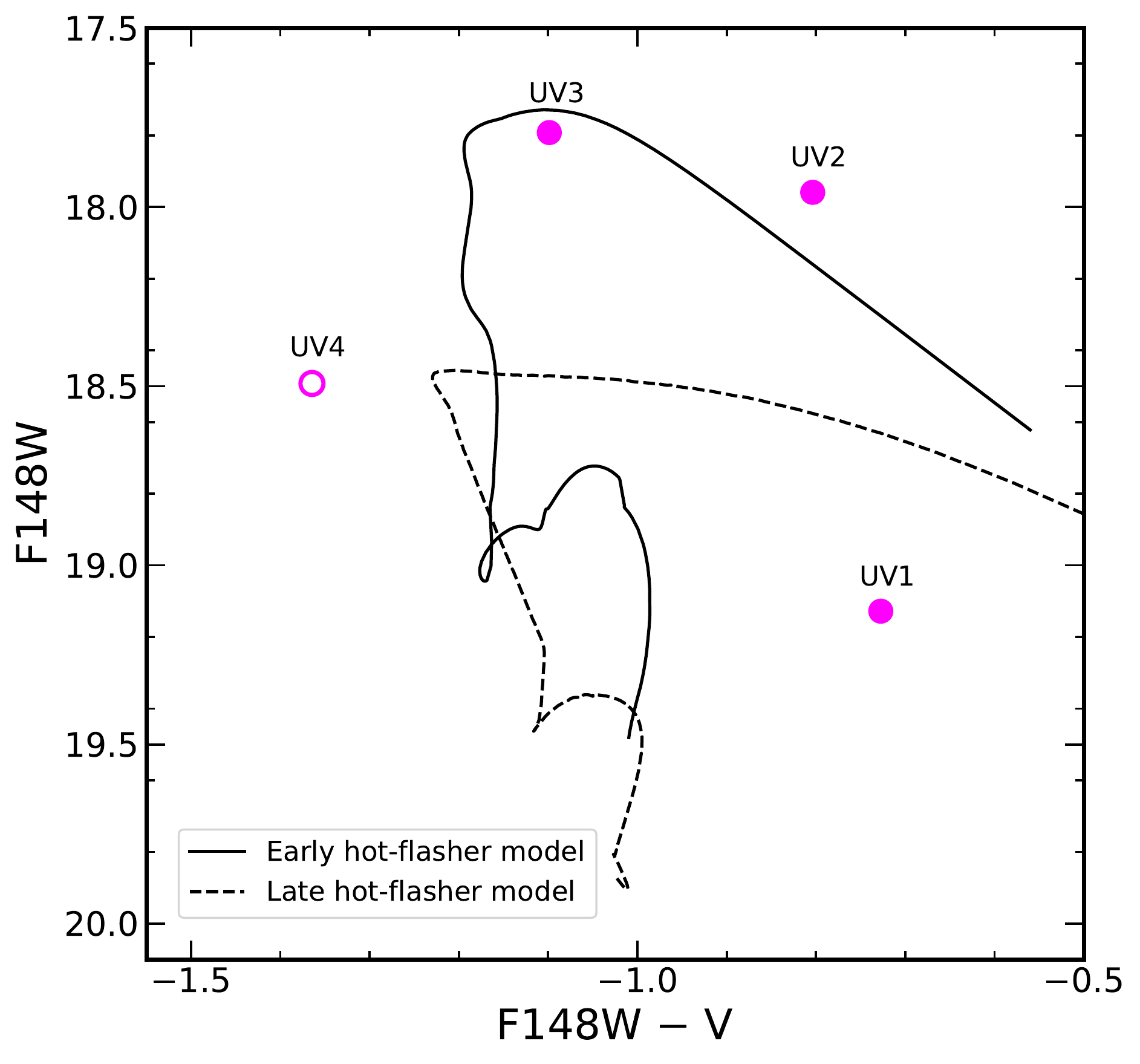}
	\includegraphics[width=0.71\columnwidth]{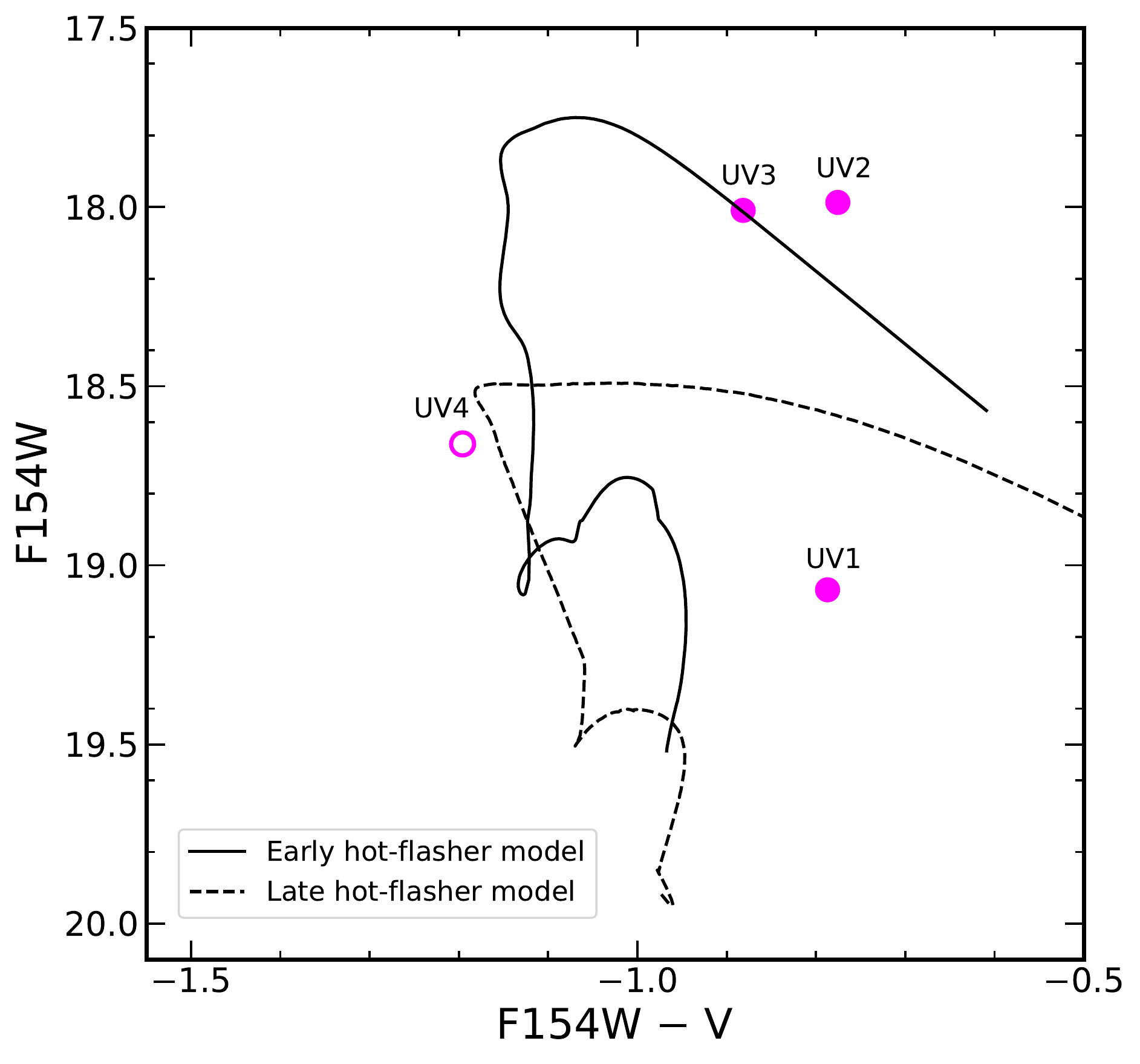}
	\includegraphics[width=0.7\columnwidth]{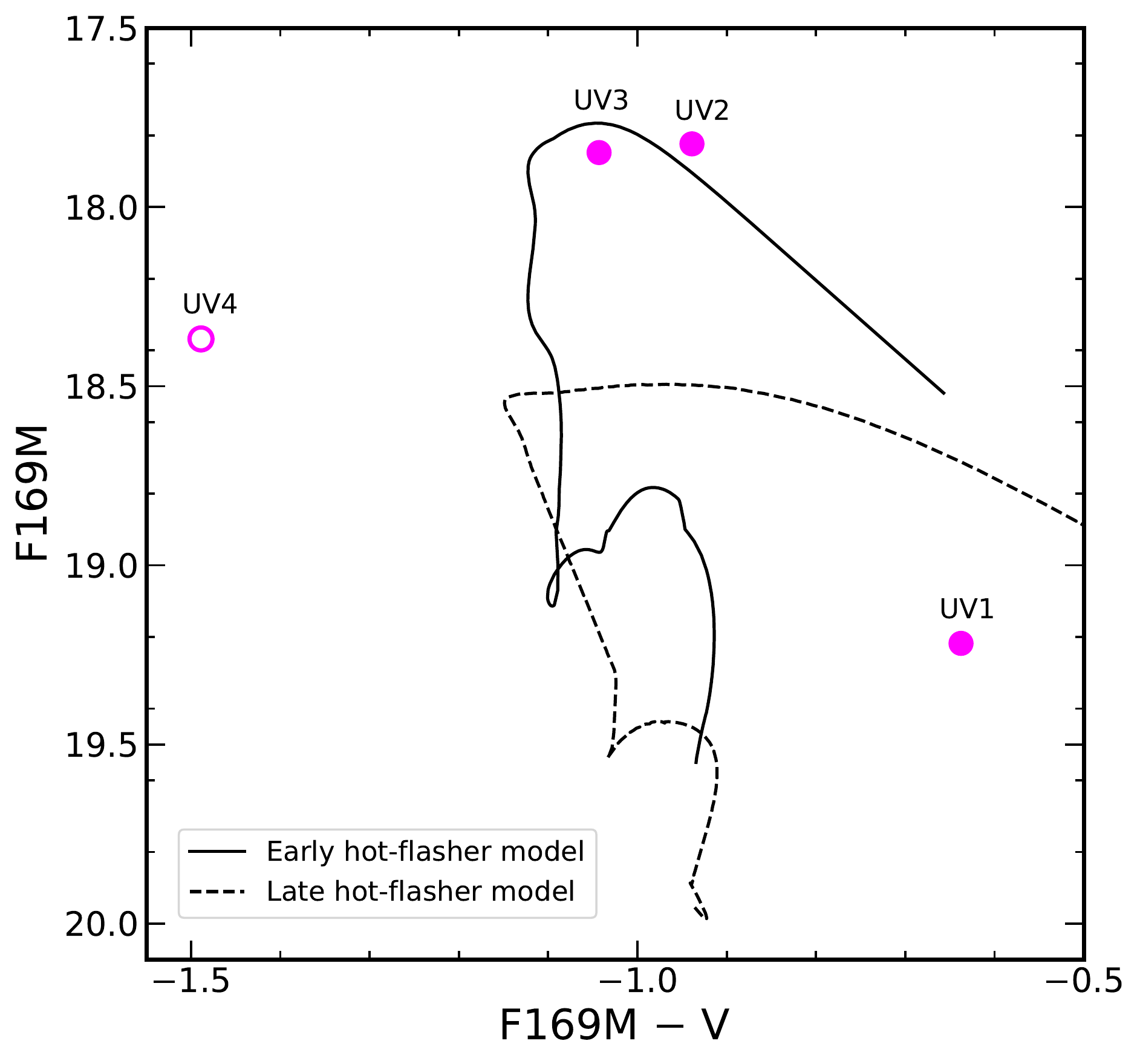}
	
     \caption{FUV-optical CMDs overplotted with hot-flasher models where solid and dashed black line correspond to the early and late hot-flasher models, respectively. The four hot HB stars are displayed with magenta symbols. The filled symbols represent stars within the HST FOV and the open symbol for the star in the outer region.}
    \label{flashermodels}
\end{figure*}

\begin{figure*}[!htb]
\centering
\includegraphics[width=0.89\columnwidth]{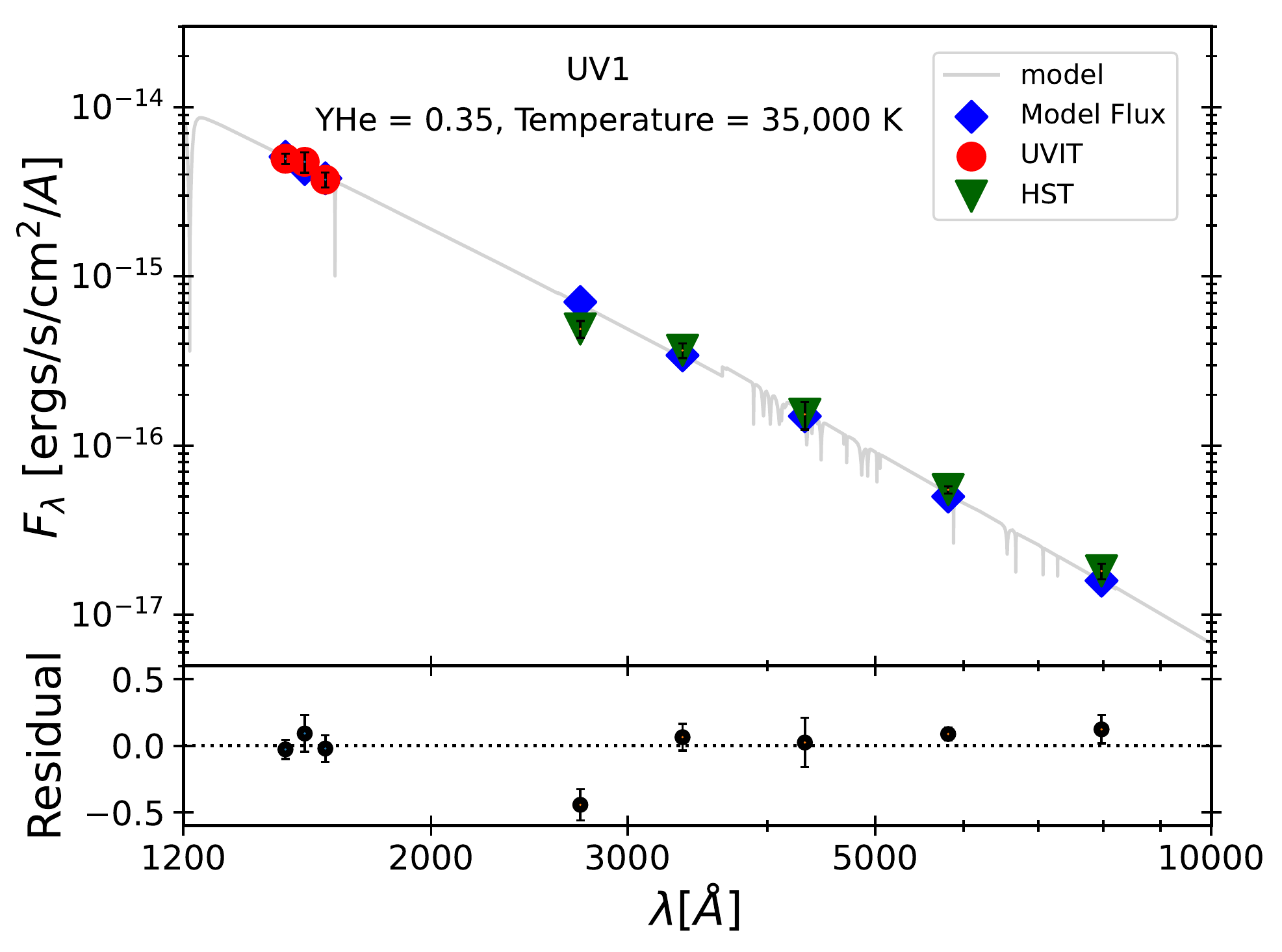}
\includegraphics[width=0.89\columnwidth]{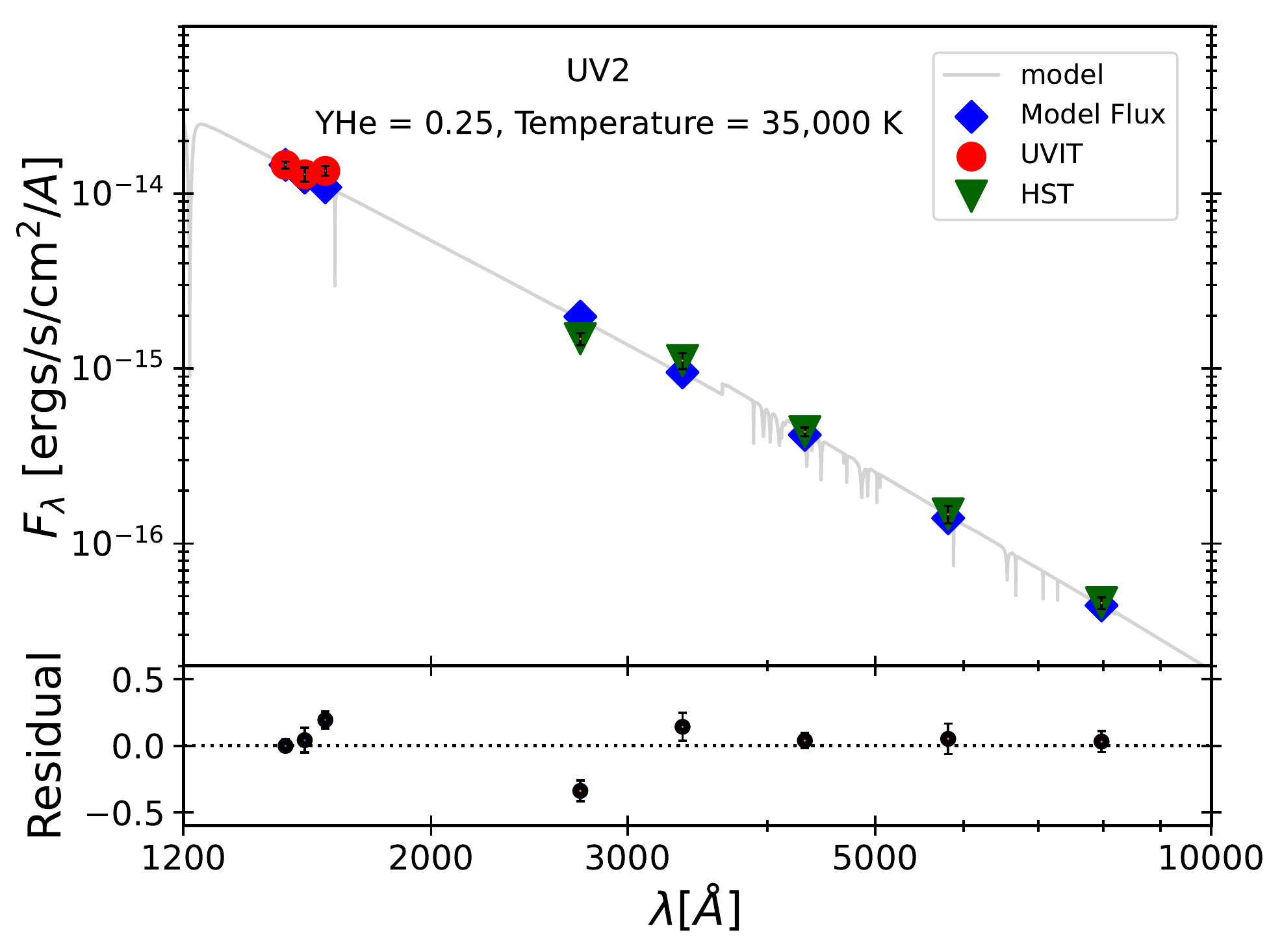}
\includegraphics[width=0.89\columnwidth]{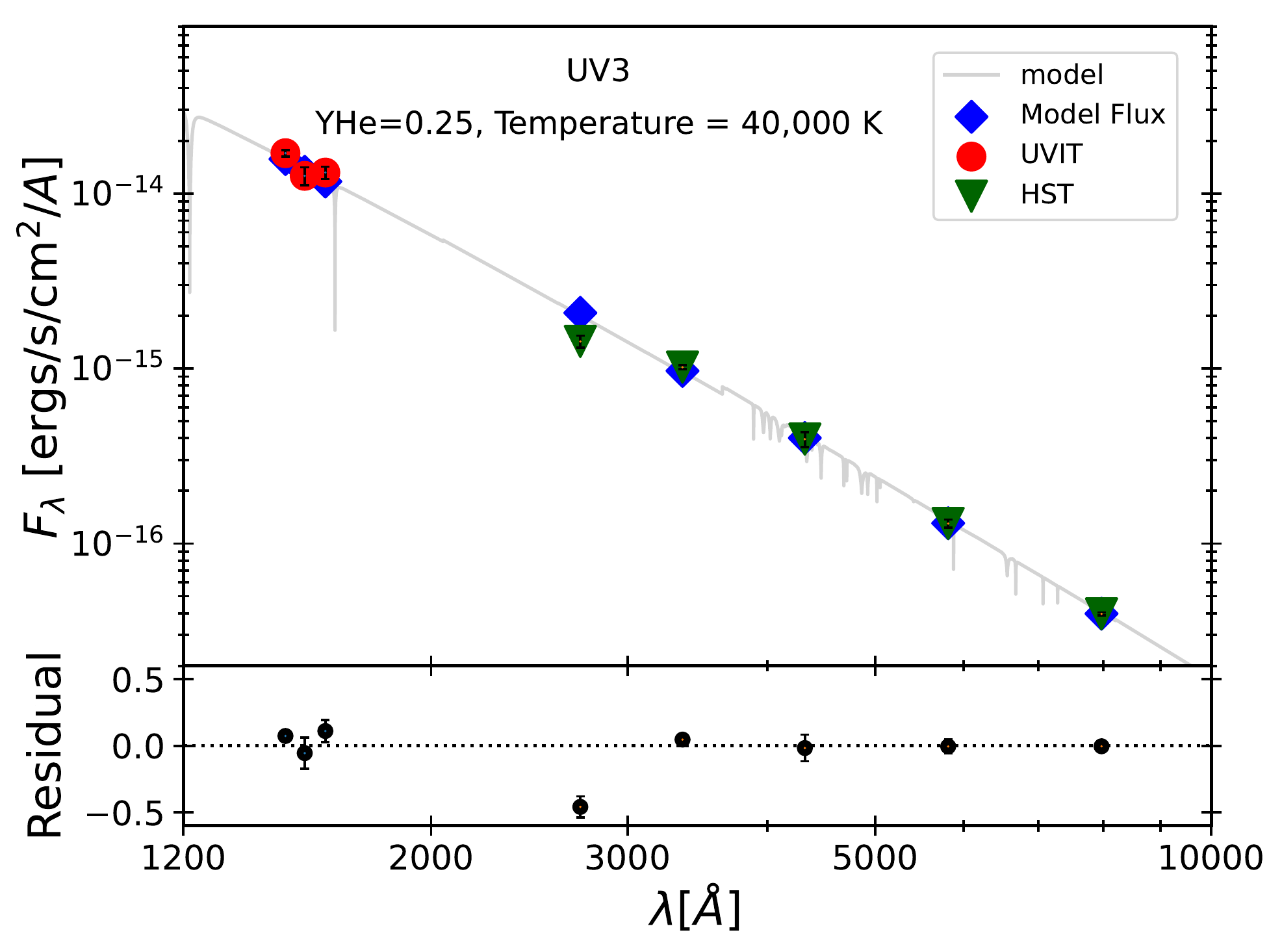}
\includegraphics[width=0.89\columnwidth]{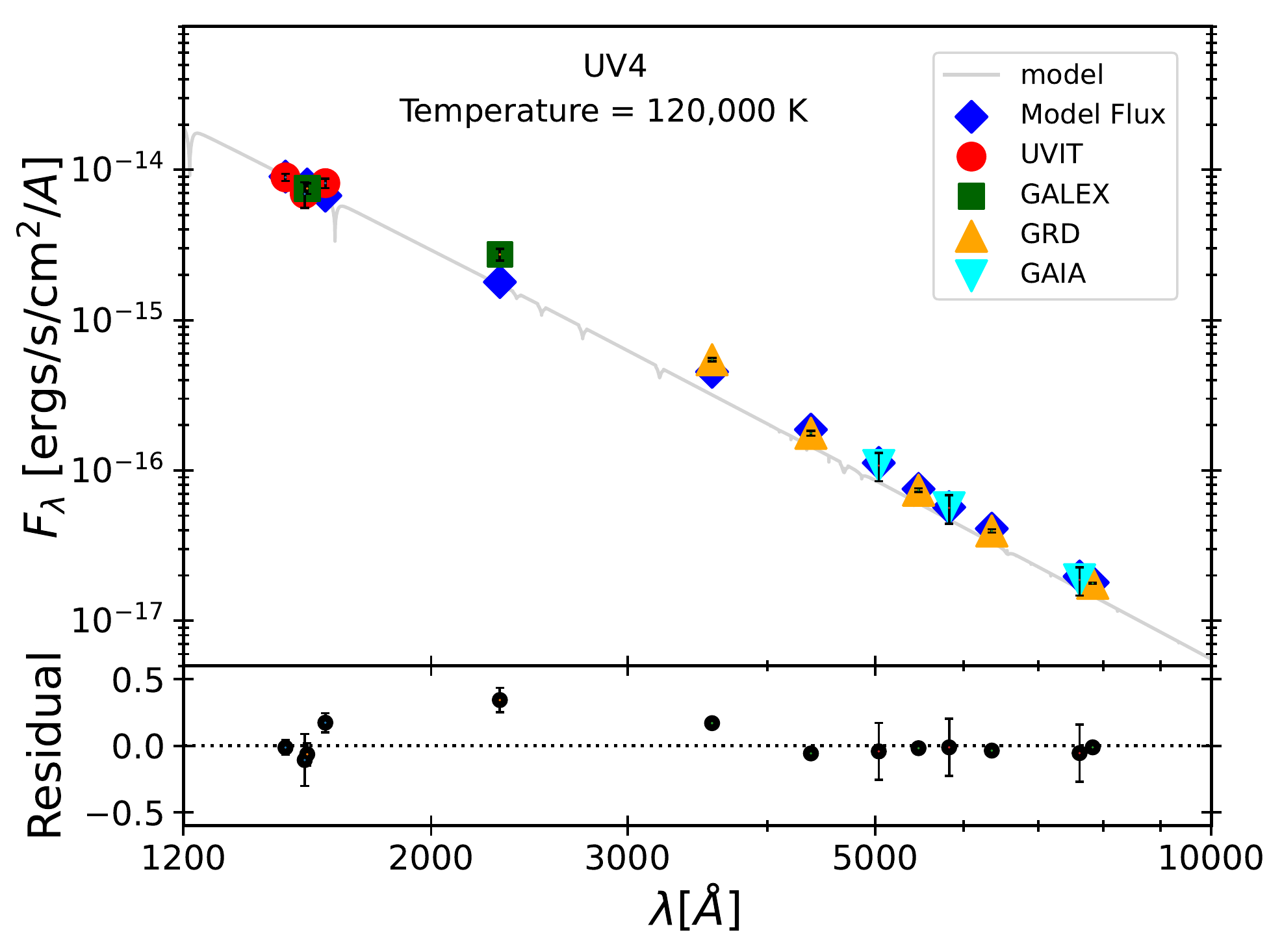}
 \caption{SEDs of four hot HB stars detected with UVIT after correcting for extinction. The stars, denoted with UV1, UV2 and UV3, are observed in the inner region of the cluster and UVIT (Red) and \textit{HST} (Green) photometric data points are used to create and fit the SED, whereas for UV4, apart from UVIT, \textit{GALEX}, \textit{Gaia} EDR3 and ground-based photometric data points are utilized as it is identified in the outer region. Blue diamonds represent the synthetic flux from the helium-rich Husfeld model used to fit the observed SED of UV1, UV2 and UV3 stars, whereas in UV4, they correspond to TMAP (Grid4) model. The best fit atmospheric parameters are mentioned in the figure. In UV1, UV2, and UV3 stars, the light gray solid line represents the theoretical helium-rich Husfeld model spectra, and TMAP (Grid4) model spectra in UV4. The residuals of the SED fit are presented in the bottom panel of all plots.}
 \label{ehbseds}
\end{figure*}

\section{Comparison with Models}
\label{sec:models}
\subsection{Synthetic Helium HB Models}
\label{sec:synchbmodels}
We constructed the synthetic HB stellar populations for enhanced \big[$\alpha$/Fe\big]=0.3 and different initial helium abundances to check for the spread in helium abundance as well as evolution. \cite{2017ApJ...842...91C} demonstrated the implications and prospects for the helium-enhanced populations in relation to the second-generation populations found in the Milky Way GCs using Yonsei Evolutionary Population Synthesis (YEPS) model. The synthetic HB models presented here are based on Yonsei–Yale ($Y^{2}$) stellar evolutionary tracks with enhanced initial helium abundance \citep{2015HiA....16..247L}. We choose three values for \textit{$Y_{ini}$} as 0.23, 0.28, and 0.33, at fixed Z value of 0.0002 (\big[Fe/H\big]=-1.9 dex) and age of 13 Gyr (close to the cluster age of 13.2 Gyr). Evolutionary  effects from ZAHB and observational photometric errors are taken into account. 
Figure~\ref{simhb} displays the synthetic CMDs for three different \textit{$Y_{ini}$} values 0.23 (gray), 0.28 (orange) and 0.33 (blue), overlaid on the observed FUV-optical CMD where observed HB stars are highlighted in red. We notice that observations match well with synthetic HB models, especially in the case of BHB stars. It is clear from the comparison of synthetic HB models with observations that all BHB stars have the same helium abundance, implying that BHBs consist of a single \textit{$Y_{ini}$}=0.23. Therefore, we suggest that color/magnitude spread among BHB stars is not caused by helium variation.
Out of four hot HB stars, one star is found to have \textit{$Y_{ini}$}=0.33, and another at the brighter
extension appears to be the product of enhanced helium, which, in turn, implies that these stars are originated from helium enhanced populations. The other two hot HB stars are brighter than the simulated stars  for \textit{$Y_{ini}$}=0.33, suggesting that they are in an evolved stage, might be the progeny of helium enhanced EHBs in this cluster. Therefore, using synthetic HB models, we estimate a \textit{$Y_{ini}$}=0.23 for BHB stars and four hot BHB stars are likely to have \textit{$Y_{ini}$}=0.33. Thus, enhanced helium abundance seems to play a role in the formation of hot HB stars in this cluster. 

\begin{figure*}[!htb]
\centering
\includegraphics[width=0.89\columnwidth]{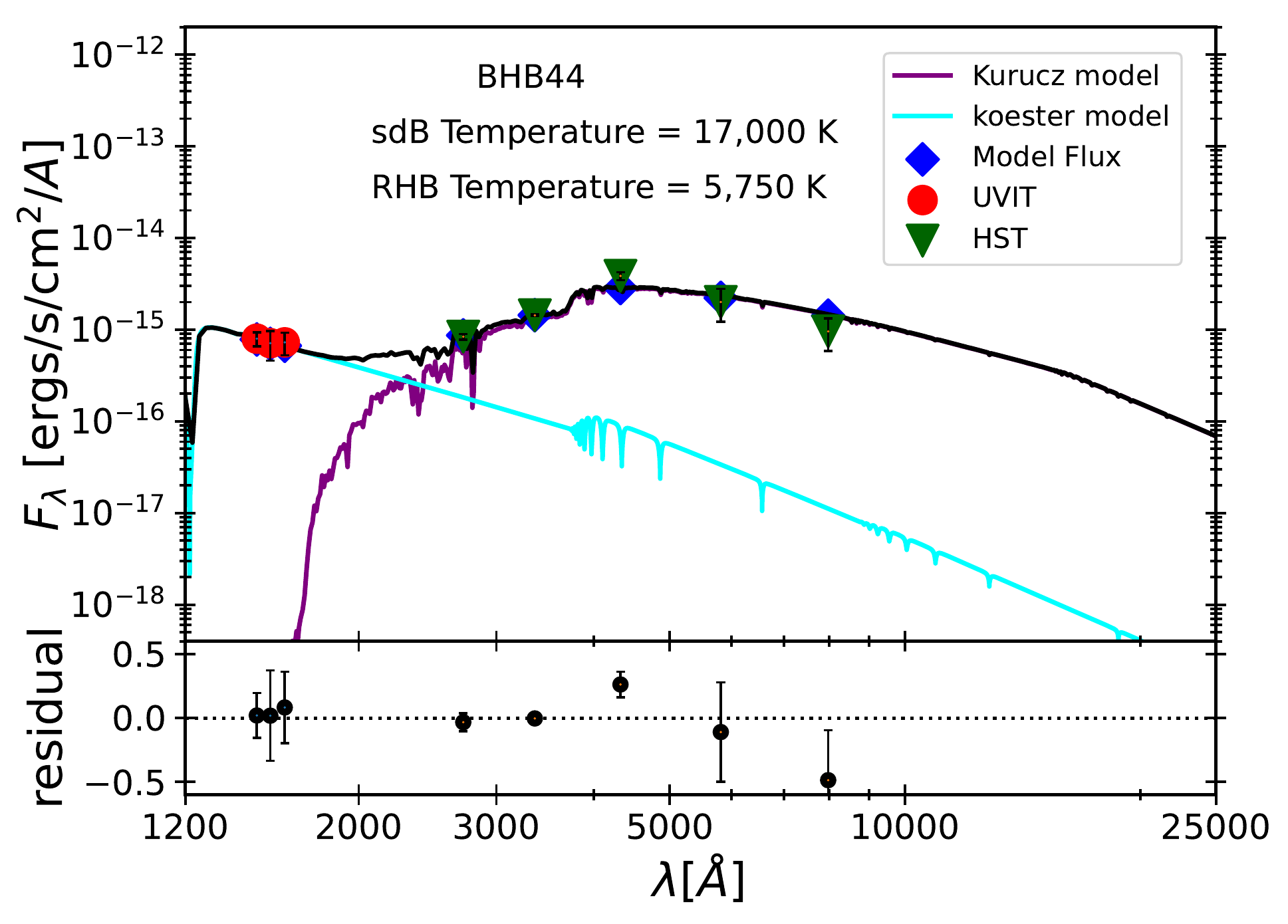}
\includegraphics[width=0.89\columnwidth]{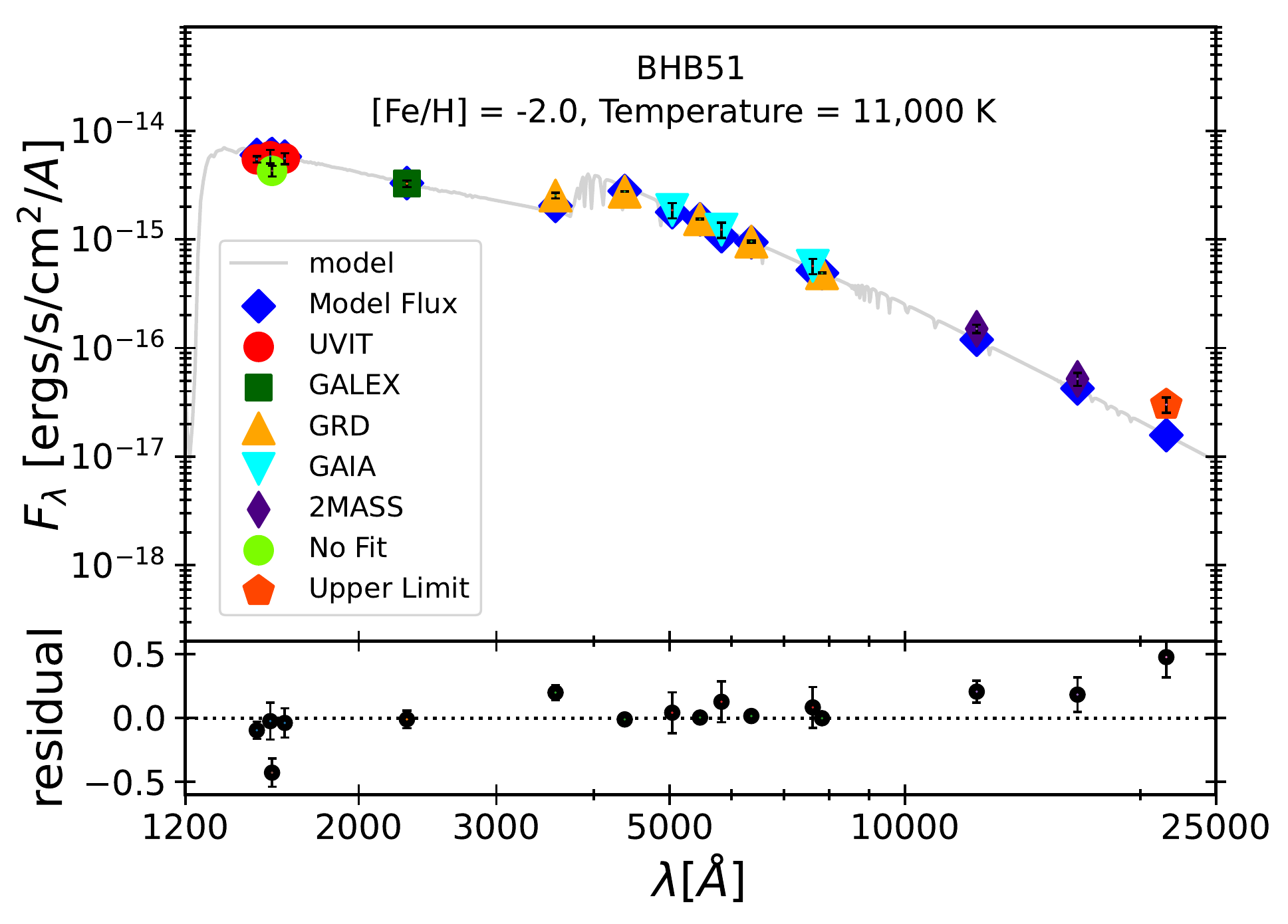}
 \caption{SEDs of BHB stars detected with UVIT in central (Left Panel) and outer (Right Panel) region after correcting for extinction. Left panel shows the composite SED fit of BHB44 star where purple and cyan color spectra indicate Kurucz and Koester model spectra as displayed in legend, respectively. Black color represents the composite spectra. The best fit parameters are displayed in the figure. The rest of the details are same as in Figure~\ref{ehbseds}.}
 \label{bhbseds}
\end{figure*}

\subsection{Hot Flasher Models}
\label{sec:hotflashermodels}
The location of very hot HB sub-populations known as BHk stars in FUV-optical CMDs is not reproduced by the canonical HB models; this sub-population is hotter than the hottest point along the ZAHBs  (See Figure 8 in \cite{2011MNRAS.410..694D}). The proposed formation scenarios for very hot HB populations such as EHB and BHk stars in clusters are extreme mass loss on RGB, helium enrichment, helium mixing, and hot-flasher. In a hot-flasher scenario, stars experience huge mass loss during the RGB phase, leave the branch before the occurrence of the helium core flash and move quickly to the He-core white dwarf cooling curve, where they experience a helium flash under conditions of strong electron degeneracy in their core  \citep{1993ApJ...407..649C}.\\

Further, depending upon the location of ignition of helium flash along the WD cooling sequence, hot-flasher models are classified into two types: early hot-flasher and late hot-flasher. The details of these scenarios are well described in \cite{2001ApJ...562..368B}. The progeny of hot-flashers end up on the hotter and bluer side of the normal BHB stars. The stars, which are products of the early hot-flasher scenario, are expected to be hotter and of similar UV magnitudes compared to BHB stars in UV CMDs known as EHB stars, whereas the stars that are hotter and fainter than normal EHB stars in UV CMDs are expected to be late hot flashers. Although we did not detect any BHk star in this cluster, there are three hot HB stars that are brighter than synthetic HB populations, and their location is not reproduced by ZAHB and TAHB tracks.

In order to check whether the hot HB stars are the product of 
strong mass loss, we have compared observations with hot-flasher models, i.e., early hot-flasher and late hot-flasher models. The hot-flasher models were generated in UVIT and \textit{Gaia} EDR3 filters \citep{2003ApJ...582L..43C}(Cassisi 2021, private comm.). The hot-flasher models superimposed on the observed hot HB stars are presented in Figure~\ref{flashermodels}. These four stars are marked with the names UV1, UV2, UV3, and UV4 in this study, as shown in  Figure~\ref{flashermodels}. Two stars, namely UV2 and UV3 are found to lie close to the early hot-flasher model in all three FUV-optical CMDs, suggesting that they are the progeny of EHB stars formed through an early hot-flasher scenario. We notice that UV2 is slightly brighter than the hot-flasher model. UV4 is lying close to the late hot-flasher model (Dashed black line), only in (F154W, F154W$-$V) CMD, suggesting that it may be an evolved product of a BHk star. However, in the rest of the FUV-optical CMDs, its location is not reproduced by hot-flasher models. From this comparison, the nature of the UV4 star is not very clear. Any of the hot-flasher models do not reproduce the position of one EHB star designated with UV1, but it is well produced by helium enhanced synthetic HB models, suggesting that it could be a result of helium enrichment in the cluster. As the stars tend to shift location among the CMDs, locating them in the H-R diagram will be ideal for evaluating their evolutionary status. Further estimation of atmospheric parameters and comparison with theoretical models in the H-R diagram is required to probe these stars' evolutionary status in detail and shed more light on their nature.\\

\section{Spectral Energy Distribution fitting}
\label{sec:SEDs}
We have detected three hot HB stars in the central region and one in the outer region. As these stars are well separated from the rest of the HB, we aim to check the evolutionary status of these stars by estimating their stellar parameters. In order to estimate the physical parameters like effective temperature ({\it $T_{eff}$}), luminosity ($\frac{L}{L_{\odot}}$), and radius ($\frac{R}{R_{\odot}}$) of the hot HB stars, we constructed their SEDs. SEDs are generated with the observed photometric data points spanning a wavelength range from FUV-to-IR and fitted with selected theoretical models. We made use of the virtual observatory tool, VOSA (VO Sed analyzer, \citealp{2008A&A...492..277B}) for SED analysis. 
VOSA utilizes the filter transmission curves to calculate the synthetic photometry of the selected theoretical model. By using the fixed distance to the cluster, synthetic fluxes are scaled with the observed fluxes. After constructing the SED, it performs a $\chi^2$ minimization test to compare the observations with the synthetic photometry to derive the best-fit parameters. We computed reduced $\chi^2_{red}$ through the following expression:

\begin{equation}
\hspace*{0.0cm}
  \chi^2_{red} = \frac{1}{N-N_{f}} \displaystyle \sum_{i=1}^{N} {\frac{(F_{o,i} - M_{d}F_{m,i})^2}{\sigma_{o,i}^2}}
\end{equation}

where N is the number of photometric data points; $N_{f}$ is the number of free parameters in the model; $F_{o,i}$ is the observed flux; $M_{d}F_{m,i}$ is the model flux of the star;
$M_{d} = \big(\frac{R}{D}\big)^2$ is the scaling factor corresponding to the star (where R is the radius of the star and D is the distance to the star) and $\sigma_{o,i}$ is the error in the observed flux. The number of photometric data points (N) for stars varies from 9 to 13 depending upon their detection in different available filters. The number of free parameters ($N_{f}$) used to fit SED are $Y_{He}$, log{\it g} and effective temperature {\it $T_{eff}$}. The radii of the stars were calculated using scaling factor, $M_{d}$.

The Kurucz stellar atmospheric models are employed to construct SEDs \citep{1997A&A...318..841C, 2003IAUS..210P.A20C} for HB stars, which have photometry ranging from UV to IR wavelengths. We fixed the value of metallicity \big[Fe/H\big] = $-$2.0, close to the cluster metallicity, and gave the range of {\it $T_{eff}$} from 5,000-50,000 K and log{\it g} from 3.5-5 dex in the Kurucz models to fit the SED of HB stars. The locus of ZAHB reflected in the {\it $T_{eff}$} versus log{\it g} plane is used to constrain the range of log{\it g} for HB stars.

We combined 3 FUV UVIT photometric data points  with 5 \textit{HST} photometric data points from \cite{2018MNRAS.481.3382N} to create SED for HB stars detected in the inner region of the cluster. For those detected in the outer region of the cluster, we combined the photometric data points of UVIT (3 passbands) with \textit{Gaia} EDR3 (3 passbands) \citep{2018A&A...616A..12G}, ground-based photometry (5 passbands) \citep{2019MNRAS.485.3042S}, \textit{GALEX} (2 passbands) and 2MASS (3 passbands). VOSA makes use of Fitzpatrick reddening law \citep{1999PASP..111...63F, 2005ApJ...619..931I} to correct for extinction in observed data points.

It is a well-known fact that HB stars hotter than 11,500 K are affected by atmospheric diffusion, that increases the atmospheric abundances of heavy elements like iron and reduces the atmospheric abundances of light elements. To take this effect into account, we fitted the SEDs of BHB stars with {\it $T_{eff}$} $>$ 11,500 K with solar metallicity models and determined their atmospheric parameters. 
As the late hot flasher scenario predicts enrichment in helium, the non-LTE helium-rich Husfeld models \citep{1989A&A...222..150H} are used to fit the observed SEDs of four hot HB stars. The model grid covers the range of stellar parameters typical of extremely helium-rich sdO stars: $35,000 K \leq T_{eﬀ} \leq 80,000 K$, $4.0 \leq log{\it g} \leq 7.0$, and $0.01 \leq Y_{He} \leq 0.7$. In the case of UV4 star, we have noticed that {\it $T_{eff}$} derived using the Kurucz and helium-rich Husfeld model fits to the observed SED corresponds to their upper limit, which indicates that this star is likely to be hotter than the estimated temperature from these models. In order to compute the accurate {\it $T_{eff}$} of this star, we have fitted its SED with the T\"ubingen NLTE Model Atmosphere Package (TMAP) (Grid4) model used for hot stars \citep{2003ASPC..288..103R, 2003ASPC..288...31W}. This model grid spans a range of atmospheric parameters such as $20,000 K \leq T_{eﬀ} \leq 150,000 K$, $4.0 \leq log{\it g} \leq 9.0$, and $0 \leq X_{H} \leq 1$. Out of three models used for the SED fit of UV4 star, the TMAP (Grid4) model gives the best fit as indicated from the smaller $\chi^2_{red}$ value tabulated in Table~\ref{tab3}.

\begin{table*}
    \begin{center}
    \small
     \caption{Atmospheric parameters derived from SED fitting of four hot HB stars detected with UVIT in NGC\,2298. Column 1 lists the star ID used in this work. Columns 2 and 3 display the RA, DEC of all the stars considered for fitting, respectively. Column 4 presents the different model used for SED fit of these stars. Column 5 and 6 lists the obtained helium mass fraction and {\it $T_{eff}$} from SED fitting using different theoretical models. The luminosities and radii of these stars along with errors are tabulated in columns 7 and 8, respectively. The column 9, and 10 lists the reduced $\chi^{2}$ value corresponding to the best fit and ratio of the number of photometric data points ($\frac{N_{fit}}{N_{tot}}$) used for the fit to the total number of available data points.}
    \label{tab3}
	\begin{tabular}{cccccccccc} 
		\hline
		\hline
		 Star ID & RA & DEC & Model Used & $Y_{ini}$ &{\it $T_{eff}$} & $\frac{L}{L_{\odot}}$  & $\frac{R}{R_{\odot}}$ & ${\chi}_{red}^2$ & $\frac{N_{fit}}{N_{tot}}$\\
		 & (deg) & (deg) & & (dex) & (K) & & & &\\
		\hline
		 UV1* &	102.2541 & -35.99134 & Husfeld & 0.35 & 35,000 (35,000-40,000) & $36.14 \pm 0.12$ &	$0.17 \pm 0.003$ & 5.1 & 8/8\\
		  & & & Kurucz & & $29,000 \pm 2,000$ & $27.58 \pm 0.12$ &	$0.2 \pm 0.004$ & 2.8 & 8/8\\
         UV2* & 102.2512 & -36.00361 & Husfeld & 0.25 & 35,000 (35,000-40,000) & $103.9 \pm 0.24$ & $0.28 \pm 0.005$ & 7.8 &	8/8\\
         & & & Kurucz & & 32,000$^{+4,000}_{-2,000}$ & $86.39 \pm 1.01$ & $0.3 \pm 0.006$ & 5.7 &	8/8\\
         UV3 &	102.2446 & -35.99501 & Husfeld & 0.25 & 40,000 (40,000-60,000) & $128.3 \pm	0.26$ & $0.24 \pm	0.004$ & 10.4 & 8/8\\
         & & & Kurucz & & 50,000$^{}_{-7,000}$ & $235.7 \pm	1.06$ & $0.2 \pm 0.004$& 8.7 & 8/8\\
         UV4* & 102.2379 & -36.03683 & TMAP (Grid4) & & 120,000 (120,000-150,000) & $1148 \pm 303.4$ &	$0.08 \pm 0.002$ & 7.8 & 13/13\\
           & 102.2379 & -36.03683 & Husfeld & 0.35 & 80,000 (75,000-80,000) & $469.2	\pm	17.9$ &	$0.11 \pm 0.002$ & 10.7 & 13/13\\
         & & & Kurucz & & 50,000$^{}_{-9,000}$ & $108.5 \pm	5.05$ &	$0.14 \pm 0.003$ & 17.4 & 11/13\\
    \hline
    \multicolumn{10}{p{\textwidth}}{Note that stars marked with $^{*}$ symbols have their estimated temperature corresponding to the best-fit SEDs and equal to the helium-rich model's lower or upper limit. The range of temperatures for ten best fits is also mentioned in the parentheses. The other atmospheric parameters are listed according to the best-fit model.}
	\end{tabular}
	 \end{center}
\end{table*}

\begin{table*}
    \begin{center}
    \large
     \caption{Atmospheric parameters derived from SED fit of BHB stellar populations detected with UVIT in NGC\,2298. The notation of all columns is same as in Table ~\ref{tab3}. 
     }
    \label{tab4}
	\begin{tabular}{ccccccccc} 
		\hline
		\hline
		 Star ID & RA (deg) & DEC (deg) & {\it $T_{eff}$} (K) & $\frac{L}{L_{\odot}}$  & $\frac{R}{R_{\odot}}$ & ${\chi}_{red}^2$ & $\frac{N_{fit}}{N_{tot}}$\\
		\hline
         BHB1 & 102.2435 & -36.01081 & $10,750 \pm 250 $ & $70.15 \pm 1.25$ & $2.38 \pm 0.04$ & 3.8 & 7/8\\
         BHB2 &	102.2357 & -36.01813 & $12,500 \pm 500$ & $43.64	\pm	0.59$ &	$1.3 \pm 0.03$ &	12.7 & 8/8\\
         BHB3 & 102.2252 & -36.02032 & $11,500 \pm 250$ & $50.83	\pm	0.39$ &	$1.78	\pm	0.03$ & 14.3 & 8/8\\
         BHB4 &	102.2281 & -36.01802 & $12,250 \pm 500$ & $48.67 \pm 1.08$ & $1.5 \pm 0.03$ & 4.7 & 8/8\\
         BHB5 &	102.2501 & -36.00238 & $10,250 \pm 250$ & $73.27	\pm	2.46$ &	$2.78	\pm	0.05$ & 9.2	& 8/8\\
         BHB6 &	102.2629 & -35.98792 & $10,500 \pm 250$ & $57.85 \pm	2.0$ & $2.3	\pm	0.04$ &	4.9	& 8/8\\
         BHB7	&	102.2271	&	-36.01273	&	$10,500 \pm 250$	&	$54.65	\pm	1.92$	&	$2.23	\pm	0.04$	&	8.4	&	8/8\\
        BHB8	&	102.2475	&	-36.00676	&	$10,250 \pm 250$	&	$44.37	\pm	1.59$	&	$2.04	\pm	0.04$	&	14.1	&	8/8\\
        BHB9	&	102.2552	&	-36.01549	&	$12,000 \pm 500$	&	$30.79	\pm	0.39$	&	$1.2	\pm	0.02$	&	11.4	&	8/8\\
        BHB10	&	102.2459	&	-36.00806	&	$11,000 \pm 250$	&	$42.73	\pm	1.51$	&	$1.79	\pm	0.03$	&	6.1	&	8/8\\
        BHB11	&	102.2508	&	-36.0122	&	$12,250 \pm 500$	&	$33.13	\pm	0.47$	&	$1.23	\pm	0.02$	&	8.9	&	8/8\\
        BHB12	&	102.2564	&	-36.00318	&	$9,75^{+250}_{-500}$	&	$43.34	\pm	1.95$	&	$2.26	\pm	0.04$	&	35.1	&	8/8\\
        BHB13	&	102.2595	&	-36.0062	&	$11,000 \pm 250$	&	$39.01	\pm	1.63$	&	$1.69	\pm	0.03$	&	7.4	&	8/8\\
        BHB14	&	102.2239	&	-36.01842	&	$10,250 \pm 500$	&	$56.95	\pm	4.86$	&	$2.14	\pm	0.04$	&	7.6	&	8/8\\
        BHB15	&	102.2633	&	-35.99884	&	$12,000^{+250}_{-500}$	&	$32.35	\pm	0.51$	&	$1.31	\pm	0.03$	&	3.6	&	8/8\\
        BHB16	&	102.2467	&	-36.00291	&	$12,500^{+250}_{-500}$	&	$33.43	\pm	0.37$	&	$1.23	\pm	0.02$	&	7.1	&	8/8\\
        BHB17	&	102.2439	&	-36.00493	&	$9,750^{+500}_{-250}$	&	$48.07	\pm	1.44$	&	$2.42	\pm	0.05$	&	13.2	&	8/8\\
        BHB18	&	102.2486	&	-36.00619	&	$10,250^{+500}_{-250}$	&	$42.13	\pm	2.43$	&	$1.99	\pm	0.04$	&	5.4	&	8/8\\
        BHB19	&	102.2498	&	-36.01167	&	$9,250^{+500}_{-250}$	&	$45.49	\pm	1.46$	&	$2.53	\pm	0.05$	&	16.7	&	8/8\\
        BHB20	&	102.2576	&	-35.98383	&	$9,500^{+500}_{-250}$	&	$53.68	\pm	2.26$	&	$2.64	\pm	0.05$	&	12.1	&	8/8\\
        BHB21	&	102.2608	&	-35.99758	&	$10,500^{+500}_{-250}$	&	$37.39	\pm	1.78$	&	$1.79	\pm	0.03$	&	6.7	&	8/8\\
        BHB22	&	102.2467	&	-36.01245	&	$10,000^{+500}_{-250}$	&	$46.03	\pm	1.82$	&	$2.23	\pm	0.04$	&	3.2	&	8/8\\
        BHB23	&	102.2292	&	-36.01033	&	$9,500 \pm 250$	&	$51.97	\pm	2.09$	&	$2.66	\pm	0.05$	&	5.5	&	8/8\\
        BHB24	&	102.237	&	-36.00386	&	$9,250 \pm 500$	&	$49.44	\pm	2.62$	&	$2.68	\pm	0.05$	&	10.8	&	8/8\\
        BHB25	&	102.2694	&	-36.00772	&	$10,250^{+500}_{-250}$	&	$33.76	\pm	1.81$	&	$1.72	\pm	0.03$	&	6.9	&	8/8\\
        BHB26	&	102.2431	&	-35.97799	&	$9,500^{+500}_{-250}$	&	$46.76	\pm	3.05$	&	$2.49	\pm	0.05$	&	1.5	&	8/8\\
        BHB27	&	102.2673	&	-36.02095	&	$9,750^{+500}_{-250}$	&	$42.58	\pm	3.83$	&	$2.09	\pm	0.04$	&	2.7	&	8/8\\
        BHB28	&	102.2465	&	-36.01901	&	$9,000 \pm 500$	&	$54.19	\pm	2.11$	&	$2.95	\pm	0.06$	&	8.0	&	8/8\\
        BHB29	&	102.248	&	-35.99478	&	$9,500^{+250}_{-500}$	&	$44.69	\pm	2.14$	&	$2.41	\pm	0.05$	&	5.2	&	8/8\\
        BHB30	&	102.2554	&	-36.01623	&	$9,250^{+750}_{-250}$	&	$41.46	\pm	3.01$	&	$2.24	\pm	0.04$	&	7.9	&	8/8\\
	\hline
	\end{tabular}
	 \end{center}
\end{table*}

\begin{table*}
     \begin{center}
	\large
	\caption{Continued.}
    \label{}
	 \begin{tabular}{ccccccccc} 
		\hline
		\hline
		 Star ID & RA (deg) & DEC (deg) & {\it $T_{eff}$} (K) & $\frac{L}{L_{\odot}}$  & $\frac{R}{R_{\odot}}$ & ${\chi}_{red}^2$ & $\frac{N_{fit}}{N_{tot}}$\\
		\hline
		BHB31	&	102.2475	&	-36.00208	&	$9,250^{+500}_{-250}$	&	$45.92	\pm	2.18$	&	$2.52	\pm	0.05$	&	8.6	&	8/8\\
        BHB32	&	102.2441	&	-36.00356	&	$9,250^{+1000}_{-250}$	&	$38.76	\pm	2.55$	&	$2.08	\pm	0.04$	&	12.5	&	8/8\\
        BHB33	&	102.25	&	-36.01609	&	$9,000^{+500}_{-250}$	&	$46.43	\pm	3.92$	&	$2.75	\pm	0.05$	&	2.7	&	8/8\\
        BHB34	&	102.2466	&	-35.99503	&	$9,000^{+750}_{-250}$	&	$41.83	\pm	3.11$	&	$2.36	\pm	0.04$	&	6.1	&	8/8\\
        BHB35	&	102.2407	&	-36.00629	&	$8,750^{+500}_{-750}$	&	$58.94	\pm	4.15$	&	$3.19	\pm	0.06$	&	9.5	&	8/8\\
        BHB36	&	102.2623	&	-36.01596	&	$9,000 \pm 500$	&	$43.43	\pm	2.52$	&	$2.53	\pm	0.05$	&	7.5	&	8/8\\
        BHB37	&	102.2406	&	-36.0042	&	$8,750^{+500}_{-250}$	&	$48.16	\pm	2.15$	&	$2.97	\pm	0.06$	&	12.2	&	8/8\\
        BHB38	&	102.2408	&	-36.01402	&	$8,500^{+500}_{-1000}$	&	$77.59	\pm	3.12$	&	$4.03	\pm	0.08$	&	1.8	&	7/8\\
        BHB39	&	102.2499	&	-35.99423	&	$8,750^{+500}_{-750}$	&	$44.66	\pm	2.54$	&	$2.81	\pm	0.05$	&	11.7	&	8/8\\
        BHB40	&	102.2469	&	-36.00461	&	$8,500 \pm 500$	&	$46.42	\pm	6.17$	&	$2.97	\pm	0.06$	&	9.9	&	8/8\\
        BHB41	&	102.2284	&	-36.00907	&	$8,500 \pm 500$	&	$52.18	\pm	6.35$	&	$3.01	\pm	0.06$	&	5.1	&	8/8\\
        BHB42	&	102.2741	&	-35.99637	&	$8,750^{+250}_{-500}$	&	$40.41	\pm	2.48$	&	$2.73	\pm	0.05$	&	2.0	&	8/8\\
        BHB43	&	102.2631	&	-36.00405	&	$8,250^{+500}_{-750}$	&	$48.52	\pm	3.31$	&	$3.34 \pm 0.06$	&	10.6	&	8/8\\
        BHB45	&	102.2548	&	-36.00226	&	$8,500^{+500}_{-1000}$	&	$49.53	\pm	3.74$	&	$3.15	\pm	0.06$	&	12.4	&	8/8\\
        BHB46	&	102.241	&	-35.99525	&	$8,250^{+500}_{-750}$	&	$54.8	\pm	3.61$	&	$3.52	\pm	0.07$	&	5.3	&	8/8\\
        BHB47	&	102.2486	&	-35.99375	&	$9,500^{+500}_{-250}$	&	$41.89	\pm	2.25$	&	$2.27	\pm	0.04$	&	8.6	&	8/8\\
        BHB48	&	102.2521	&	-36.01067	&	$7,750 \pm 500$	&	$51.41	\pm	2.84$	&	$3.8	\pm	0.07$	&	14.0	&	6/6\\
		BHB49	&	102.2503	&	-36.01369	&	$8,000 \pm 125$	&	$51.16	\pm	4.78$	&	$3.65	\pm	0.07$	&	32.9	&	8/8\\
        BHB50	&	102.1974	&	-36.0043	&	$12,250 \pm 250$	&	$53.02	\pm	2.62$	&	$1.59	\pm	0.03$	&	1.2	&	15/16\\
        BHB51	&	102.302	&	-36.00933	&	$11,000^{+250}_{-500}$	&	$62.62	\pm	3.29$	&	$2.15	\pm	0.04$	&	2.0	&	15/16\\
        BHB52	&	102.2673	&	-36.06088	&	$10,000^{+250}_{-500}$	&	$46.22	\pm	6.76$	&	$2.22	\pm	0.04$	&	3.9	&	15/16\\
        BHB53	&	102.2418	&	-36.04874	&	$10,250^{+250}_{-500}$	&	$54.08	\pm	8.19$	&	$2.32	\pm	0.04$	&	3.9	&	15/16\\
        BHB54	&	102.1568	&	-36.10535	&	$9,500 \pm 250$	&	$66.63	\pm	4.47$	&	$2.99	\pm	0.06$	&	0.74 &	15/16\\
        BHB55	&	102.2196	&	-35.98094	&	$10,000 \pm 500$	&	$52.23	\pm	6.51$	&	$2.39	\pm	0.05$	&	2.6	&	15/16\\
        BHB56	&	102.2171	&	-35.97486	&	$9,750^{+250}_{-750}$	&	$43.27	\pm	6.56$	&	$2.27	\pm	0.04$	&	1.4	&	15/16\\
        BHB57	&	102.2633	&	-36.03276	&	$8,750^{+250}_{-500}$	&	$50.34	\pm	10.56$	&	$3.01	\pm	0.06$	&	1.7	&	15/16\\
        BHB58	&	102.2021	&	-36.01866	&	$8,750^{+250}_{-500}$	&	$64.73	\pm	11.11$	&	$3.44	\pm	0.06$	&	1.8	&	15/16\\
        BHB59	&	102.2339	&	-36.04152	&	$8,000^{+250}_{-500}$	&	$65.59 \pm	5.47$	&	$4.16	\pm	0.08$	& 3.5 &	11/11\\
        BHB60 & 102.2492 & -36.00759 & $10,750^{+250}_{-500}$ & $35.41 \pm 2.0$ & $1.62 \pm 0.03$ & 7.9 & 8/8\\
        BHB61 & 102.2539 & -36.00547 & $9,250^{+500}_{-250}$ & $47.15 \pm 2.99$ & $2.61 \pm 0.05$ & 3.5 & 8/8\\
        BHB62 & 102.2809 & -36.01208 & $8,250 \pm 500$ & $54.26 \pm 4.36$ & $3.36 \pm 0.06$ & 2.7 & 6/6\\
        BHB63 & 102.2535 & -35.99972 & $8,000 \pm 500 $ & $49.49 \pm 3.77$ & $3.48 \pm 0.07$ & 3.0 & 7/7\\
     \hline
	\end{tabular}
	\end{center}
\end{table*}

We carried out SED fitting analysis for 63 BHB and four hot HB stars. Figures~\ref{ehbseds} and \ref{bhbseds} show the SED fit for four hot HB and two BHB stars overplotted with the corresponding best-fit models (smallest value of $\chi^2_{red}$ ) shown with light gray color. In the lower panels of all plots, we show the residuals between the observed SED and the best-fit model. The star ID used in this work, metallicity value of the fitted model spectrum, and estimated temperature are displayed in all the SED plots. As mentioned above in section~\ref{sec:fuvcmds}, variables such as RR Lyrae are observed at random phases, so we have not considered them for SED analysis. The estimated parameters of four hot HB stars using the Kurucz, Husfeld, and TMAP (Grid4) models are tabulated in Table~\ref{tab3}. The Husfeld model fits find UV1 and UV4 to be much hotter when compared to the estimates from Kurucz model fits, and also suggests these stars to be helium-rich. On the other hand, the Husfeld model fits to the SEDs of UV2 and UV3 provide normal helium values along with {\it $T_{eff}$} similar to those obtained from Kurucz model fits (within errors). The {\it $T_{eff}$} (120,000 K) and luminosity (1148$L_{\odot}$) of UV4 obtained using the TMAP (Grid4) model are much higher than estimated from the other two models. The derived values of parameters  {\it $T_{eff}$}, $\frac{R}{R_{\odot}}$, and $\frac{L}{L_{\odot}}$ corresponding to the best-fit Kurucz model spectrum along with the errors for BHB stars are listed in Table~\ref{tab4}. VOSA computes uncertainties in the effective temperatures as half the grid step, around the best-fit value. As errors estimated through VOSA are not realistic, we have reported the range in {\it $T_{eff}$} as found from the 10 best-fit values. While the log{\it g} values for these stars are also estimated using this technique, these values are not reliable as SED fits are not sensitive to this parameter. 

In Figure ~\ref{ehbseds}, it can be seen from the residuals that the observed data points are well fitted with the model spectrum. However, in most of the SED fits, {\it HST} F275W data point does not fit with the model flux. The observed flux at F275W is found to be less than the expected flux from the model, which, in turn, gives a negative residual. We note the presence of strong absorption lines such as Fe II and Mg II lines in the wavelength range covered by this filter. It might be possible that 
there is a mismatch in the strength of these spectral lines between the observations and the models. From the comparison of synthetic Kurucz spectra at different {\it $T_{eff}$}, we notice that the contribution of these lines to the integrated flux of F275W is more at cooler temperatures than the hotter ones, which results in a larger deviation from the expected flux for cooler stars than the hotter stars.
In order to establish the best fit to the observed SED, we have only considered the data points which fits well to the model spectra.
The high temperatures of UV1 (35,000 K) and UV2 (35,000 K) suggest that they may belong to the class of EHB stars as they have temperatures around 30,000 K \citep{1986A&A...155...33H}. The other two hot HB stars UV3 and UV4, with very high temperatures of 40,000 and 120,000 K, respectively, might belong to the pHB phase. Also, note that SEDs of all hot HB stars are well fitted with a single spectrum with minimum residual across wavelength, which, in turn, indicates that these are likely to be single stars.\\

The SEDs of all BHB stars are presented in Appendix~\ref{sec:appendix}. The BHB stars span an effective temperature range from 8,000-12,250 K. For one BHB star, namely BHB44, observed SED is not fitted with a single spectrum as shown in the left panel of Figure~\ref{bhbseds}. Compared to Kurucz model spectrum and synthetic flux, there seems to be a large amount of excess flux in FUV filters. We checked whether the star is well-resolved in all FUV images (See Figure~\ref{locuv}) and also ensured that the cross-identification with the \textit{HST} catalog is correct. It is located at a distance of $0\farcm68$ from the center of the cluster. It is possible that this star might be a binary star or variable star. If we check the position of this star, marked with a black outlined star symbol in Figures~\ref{optcmds} and \ref{uvcmds}, in optical as well as in UV CMDs, it is lying close to or in the variable region. In the literature, it is not reported as a variable star. The temperature derived from the single fit of BHB44 corresponds to that of an RHB star which shows significant FUV excess due to a possible hot companion, as RHB stars are too cool to be seen in FUV. In order to check what type of hot companion is present, we fitted the SED with a combination of hot and cool theoretical spectra. To fit the FUV region of the observed SED of this star, we selected a Koester WD model \citep{2010MmSAI..81..921K, 2009ApJ...696.1755T}. The free parameters of the Koester model are {\it $T_{eff}$} and log{\it g}. The value for the {\it $T_{eff}$} for the Koester model ranges  from 5,000-80,000 K and logg from 6.5-9.5 dex. We utilized VOSA to obtain the binary fit of this star. The composite SED fit of this star is presented in the left panel of Figure~\ref{bhbseds} where we can see that the hotter part of SED is well fitted with a Koester model corresponding to a temperature of 17,000 K, and cooler part is fitted with Kurucz model of temperature 5,750 K. The detailed parameters of both companions are listed in Table~\ref{tab5}. The {\it $T_{eff}$} and radius obtained for the cooler part from the best Kurucz model fit correspond to an RHB star. From the {\it $T_{eff}$} and radius,  we suggest that the hot companion might belong to the class of sub-luminous sdB stars \citep{2003A&A...411L.477H}.

Further, to confirm the nature of four hot HB stars, the comparison of the obtained parameters from the SED fit with theoretical evolutionary tracks is needed and discussed in the following section.

\section{Evolutionary Status of hot HB and BHB stars}
\label{sec:status}
\begin{figure*}[!htb]
\hspace{-0.3cm}
\centering
\makebox[\textwidth]
{
\includegraphics[width=0.98\textwidth]{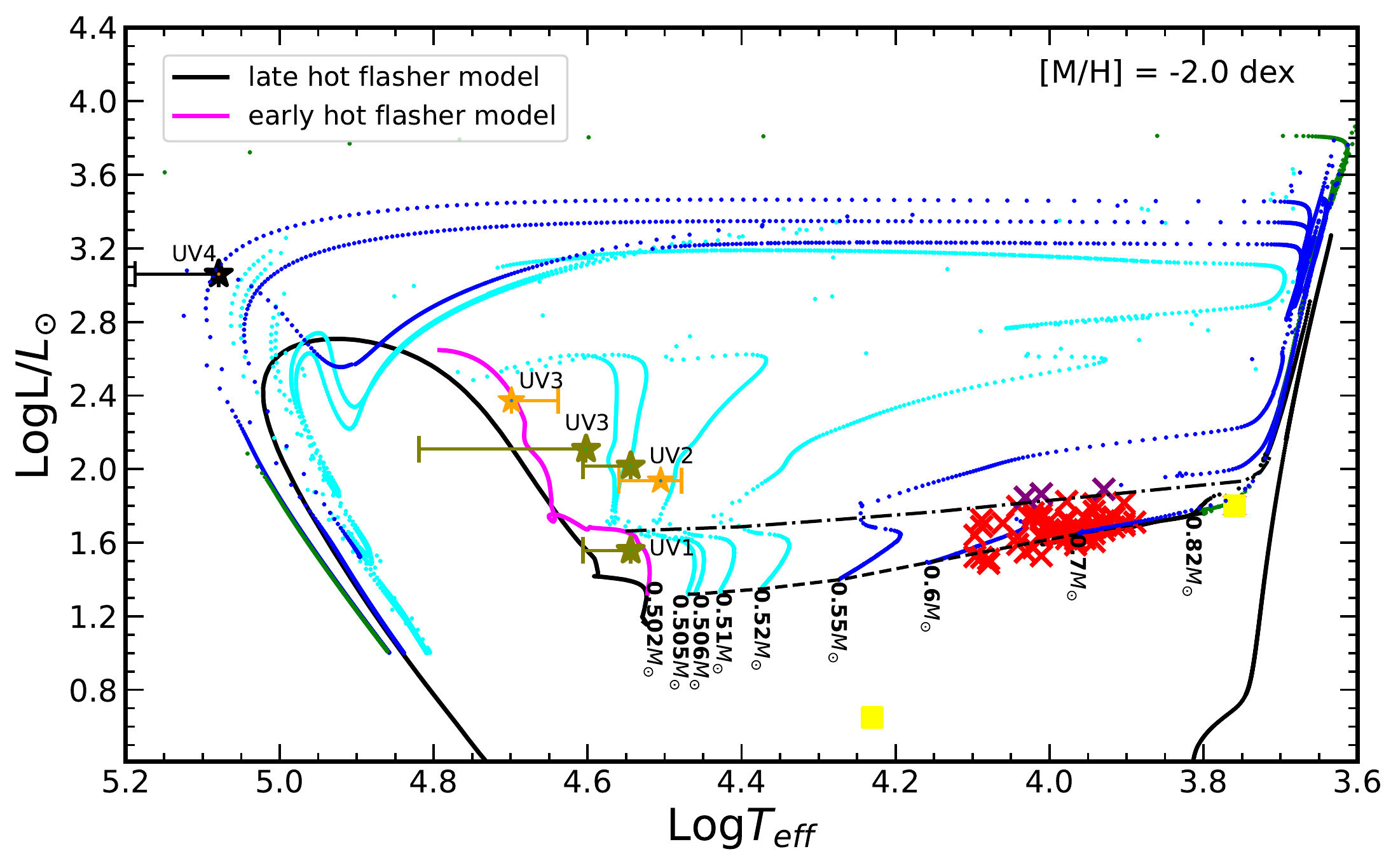}
}
\caption{Position of hot HB and BHB stars identified with UVIT in NGC\,2298 in the H-R diagram along with theoretical evolutionary tracks. The evolutionary tracks starting from MS turn-off to the moment when a star has entered to the WD cooling phase \citep{2018ApJ...856..125H} are presented in this plot. Along the HB phase, Post-ZAHB tracks span a mass range from 0.502 $-$ 0.82M$_{\odot}$. In the plot, cyan, blue and green colors correspond to the sequences populating the extreme, blue and red parts of the HB. The Black dashed and dash-dotted lines show the position of the canonical ZAHB and TAHB, respectively. The black and magenta solid lines indicate the late and early hot-flasher models, respectively. The SED fit parameters obtained from Kucurz model fit to the observed SEDs of hot HB and BHB stars are shown with orange star and red cross symbols, respectively. The olive and black filled star symbols present the location of hot HB stars corresponding to the helium-rich Husfeld and TMAP (Grid4) model fit, respectively. Yellow square symbols present the location of hot and cool companion of a BHB44 star. Several BHB stars evolving towards peAGB phase are shown with purple cross symbols.}
\label{bastievotrack}
\end{figure*}

The derived atmospheric parameters of hot HB and BHB stars from SED fit are compared with theoretical evolutionary tracks in order to check their evolutionary status. We plotted the theoretical evolutionary tracks employing the models presented by \cite{2019A&A...627A..34M}. 
The MS to RGB evolutionary track is generated using the updated BaSTI-IAC models presented by \cite{2018ApJ...856..125H}.
We selected the model with metallicity close to the cluster metallicity. We show the theoretical evolutionary tracks from MS turn-off to the tracks corresponding to different masses starting from the ZAHB through to a point late in the pHB evolution or a point on the pAGB cooling sequence in Figure~\ref{bastievotrack}. The ZAHB and TAHB, representing the end of the HB phase, are shown with a dashed and dash-dotted line in Figure~\ref{bastievotrack}. We also have plotted the early hot-flasher, and late hot-flasher tracks shown with magenta and black solid lines. The parameters estimated from the best SED fit for hot HB and BHB stars are plotted in the H-R diagram and shown with different symbols and colors. We can see in Figure~\ref{bastievotrack} that most of the BHB stars marked with red cross symbols are lying along the BHB tracks shown with blue lines. Nevertheless, there are a few BHB stars, displayed with purple cross symbols, lying above the TAHB, indicating that these stars' cores have already run out of helium and they have started evolving towards the pHB or peAGB phase.\\

To compare the estimated parameters of hot HB stars computed using different models, we plotted hot HB stars' location in the H-R diagram as shown in Figure~\ref{bastievotrack}. 
As the hot HB star, UV1 was found to be helium-rich by the synthetic HB models; we have shown only the location of this star in the H-R diagram as found from the helium-rich Husfeld model. UV1 is found to be located between the ZAHB and TAHB and slightly hotter than the EHB track for 0.502 M$_\odot$. Thus, UV1 is still in the HB evolutionary phase, which implies that it is an EHB star, as also seen from Figures~\ref{uvcmds} and~\ref{simhb}. Figure~\ref{bastievotrack} shows that the hot HB star UV2 is found along the AGB-manqu\'e evolutionary stage, corresponding to the initial EHB mass range $0.506 - 0.51 M_{\odot}$, which is likely to evolve from the EHB phase. Stars UV3 and UV4 are found to be much hotter and brighter than the EHB tracks. UV3 is found to be located close to the evolutionary track corresponding to an early hot-flasher. 
We infer from here that UV3 is the progeny of an EHB star (with $\sim0.502 M_{\odot}$) formed through an early hot-flasher scenario. The position of UV4 star indicates that it might have crossed the post-early-AGB (p(e)AGB) stage, and is about to enter the WD cooling stage. 
TMAP (Grid4) model parameter estimates for UV4 star suggest that it is a product of BHB star with the mass of $\sim0.7 M_{\odot}$. 
From the model, the mass of EHB star UV1, turns out to be 0.502 M$_{\odot}$. The rest of the hot stars are likely to be evolved from the EHB stars with a ZAHB mass of $\sim0.5M_{\odot}$. In comparison, the BHB stars have masses in the range 0.6 - 0.75 M$_{\odot}$. This suggests that the hot HB stars have lost $\sim$ 0.1 - 0.2 M$_{\odot}$ more envelope mass due to mass loss in the RGB. The reason for this enhanced mass loss could be many, including enhanced helium due to mixing, binary interactions, high rotation, etc. 
In the case of BHB44 star, a comparison of the L and T$_{eff}$ of the hot companion with the theoretical evolutionary sequences for extremely low-mass (ELM) WDs computed by \citep{2013A&A...557A..19A} suggests that it has a low mass of $\sim 0.187 M_{\odot}$. As the mass of the sdB is too low to support the core helium burning, it might be evolving into a helium core WD and is likely to be an ELM WD candidate. Therefore, this binary is most likely to be a post mass-transfer system, consisting of an ELM-WD candidate and an RHB star.\\
The expected number of pHB stars in the cluster is estimated based on the fact that the number of stars in two post-MS phases, in general, will be proportional to the ratio of the lifetimes of these phases \citep{2002ApJ...579..752K}. The following relation is used to calculate the expected number of pHB stars:
\begin{equation*}
    N_{pHB} = N_{HB}\left( \frac{\tau_{pHB}}{\tau_{HB}} \right)
\end{equation*}

Where N$_{pHB}$ is the expected number of pHB stars in NGC\,2298; N$_{HB}$ is the number of HB stars in the cluster; $\tau_{pHB}$ and $\tau_{HB}$ represents the lifetimes of pHB and HB evolutionary phases of a low mass star, respectively. We have taken the duration of the HB phase as $\tau_{HB}$ = $\sim 10^{8}$ years \citep{1992ApJS...81..221D}, and that of pHB phase as $\sim 10^{7}$ years from BaSTI pHB tracks. We therefore estimate the number of expected pHB stars to be $\sim 7$. The observed number of pHB stars are six in this cluster, which is in fair agreement with the theoretically expected number.

\begin{table*}
  	\begin{center}
	\small
	\caption{Derived parameters of BHB44 star from composite SED fit.}
	\label{tab5}
\begin{tabular}{cccccccccc} 
		\hline
		\hline
		 Star ID & RA (deg) & Dec (deg) & Type &  Model Used & {\it $T_{eff}$} (K) & $\frac{L}{L_{\odot}}$  & $\frac{R}{R_{\odot}}$ & ${\chi}_{red}^2$ & $\frac{N_{fit}}{N_{tot}}$ \\
		\hline
		 BHB44 & 102.2532 &	-36.01526 & sdB & Koester & $17000 \pm 1000$ & $4.46 \pm 0.23$ & $0.24 \pm 0.005$ & 2.9 & 8/8\\ 
		 & & & RHB & Kurucz & $5750 \pm 125$ & $63.38 \pm 5.7$ & $8.17 \pm 0.15$ & & \\ 
		\hline
	\end{tabular}
	\end{center}
\end{table*} 

\section{Discussion}
\label{sec:dis}
We have analyzed the UVIT data aboard the \textit{AstroSat} satellite covering the GC NGC\,2298 to characterize the hot HB population in order to understand their formation and evolution. To date, this cluster has been studied in UV, only as a part of the group, for comparative studies of HB morphology. A focussed study on the HB population of this cluster has not been done so far. This is the first time hot HB stars are characterized in this cluster using FUV photometric data combined with the \textit{HST} and ground-based data. We have combined FUV photometric data with optical photometry to generate the FUV-optical CMDs in three FUV filters (F148W, F154W, F169M) and detected the BHB, four hot HB and a few bright BS stars. 

\cite{2016ApJ...822...44B} analyzed the 53 GGCs using \textit{HST} UV and blue photometric data to explore the HB morphology, including NGC\,2298. They created color-color plots for all selected clusters and found two hot HB stars bluer than the gap between EHB and BHk stars in NGC\,2298. They classified these two hot HB stars as BHk stars based on their location in the color-color plot. In our FUV images, we have detected three hot HB stars in the central region and one in the outer region. In FUV CMDs, out of four hot HB stars, three are brighter than canonical EHB stars and classified as pHB stars. 

\cite{2012AJ....143..121S} presented the UV CMDs for 44 GGCs using \textit{GALEX} photometric data in NUV and FUV passbands, including NGC\,2298. They had detected HB and BS stars in NGC\,2298, but the sample had issues because of the limited spatial resolution of \textit{GALEX} and lack of membership analysis for detected stars. Our study detected more than 90 $\%$ of HB stars compared to the \textit{HST} and ground-based catalogs, and the PM membership is also confirmed. The stars which are not detected in the FUV images are fainter than the limiting magnitude of UVIT, and the exposure times in all filters are not deep enough to detect them.\\

Further, we compared the observed HB sequence with theoretical ZAHB and TAHB sequences for standard and enhancd initial helium abundances. The theoretical ZAHB tracks with distinct initial helium abundances could not reproduce the observed color spread along the BHB sequence.  \cite{2018MNRAS.481.5098M} determined the average helium difference between the second-generation (2G) and first-generation (1G) stars along RGB in a large sample of 57 GGCs and the maximum helium variation within each GC. The maximum helium variation was found to be 0.011 dex in NGC\,2298. Our study does not support a large spread in helium along the BHB sequence, based on comparison of observed HB with theoretical tracks and synthetic HB, though we detect a possible helium difference of 0.1 dex between the BHB stars and the hot HB stars, as suggested by the synthetic HB simulations.


\cite{1994PASP..106..718W} presented a spectroscopic study of an extremely blue star in NGC\,2298, classifying it as helium-rich O type subdwarf star (sdO), but they could not confirm the membership of this star. In their optical CMD shown in Figure 1, the position of this star coincides with that of EHB or BHk stars. The RA and DEC information of this star is not provided in the above paper. Therefore, we could not check whether this star is detected in our FUV images or not. 

It is well established from the photometric as well as from spectroscopic studies that most of the GCs like NGC\,2808 and $\omega$ Cen, with well-populated HBs, contain helium-rich populations showing discrete HBs, which indicate discrete helium abundances \citep{2011MNRAS.410..694D, 2014MNRAS.437.1609M, 2011A&A...526A.136M}. The four hot HB stars in NGC\,2298 are found to be helium-rich with respect to BHB stars with a standard helium abundance. This may suggest that these stars are products of helium-rich second-generation stars in this cluster. However, the helium-rich population will be hard to be detected due to their small number fraction compared to the normal- or slightly enhanced initial helium abundance in the MS to RGB stages of NGC\,2298.


Our {\it $T_{eff}$} estimation for three hot HB stars covers a range from 35,000-40,000 K, whereas BHB stars span a {\it $T_{eff}$} range from 7,500-12,250 K. We could not accurately estimate the {\it $T_{eff}$} of UV4 star, but the temperature of this star can be around $\sim 100,000 K$.
By comparing both {\it $T_{eff}$} and luminosity of hot HB stars with evolutionary tracks implies that three stars have evolved away from the HB, and one is still in EHB phase. The hot HB stars also have lost more mass in the RGB ($\sim$ 0.1-0.2 M$_\odot$ than the BHB stars). 

Many authors put forward several formation scenarios to explain the formation of EHB and BHk stars in GCs other than helium enrichment. 
\cite{2015MNRAS.449.2741L} suggested that BHk stars could be the product of binary interactions as tidally enhanced stellar wind in binary evolution might provide the substantial mass loss on the RGB and produce BHk stars. Nevertheless, our SED fits of hot HB stars do not show the signature of binarity, hence, could not support the origin through the binary interaction scenario.
The other formation channel suggested for EHB and BHk stars, such as the hot-flasher scenario, is described in detail in Section~\ref{sec:hotflashermodels}. \cite{1997ApJ...474L..23S} demonstrated that when stars undergo a late helium-core flash on the WD cooling curve, flash mixing of the hydrogen envelope with the helium core will extensively enhance the envelope helium and carbon abundances. In contrast, mixing can not occur in the early hot-flasher scenario because the large entropy barrier of a strong hydrogen-burning shell prevents the products of core helium burning from being mixed to the surface. 

\cite{2012ApJ...748...85B} presented \textit{HST} FUV spectroscopy of hot HB stars including one pHB, five BHB and three unclassified stars with blue UV colors in GC NGC\,2808. They also found enhanced helium and carbon abundances in their BHk sample, which could be the result of flash-mixing in the late hot-flasher mechanism, whereas EHB stars in their sample exhibit carbon abundances much lower than the cluster value, and helium abundances at or below the solar value, that could be the effect of  diffusion. The two hot HB stars in our study, UV2, and UV3 are found close to early hot-flasher tracks, whereas UV4 is located above the post-BHB track, indicating that UV2 and UV3 could be off-springs of EHB stars, and UV4 is a plausible progeny of BHB star.
Therefore, the surface abundances of three stars except UV4 are expected to remain the same if they are products of the early hot-flasher scenario. Nevertheless, these stars could have helium enrichment, as evident from a comparison with simulations. Further spectroscopic follow-up observations are required to confirm the nature of the evolutionary process these stars have gone through. 

\section{Summary and Conclusions}
\label{sec:summary}
The main results from this work can be summarized as follows: 
\begin{itemize}
    \item In this study, we employed UVIT observations in combination with \textit{HST}, \textit{Gaia} EDR3, and ground-based photometric data to examine the HB morphology of the GC NGC\,2298. \textit{Gaia} EDR3 data is utilized to obtain the PM members of the cluster in the outer region of the cluster.
    \item We constructed optical and FUV-optical CMDs for the member stars. Only BHB and four hot HB stars are detected in all FUV images. Very few BSSs, which are hot and bright, are detected in FUV CMDs.
    \item Optical and FUV-optical CMDs are overlaid with updated BaSTI-IAC isochrones generated for respective filters to compare the observations with theoretical predictions. The theoretical HB tracks with enhanced alpha and helium abundances could not reproduce the observed color/magnitude spread among BHB stars in FUV-optical CMDs. 
    \item From the comparison of observed HB with synthetic HB simulations, we found a helium abundance difference between BHB and hot HB stars (Helium enhanced) in this cluster. However, BHB stars have a single initial helium abundance (\textit{$Y_{ini}$} = 0.23), with probably a very small scatter. 
    \item We estimated {\it $T_{eff}$}, luminosities, and radii of 63 BHB and four hot HB stars by generating SED using multi-wavelength data. The {\it $T_{eff}$} of BHB stars ranges from 7,500-12,250 K whereas three hot HB stars span {\it $T_{eff}$} from 35,000-40,000 K. The temperature of UV4 star is found to be around $\sim 100,000 K$.
    \item The evolutionary status of HB stars is probed by comparing derived parameters with theoretical evolutionary post-ZAHB tracks. Many BHB stars are found to be located between the ZAHB and TAHB, suggesting that they are evolving off the HB into the pHB phase. These stars have mass in the range of 0.6-0.75 M$_\odot$. Some of the BHB stars are found to be evolving towards the peAGB phase.
    \item We found a sub-luminous sdB companion to an RHB star in the cluster. From the comparison with the ELM WD evolutionary tracks, the mass of the sdB turns out to be $\sim 0.187 M_{\odot}$, and likely to be evolving into a helium core WD. We suggest that it is probably an ELM WD candidate formed from mass-transfer in this binary system.
    \item Out of four hot HB stars, we find that two already evolved off the EHB phase, and they are in the AGB-manqu\'e phase. One star is located between the ZAHB and TAHB tracks, and hence it is a confirmed EHB star, likely to be helium enriched. One star is found to match with the theoretical prediction of the early hot-flasher scenario (and maybe one more), whereas another stars' has evolved off the p(e)AGB phase, and is probably evolving towards the WD cooling stage. The theoretically expected number of pHB stars found to agree well with the observed number. 
    \item As the late and early hot-flashers are supposed to have different chemical signatures, the pHB stars are targets for further spectroscopic studies in order to explore their nature to constrain their formation pathways.
\end{itemize}

\section*{Acknowledgements}
We thank the anonymous referee for a constructive report that improved the quality of our manuscript. We would like to thank Santi Cassisi for providing us the hot-flasher models for UVIT filters. C. Chung acknowledges the support provided by the National Research Foundation of Korea to the Center for Galaxy Evolution Research (No. 2017R1A2B3002919). S. Rani thanks Deepthi S. prabhu for useful discussions. This publication utilizes the data from {\it AstroSat} mission's UVIT, which is archived at the Indian Space Science Data Centre (ISSDC). The UVIT project is a result of collaboration between IIA, Bengaluru, IUCAA, Pune, TIFR, Mumbai, several centers of ISRO, and CSA. This research made use of VOSA, developed under the Spanish Virtual Observatory project supported by the Spanish MINECO through grant AyA2017-84089. This research also made use of Aladin sky atlas developed at CDS, Strasbourg Observatory, France \citep{Bonnarel2000}.

\vspace{5mm}


\software{GaiaTools \citep{2019MNRAS.484.2832V}, Topcat \citep{ 2011ascl.soft01010T}, Matplotlib \citep{2007CSE.....9...90H}, NumPy (\citealp{2011CSE....13b..22V}), Scipy \citep{2007CSE.....9c..10O, article}, Astropy \citep{2013A&A...558A..33A, 2018AJ....156..123A} and Pandas \citep{mckinney-proc-scipy-2010}}

\appendix

\section{SED Fitting for BHB Stars}
\label{sec:appendix}
The details of SED fitting technique are described in Section~\ref{sec:SEDs}. The SEDs for 61 BHB stars are shown in Figure~\ref{sed1}. 

\begin{figure*}[htb]
\centering
\includegraphics[width=0.32\columnwidth]{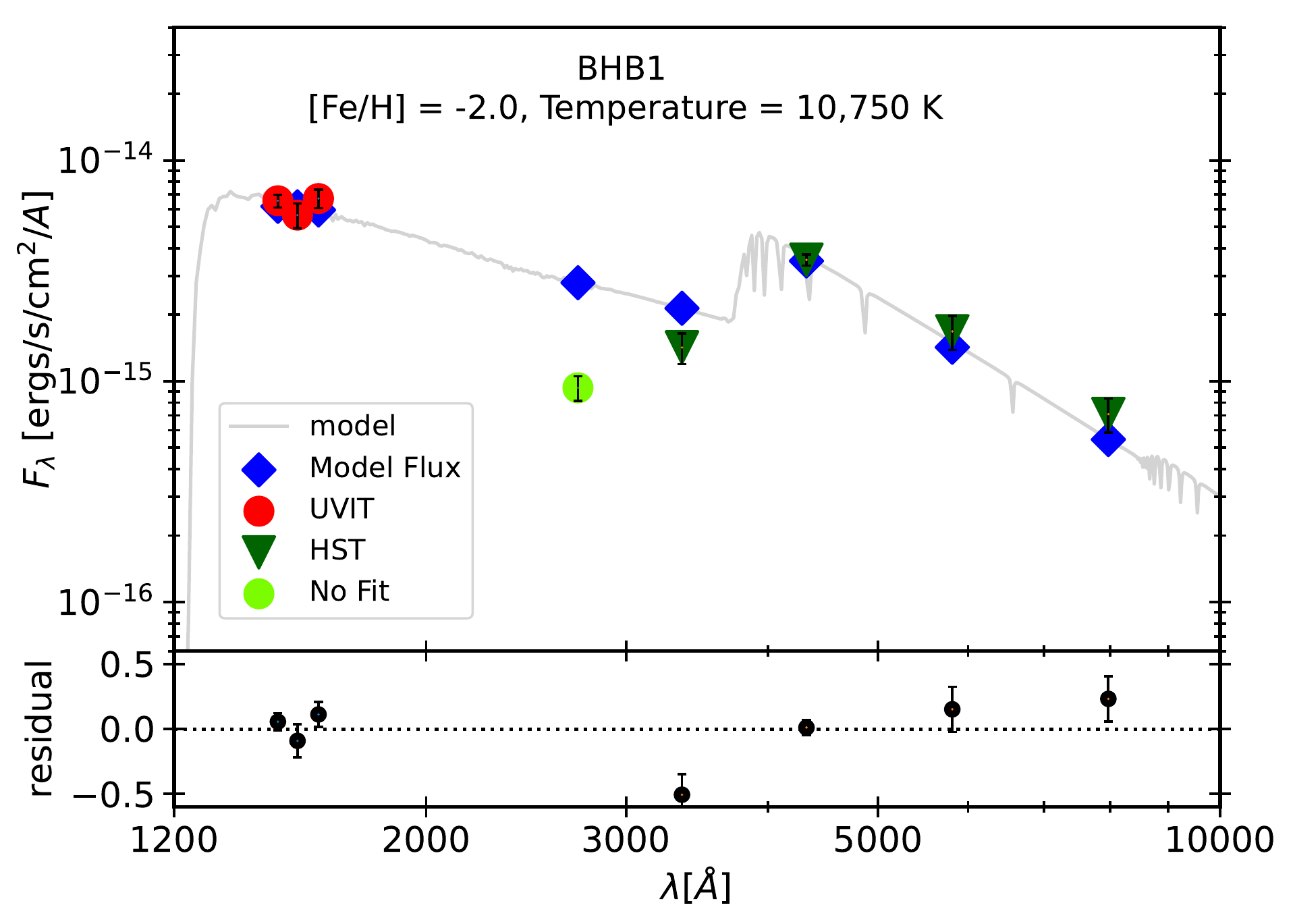}
\includegraphics[width=0.32\columnwidth]{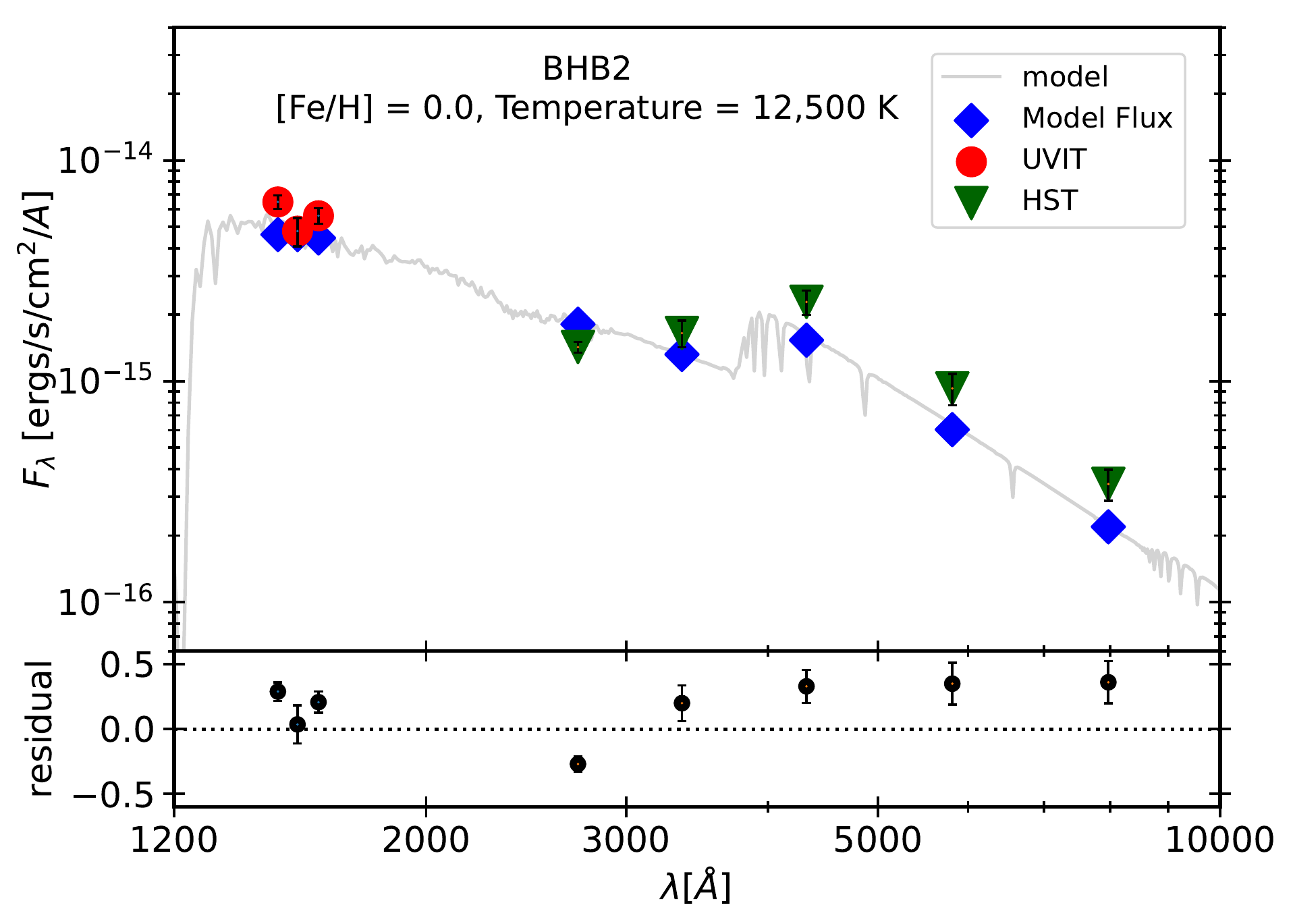}
\includegraphics[width=0.32\columnwidth]{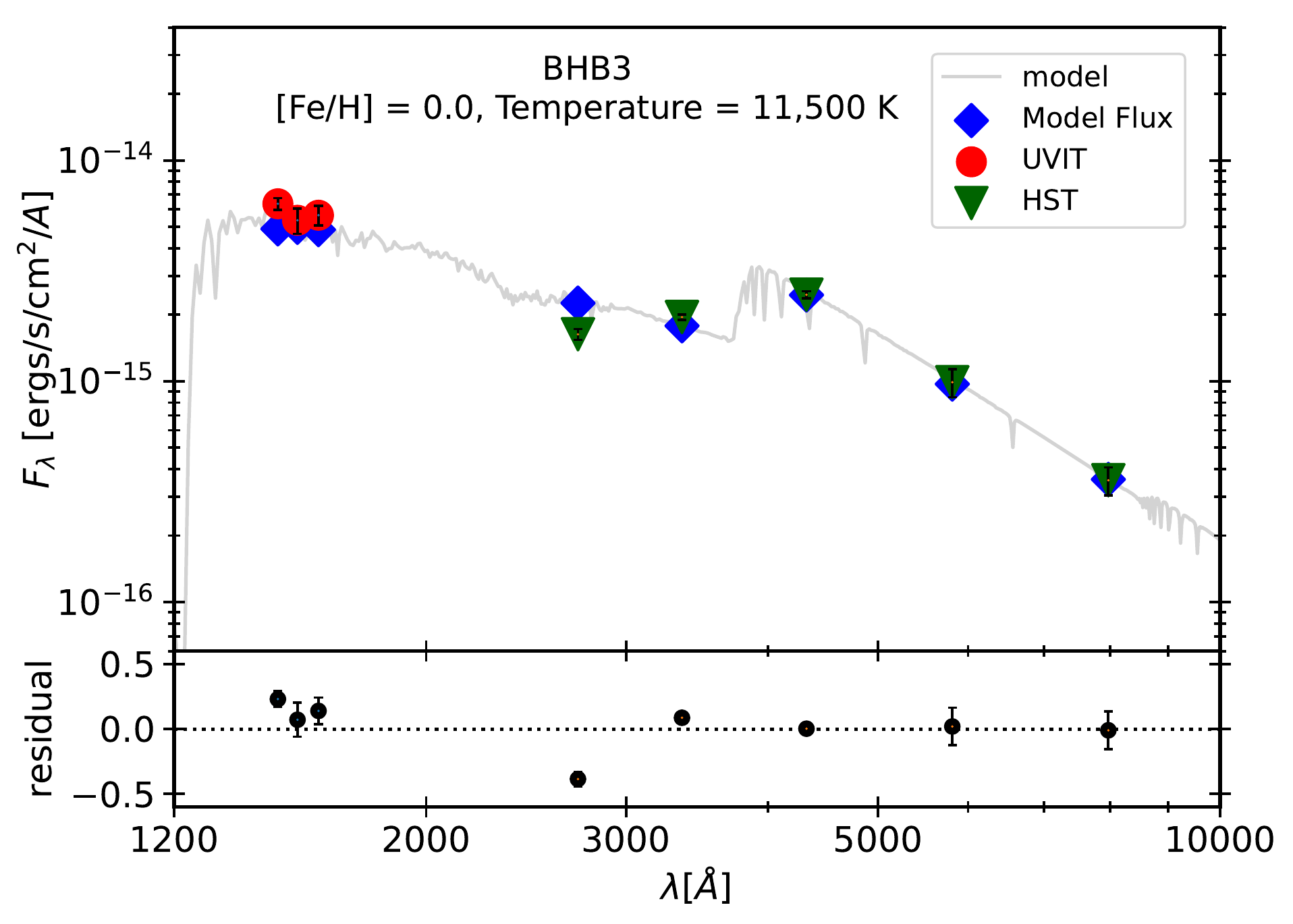}
\includegraphics[width=0.32\columnwidth]{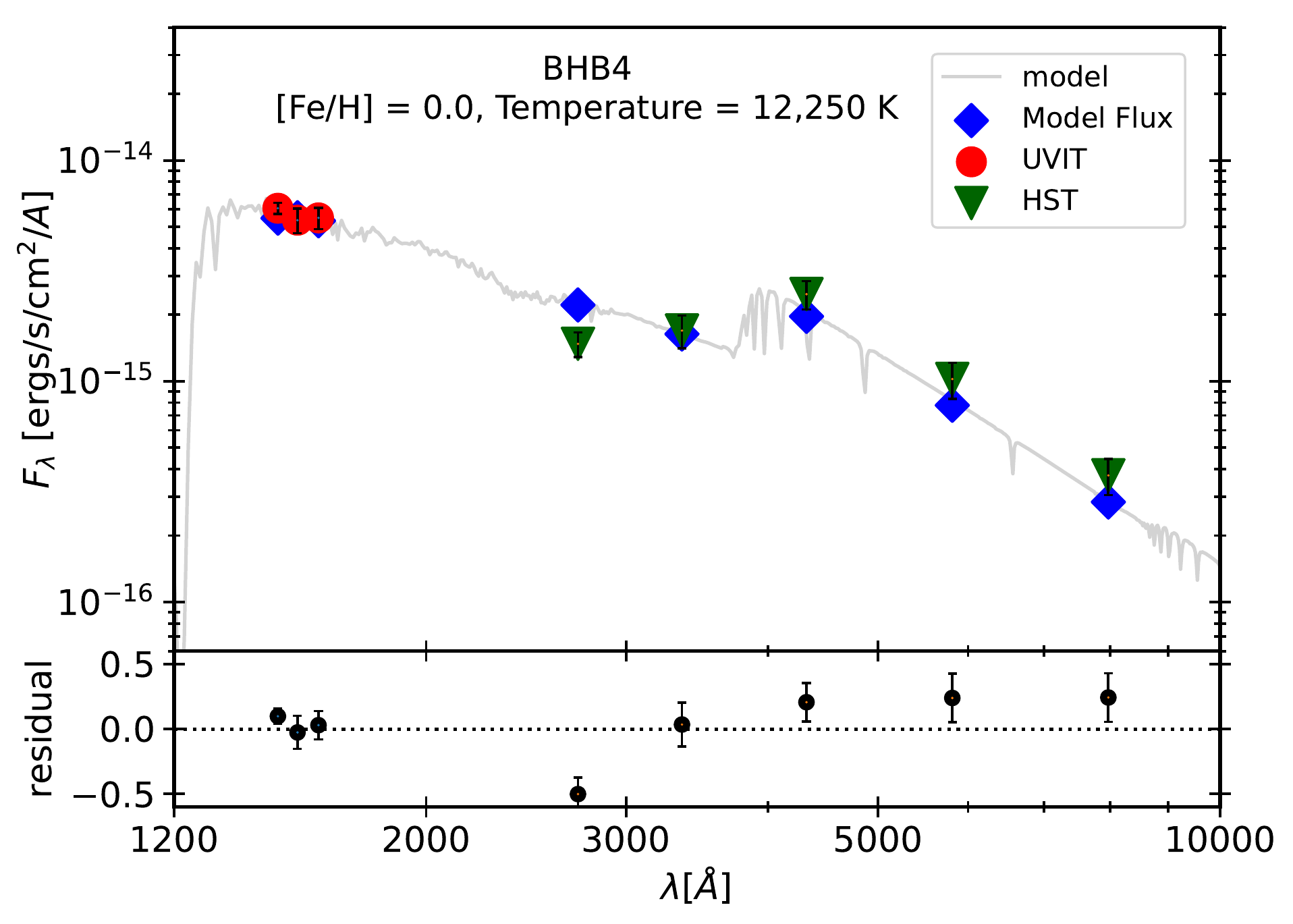}
\includegraphics[width=0.32\columnwidth]{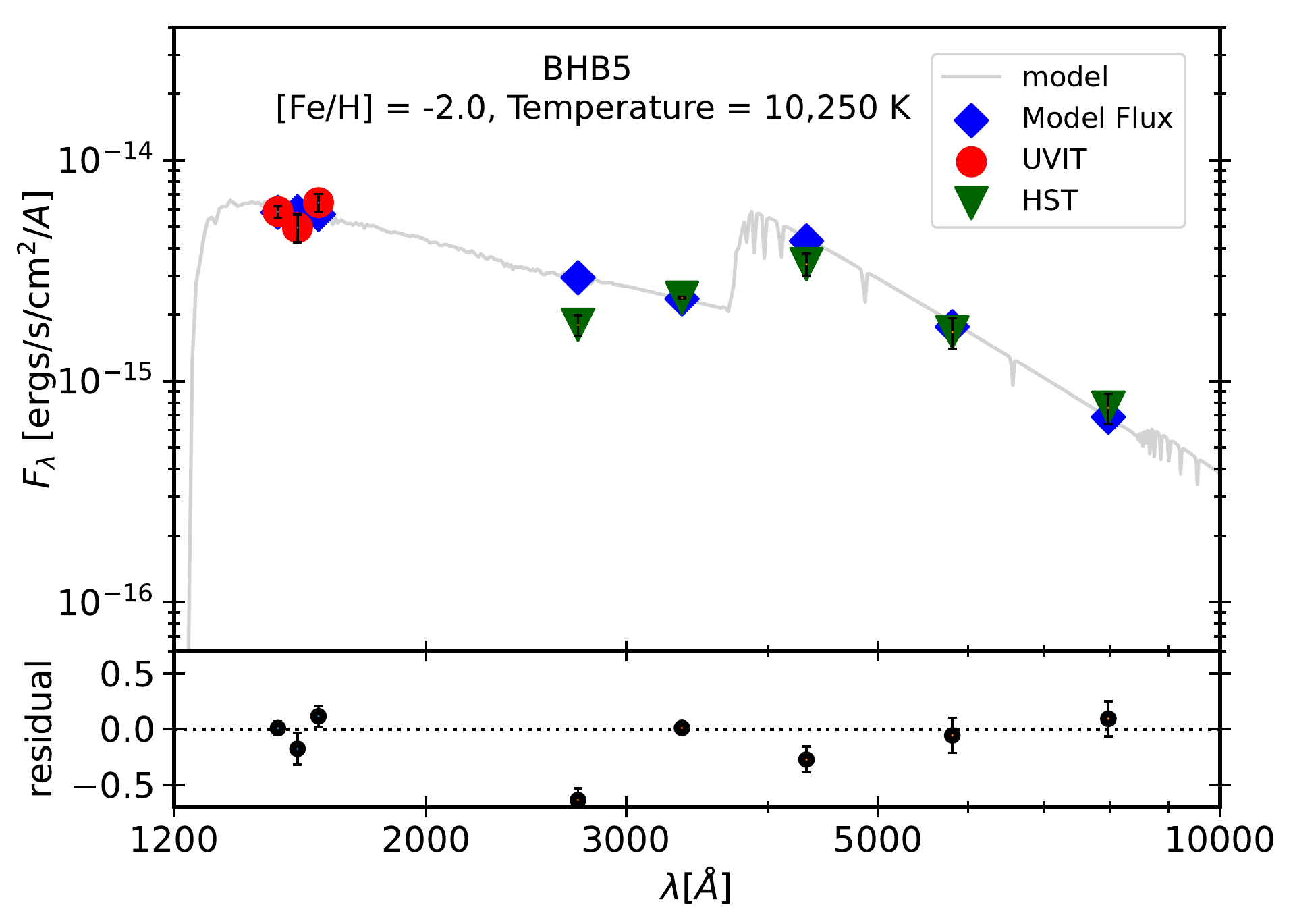}
\includegraphics[width=0.32\columnwidth]{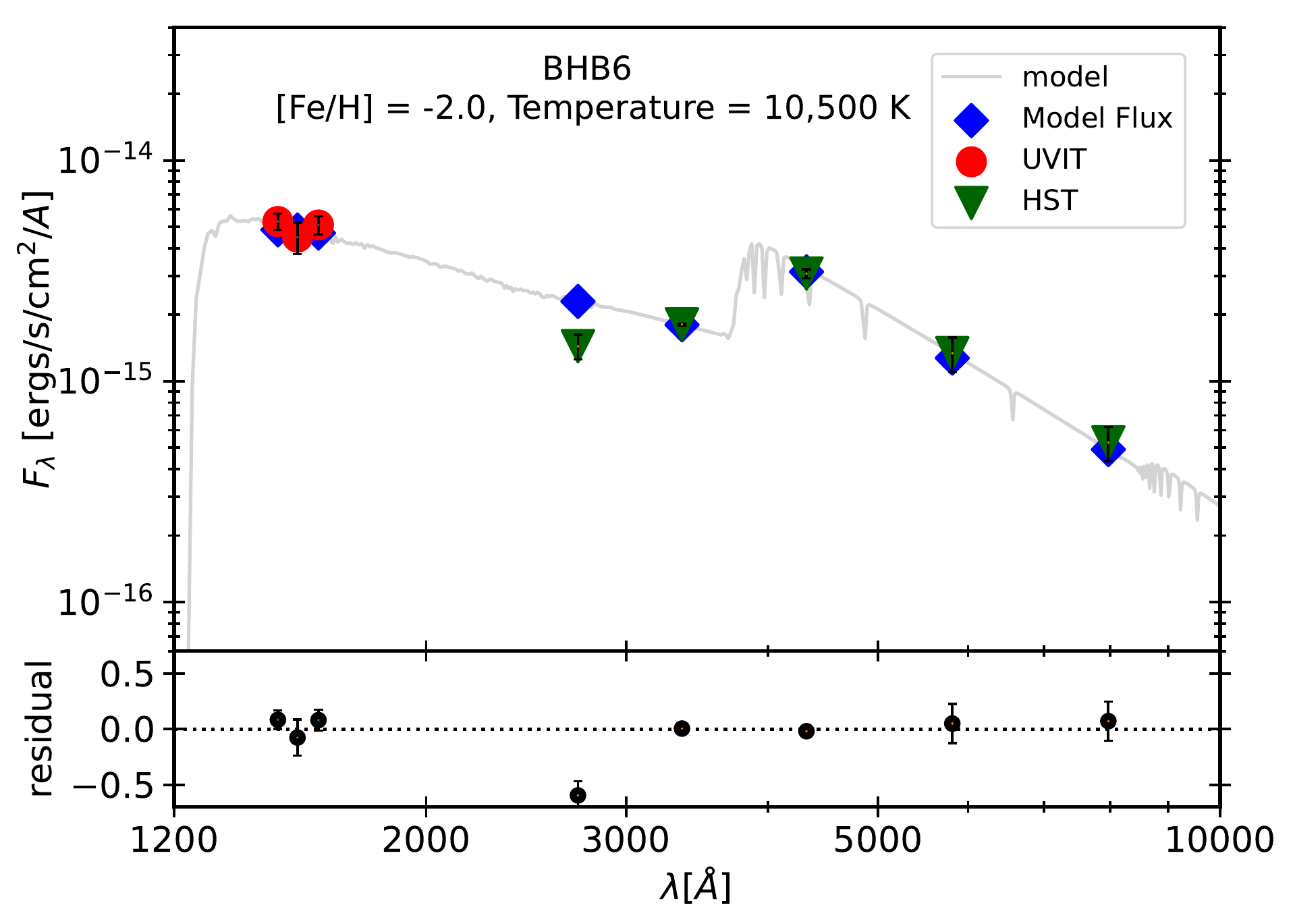}
\includegraphics[width=0.32\columnwidth]{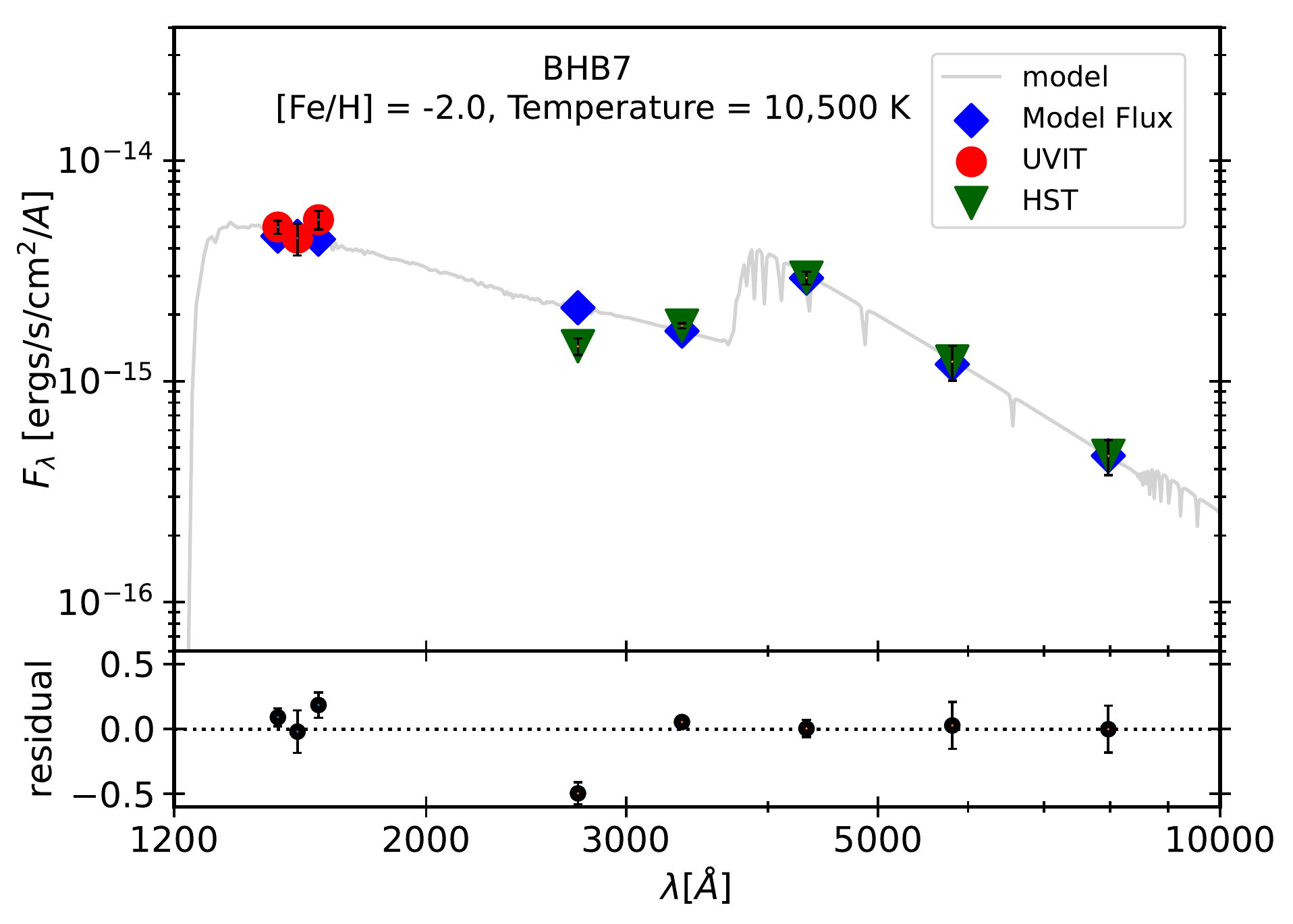}
\includegraphics[width=0.32\columnwidth]{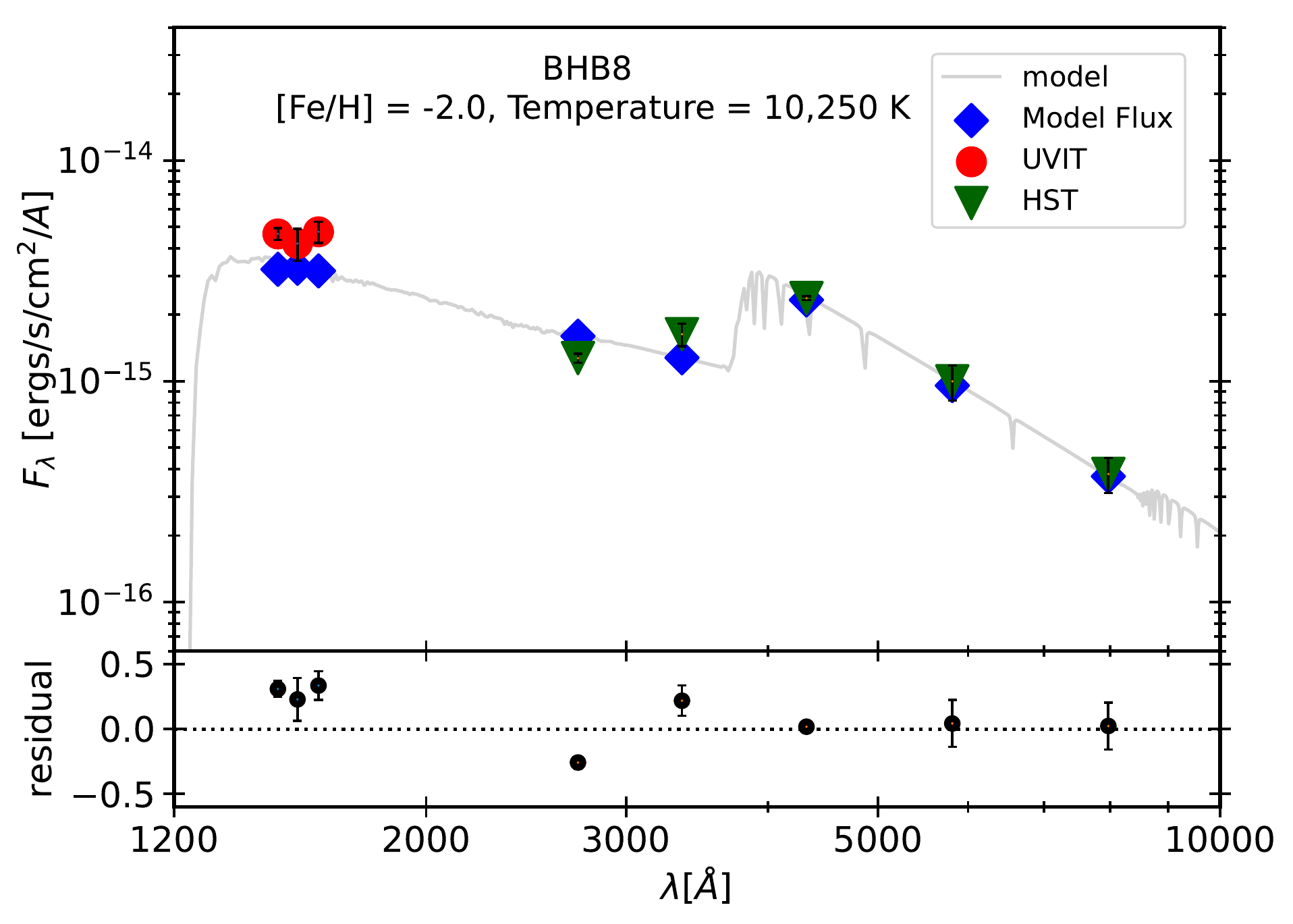}
\includegraphics[width=0.32\columnwidth]{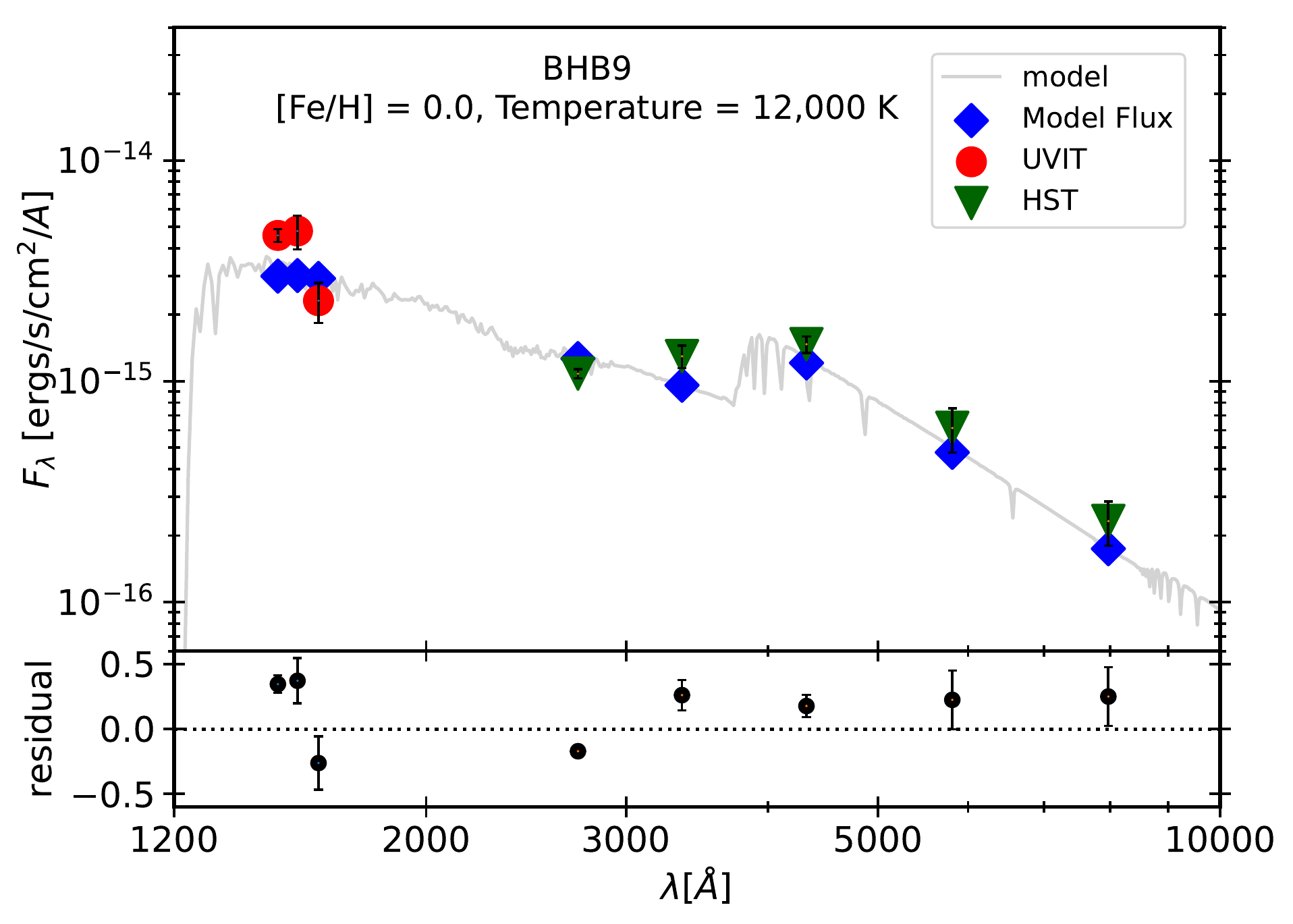}
\includegraphics[width=0.32\columnwidth]{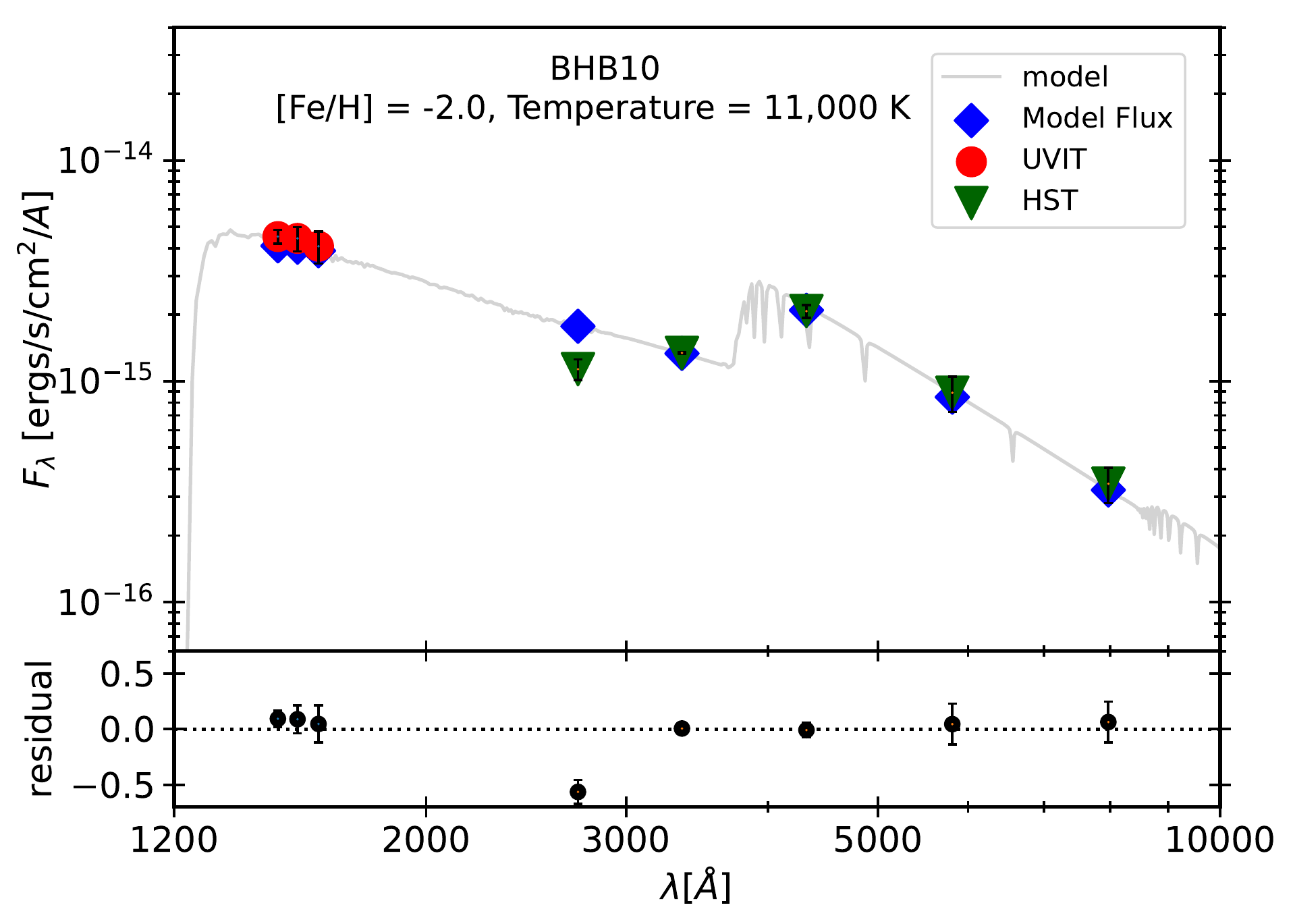}
\includegraphics[width=0.32\columnwidth]{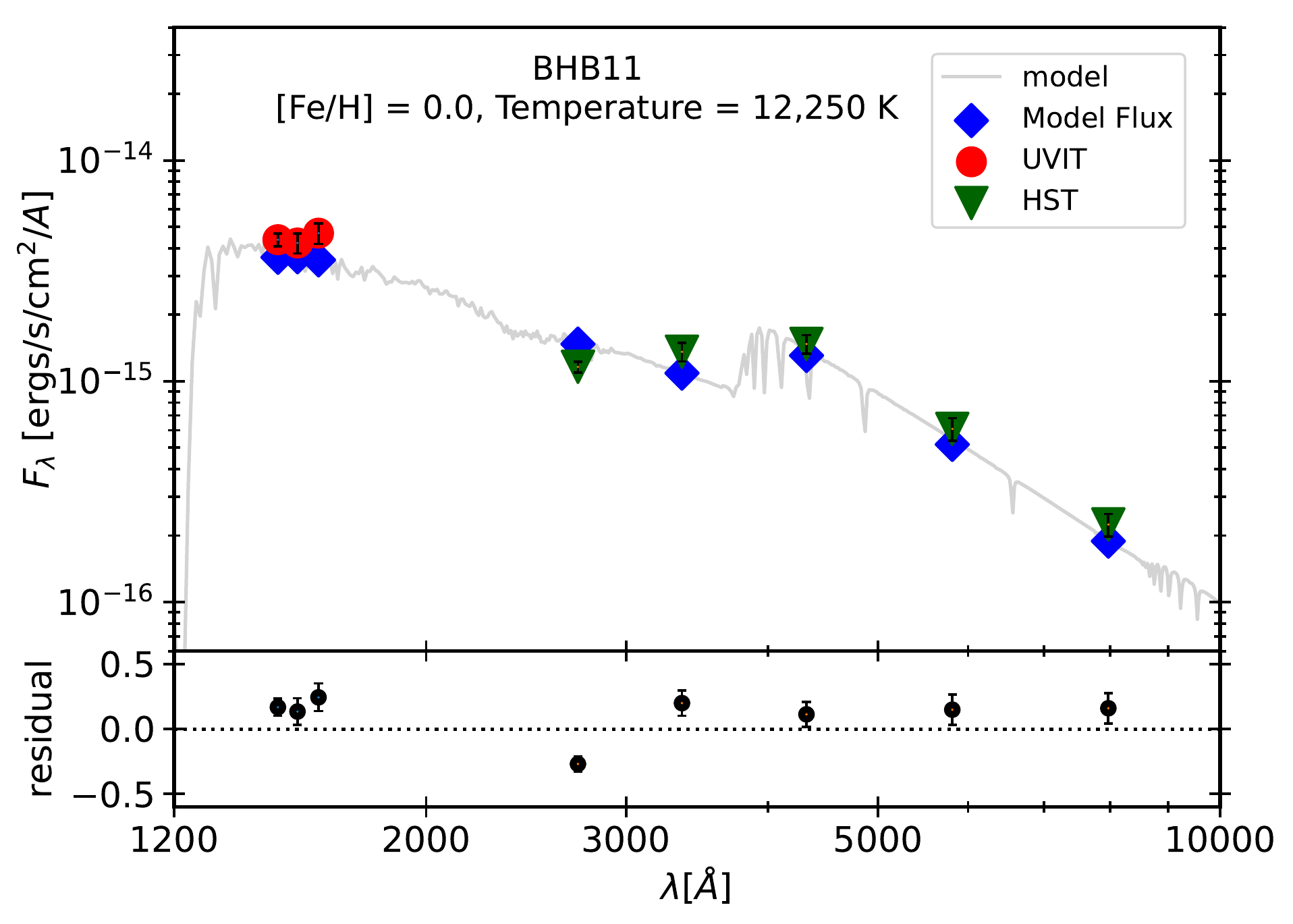}
\includegraphics[width=0.32\columnwidth]{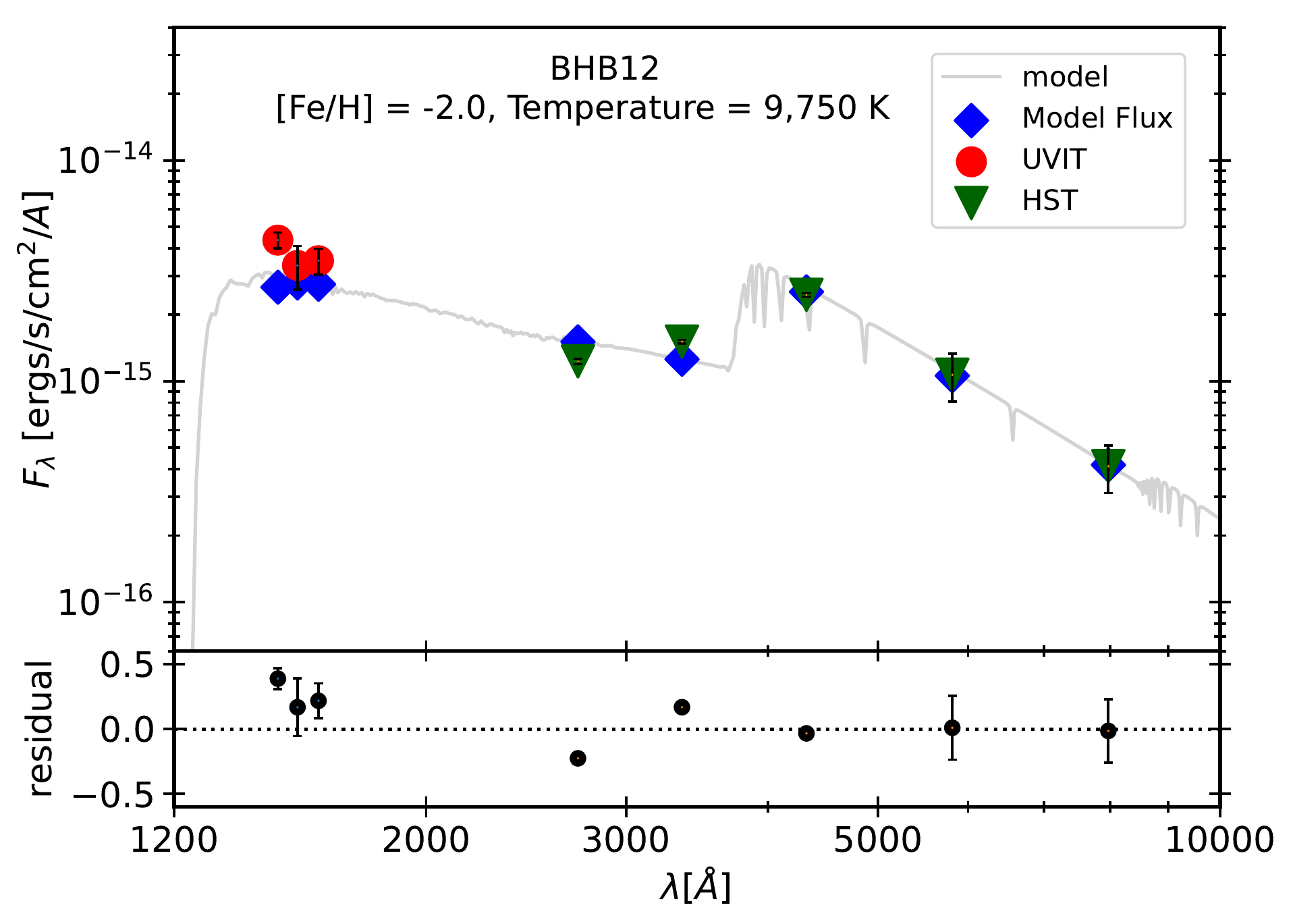}
\includegraphics[width=0.32\columnwidth]{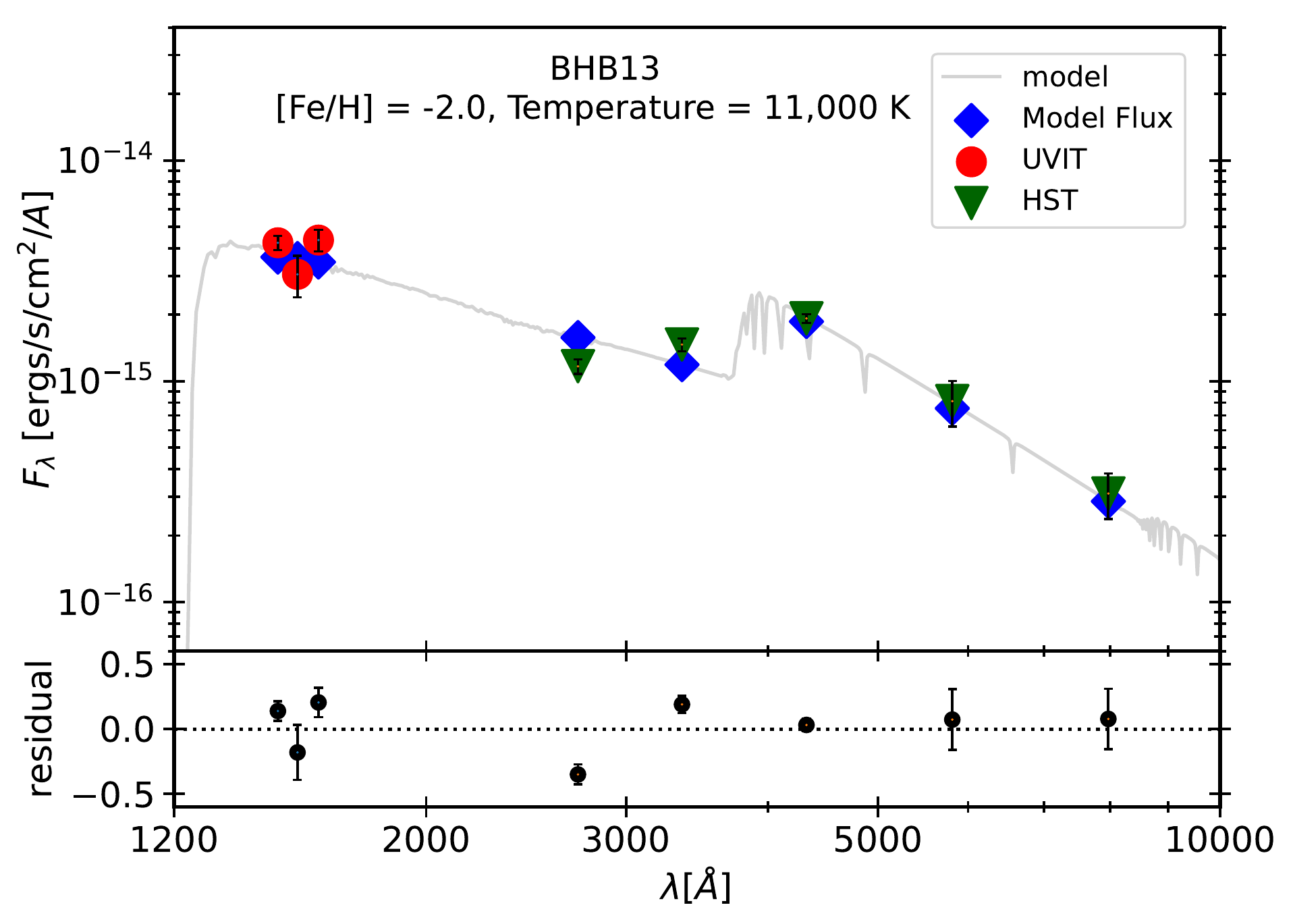}
\includegraphics[width=0.32\columnwidth]{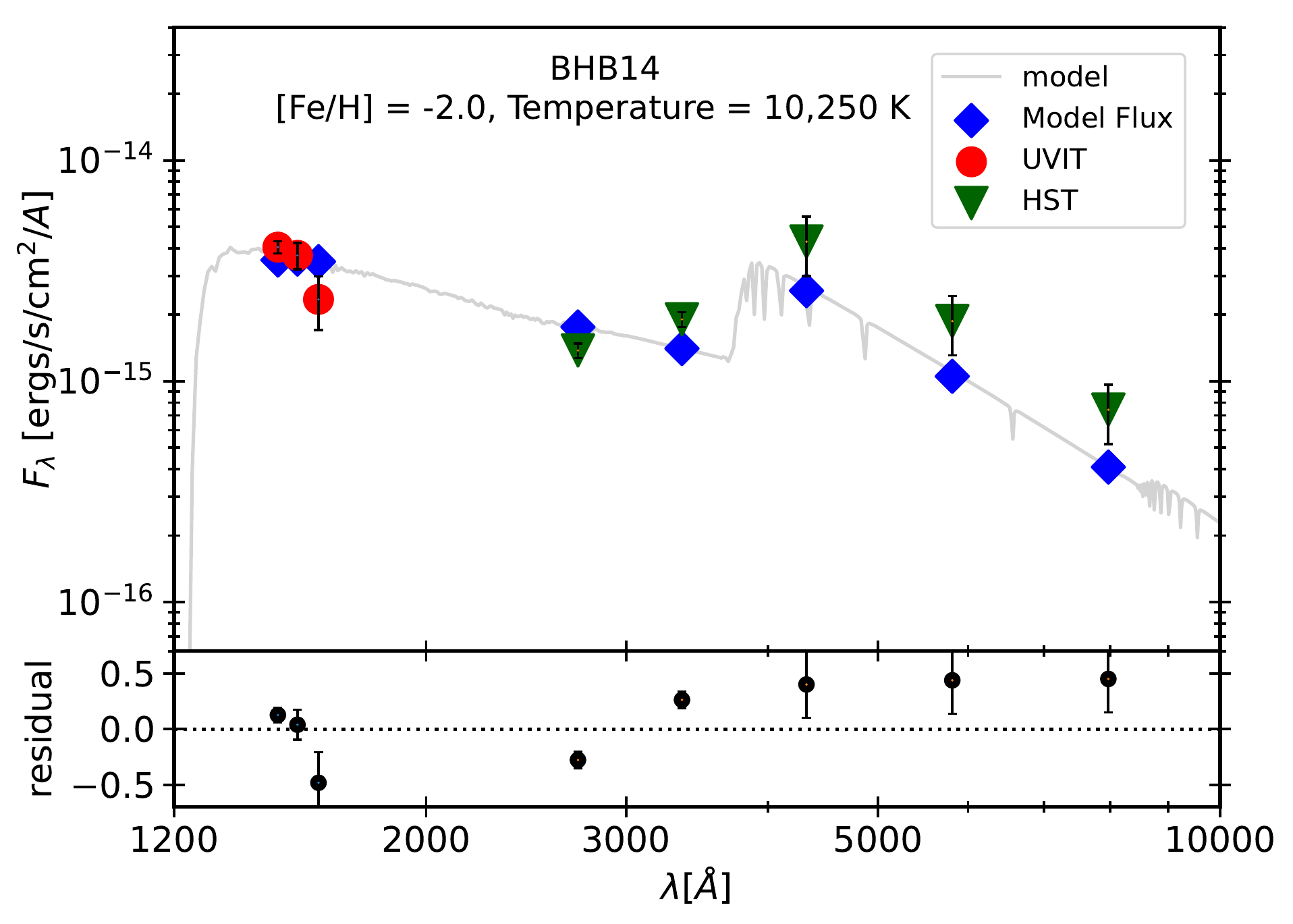}
\includegraphics[width=0.32\columnwidth]{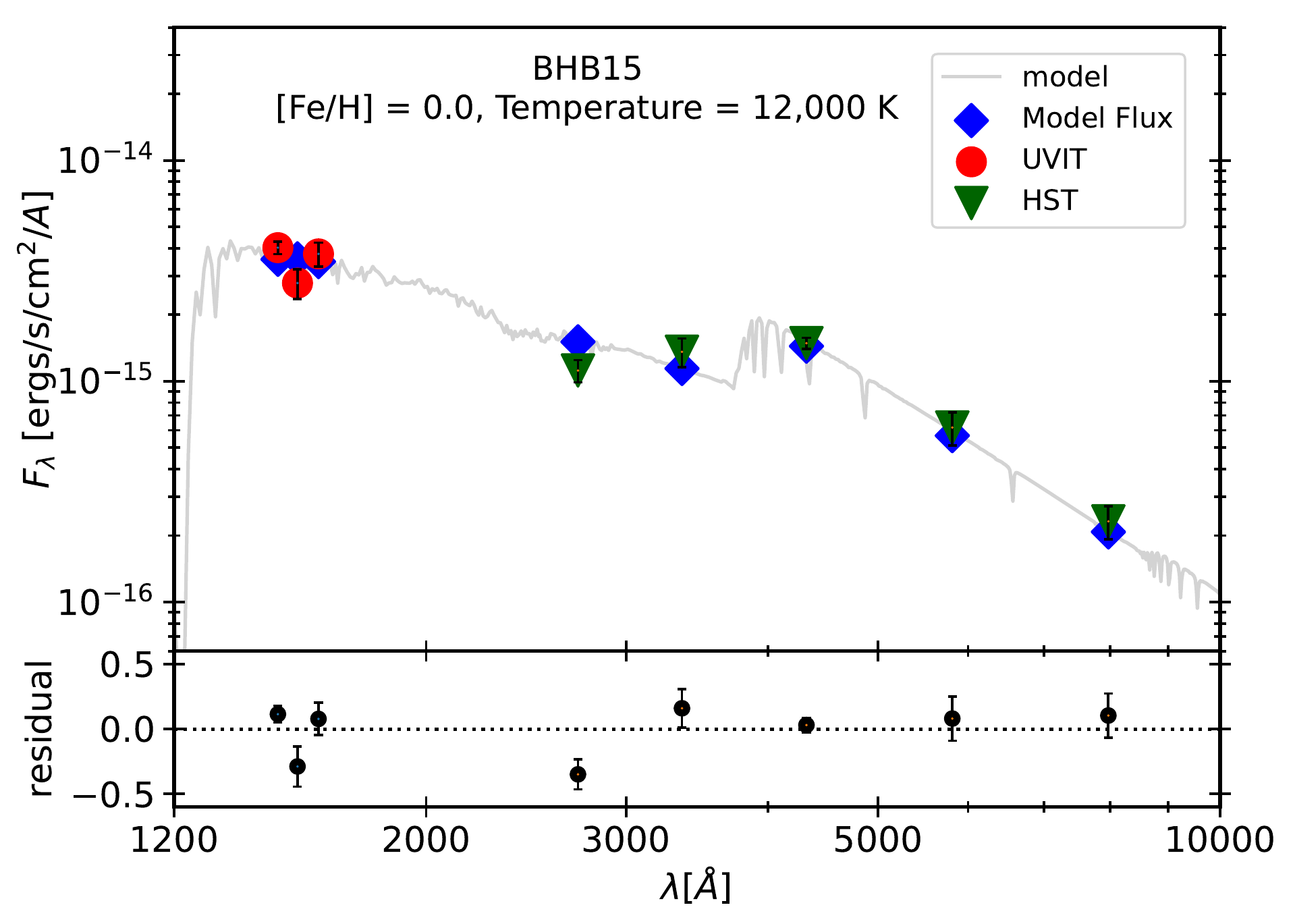}
 \caption{SEDs of rest of the BHB stars. The best-fit parameters are mentioned in the figure. The UVIT, \textit{HST} data points used to create SEDs for stars lying in the inner region are shown with red circles and green inverted triangles, respectively. For the stars lying in the outer region, UVIT, \textit{GALEX}, Ground-based photometric, \textit{Gaia} EDR3, and 2MASS data points are shown with red circles, green squares, orange triangles, cyan inverted triangles, and purple diamonds, respectively.}
 \label{sed1}
\end{figure*}

\renewcommand{\thefigure}{A\arabic{figure}}
\addtocounter{figure}{-1}
\begin{figure*}[htb]
\centering
\includegraphics[width=0.32\columnwidth]{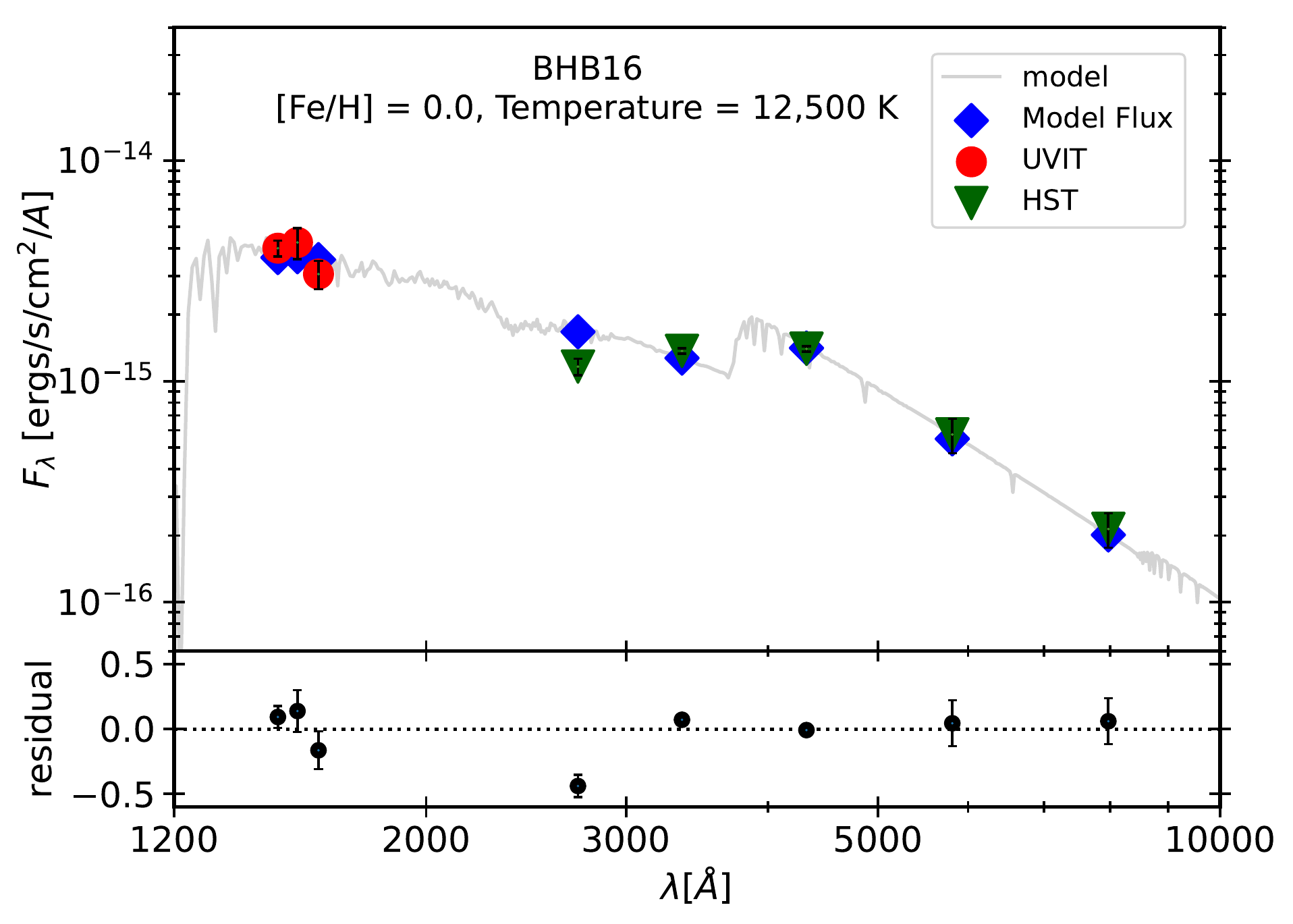}
\includegraphics[width=0.32\columnwidth]{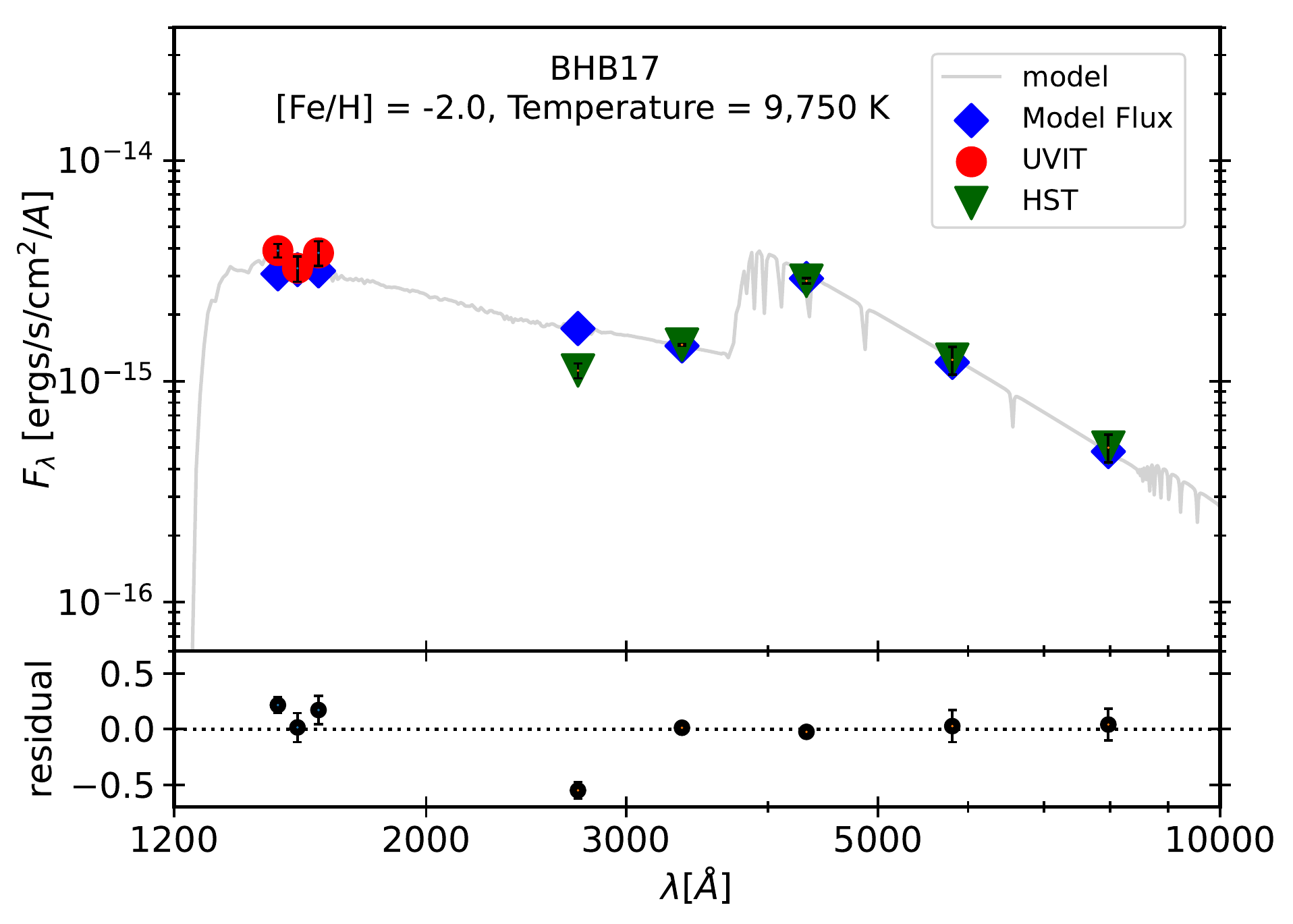}
\includegraphics[width=0.32\columnwidth]{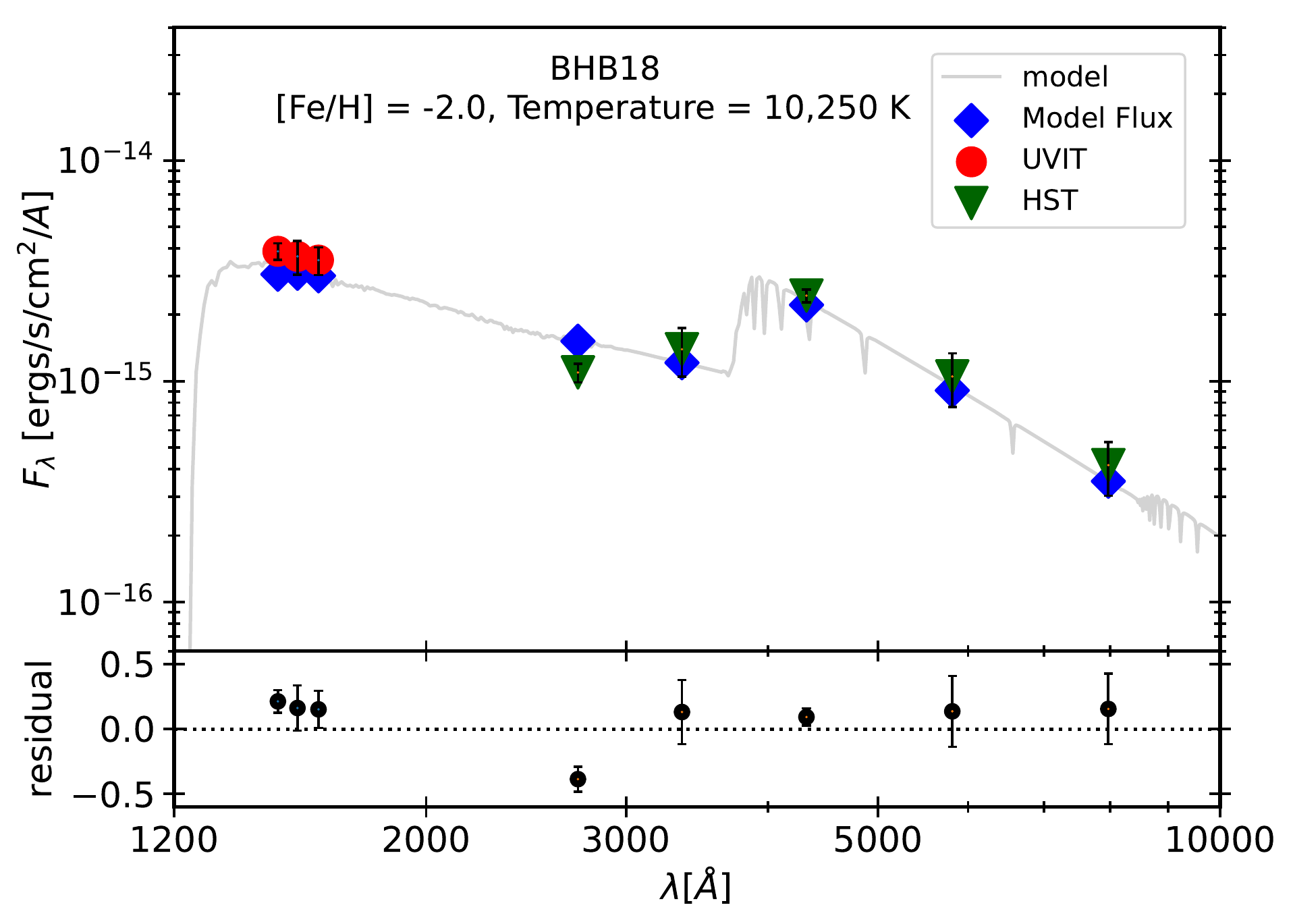}
\includegraphics[width=0.32\columnwidth]{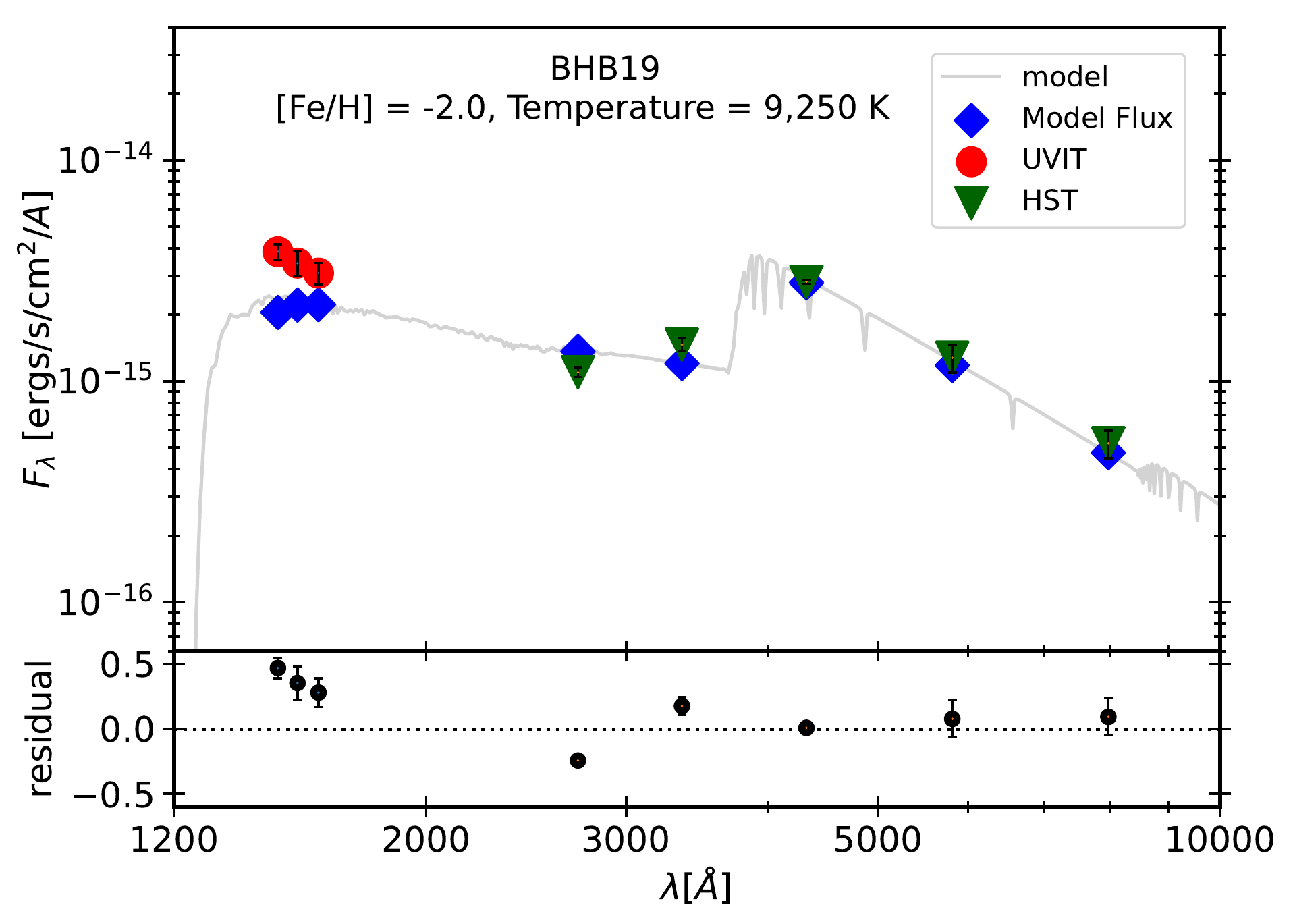}
\includegraphics[width=0.32\columnwidth]{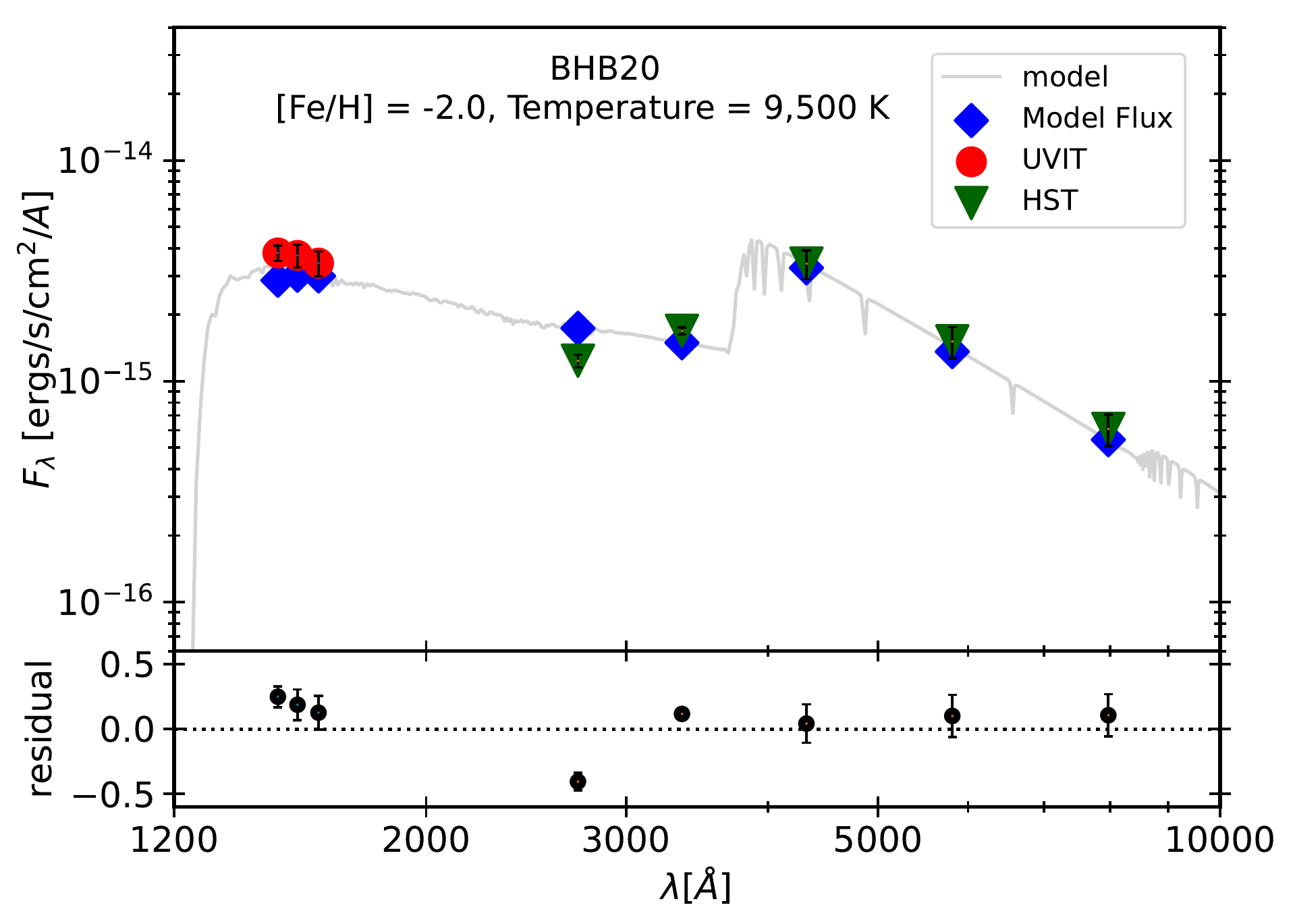}
\includegraphics[width=0.32\columnwidth]{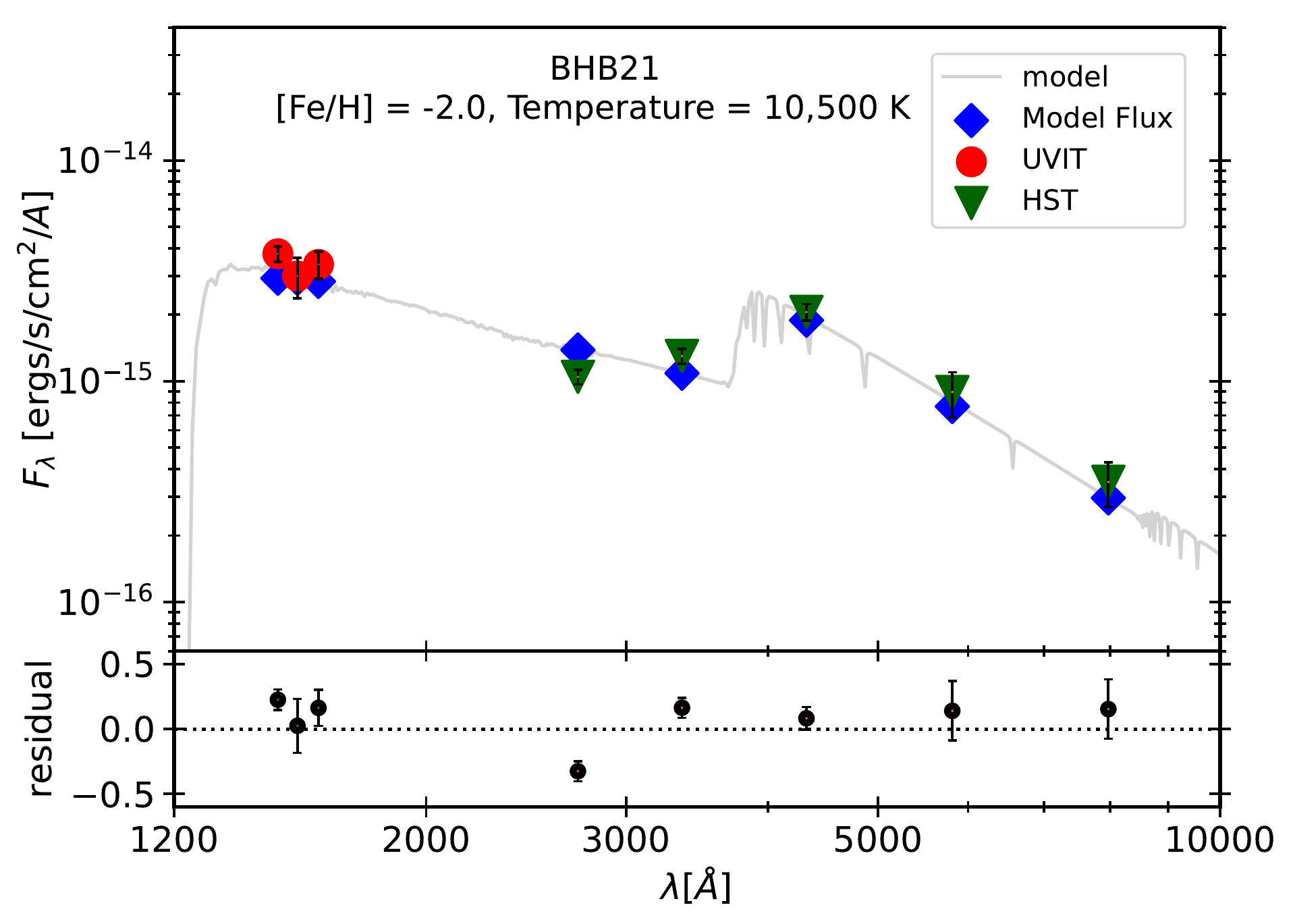}
\includegraphics[width=0.32\columnwidth]{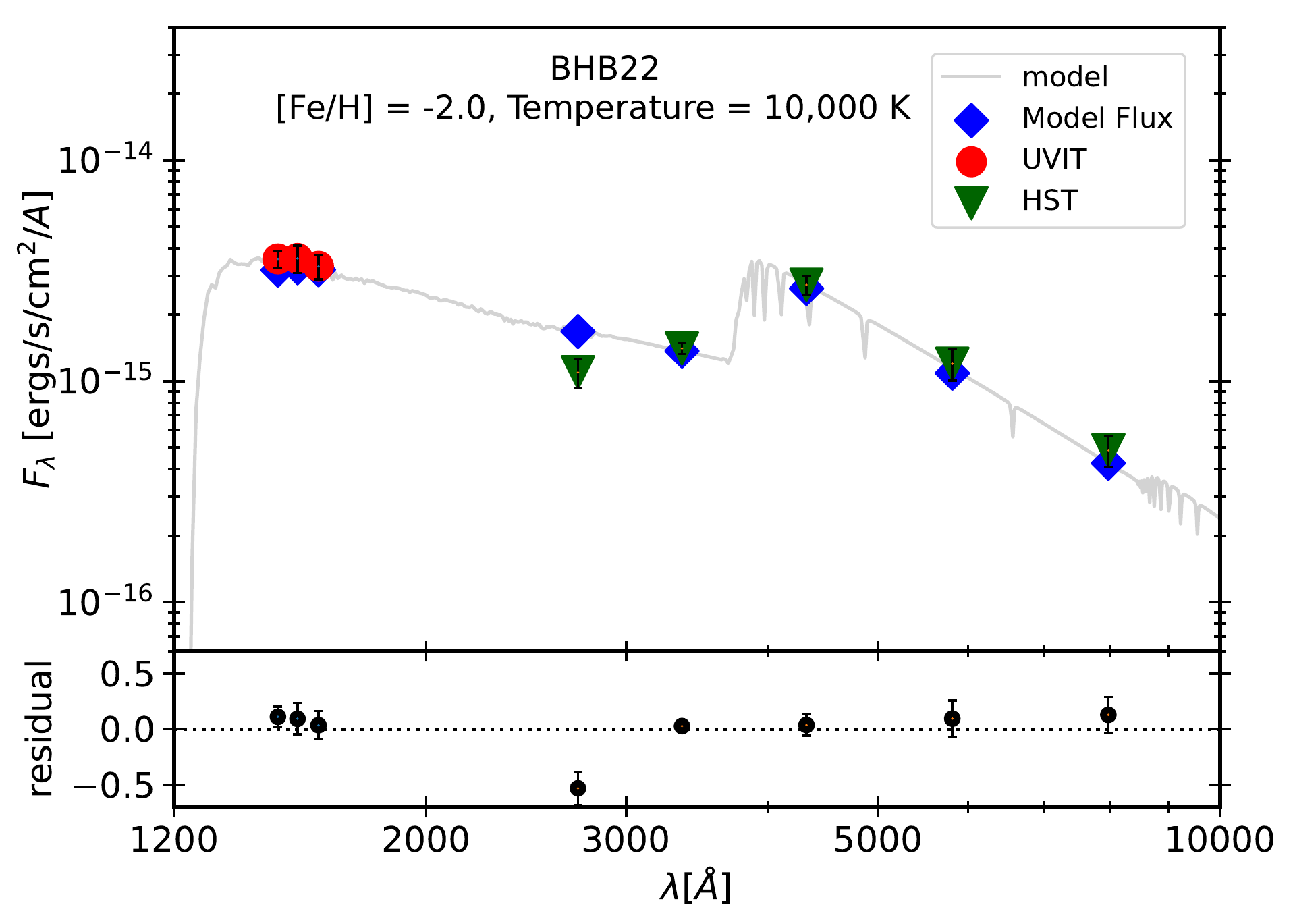}
\includegraphics[width=0.32\columnwidth]{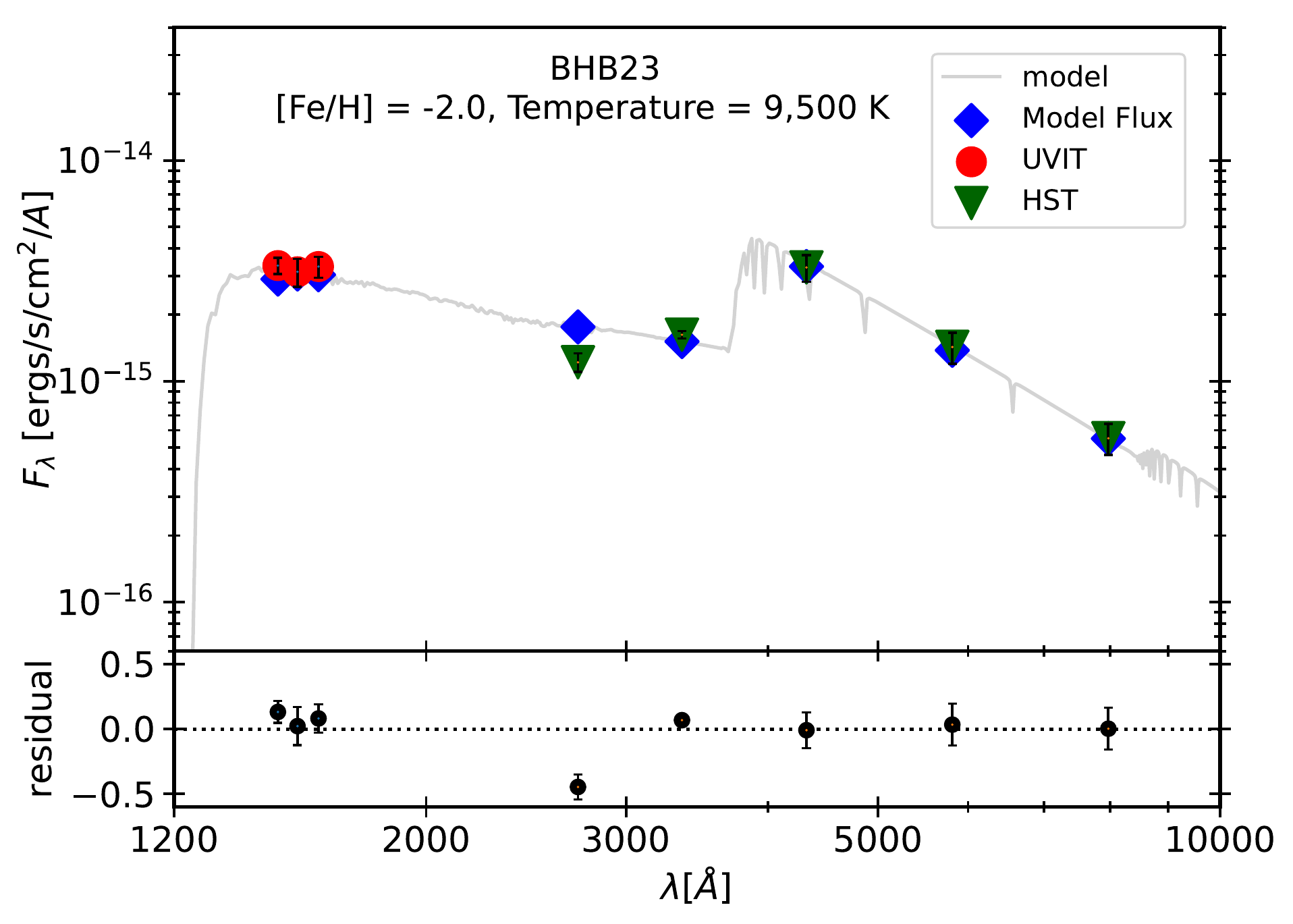}
\includegraphics[width=0.32\columnwidth]{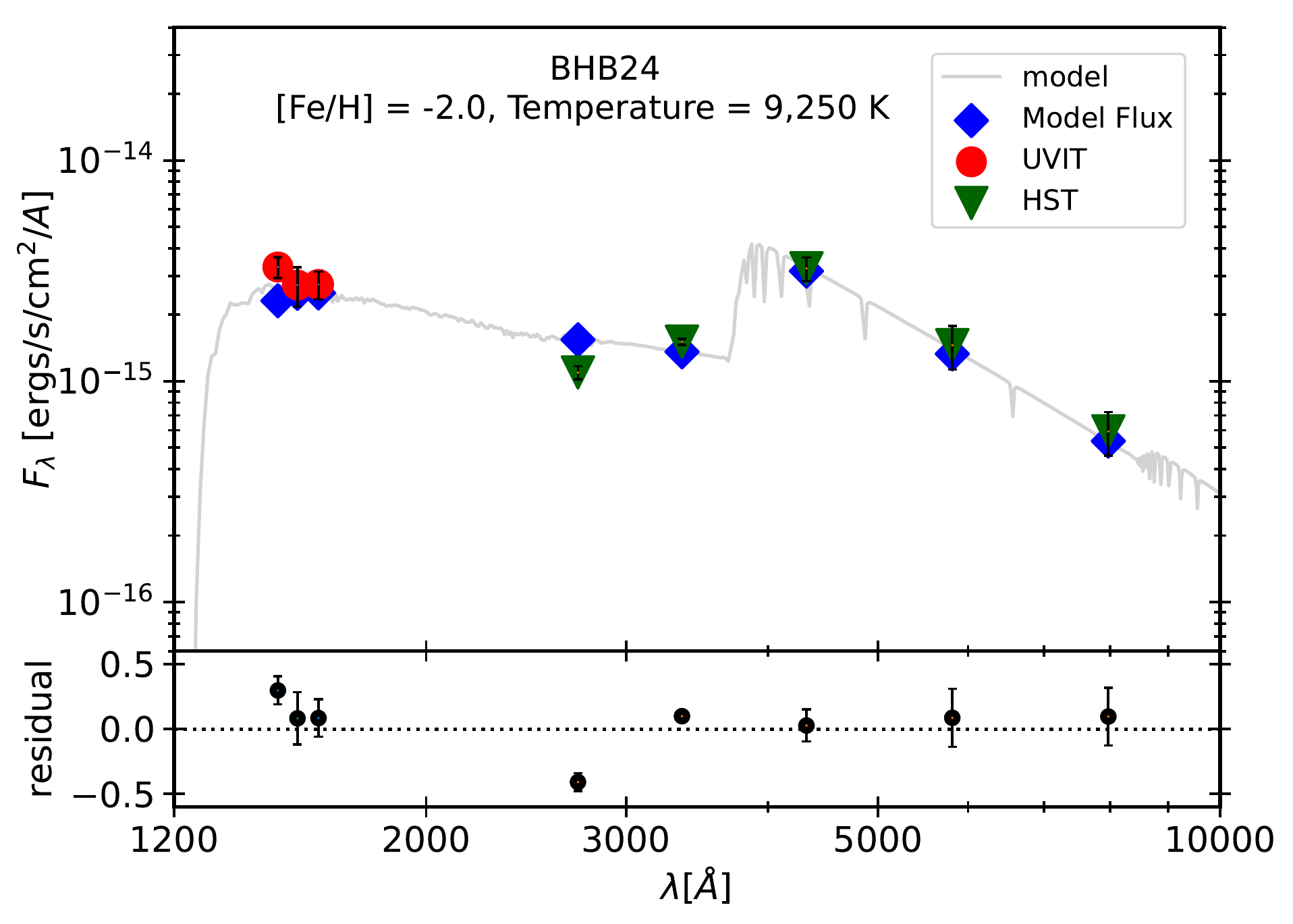}
\includegraphics[width=0.32\columnwidth]{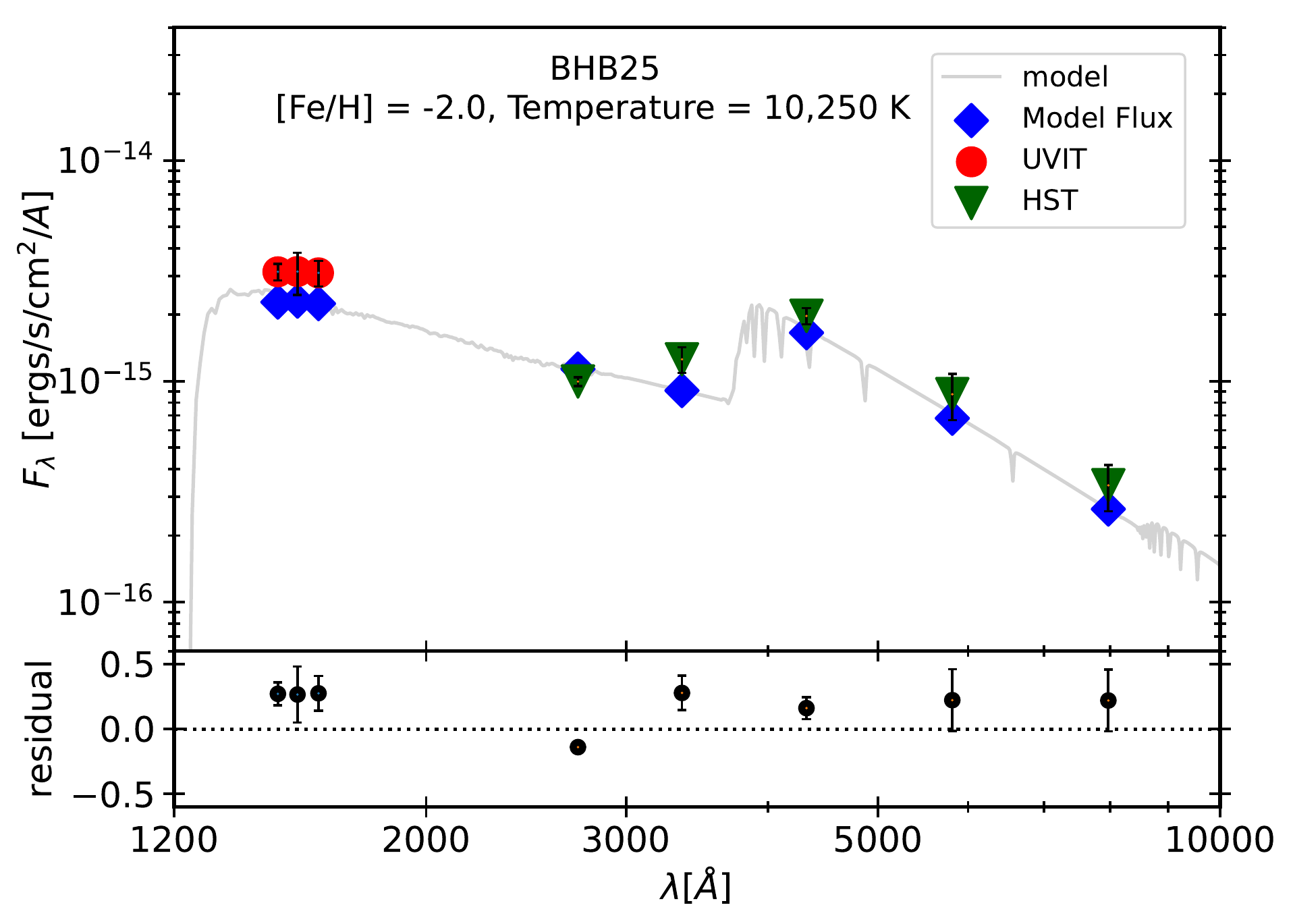}
\includegraphics[width=0.32\columnwidth]{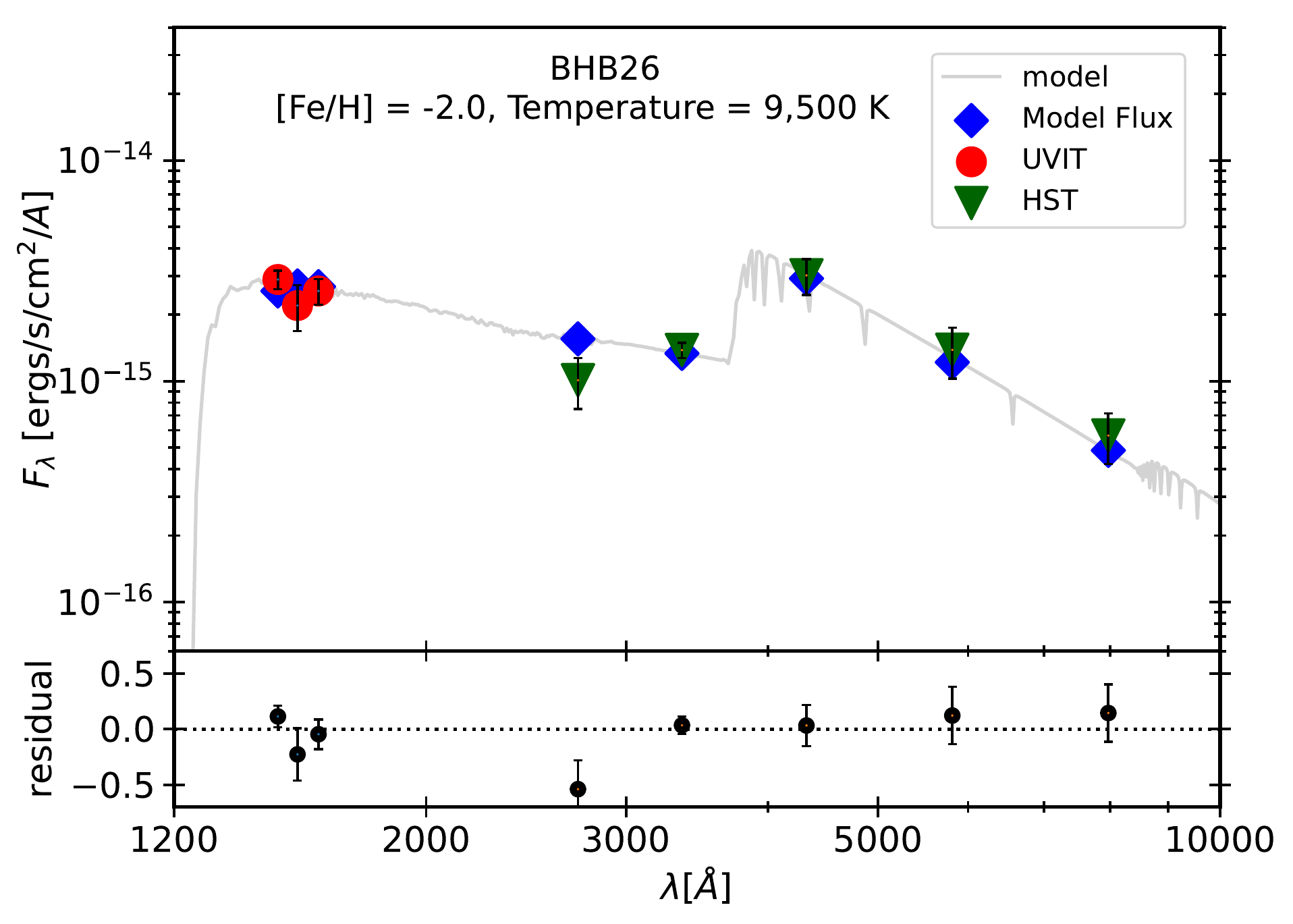}
\includegraphics[width=0.32\columnwidth]{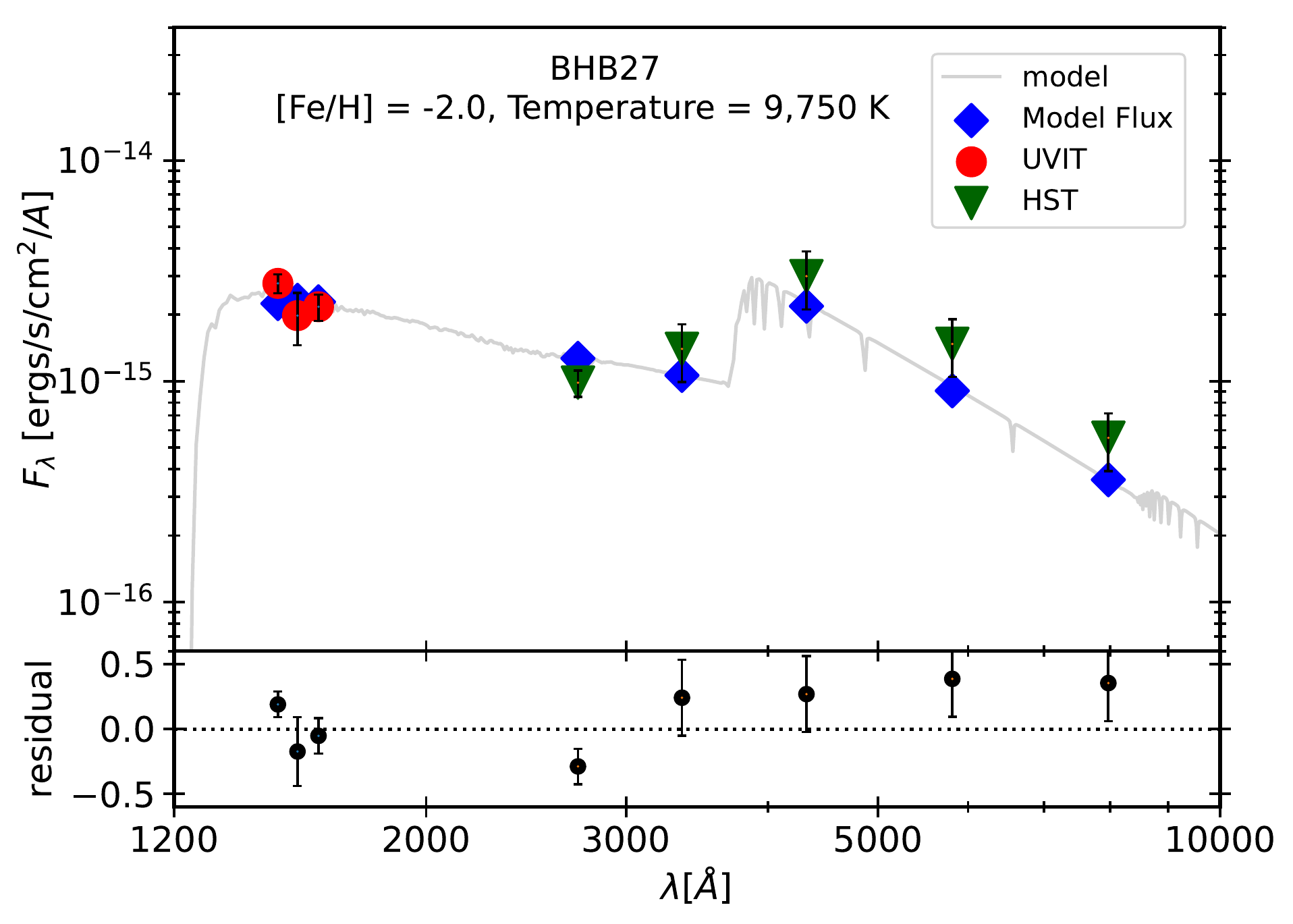}
\includegraphics[width=0.32\columnwidth]{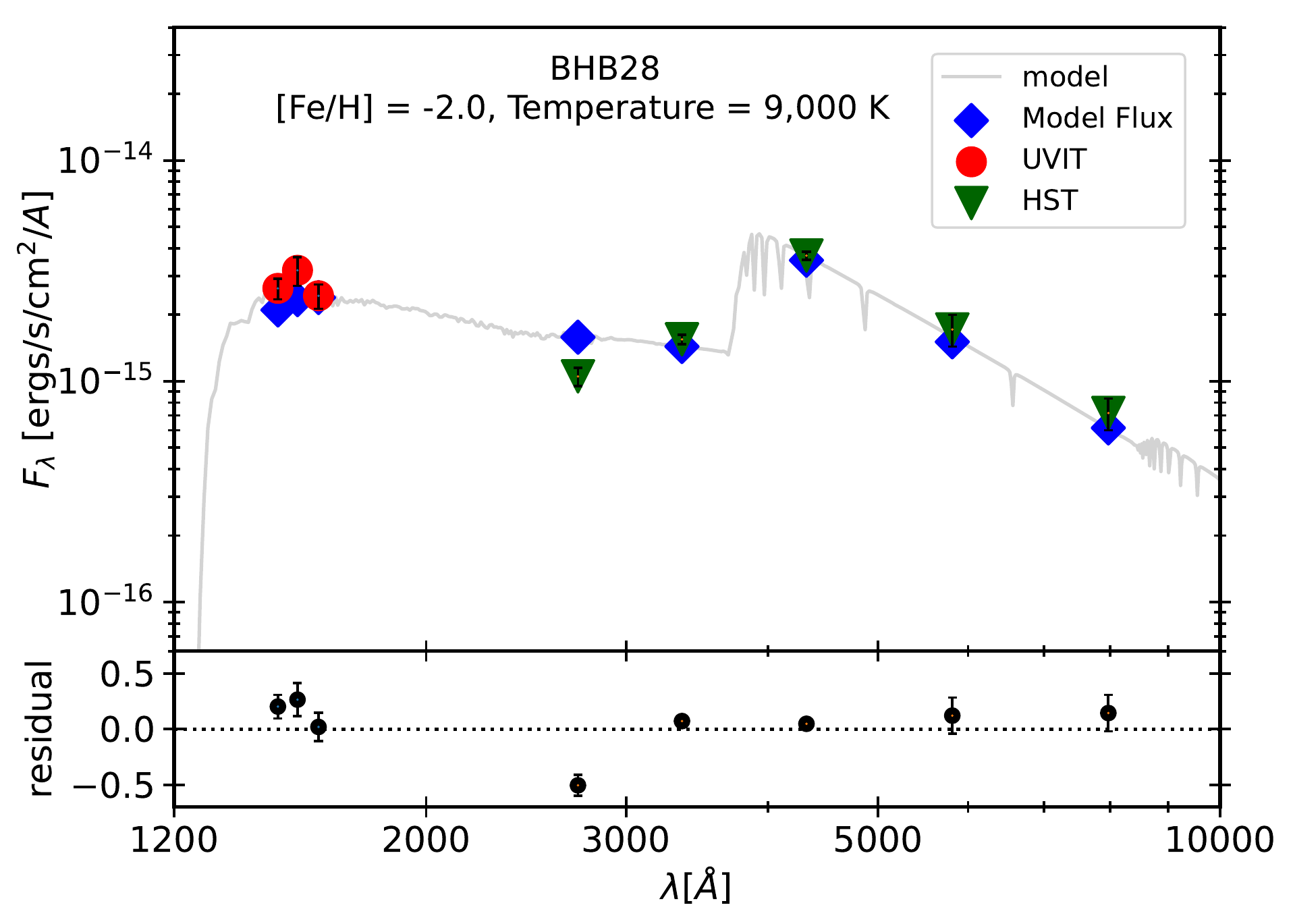}
\includegraphics[width=0.32\columnwidth]{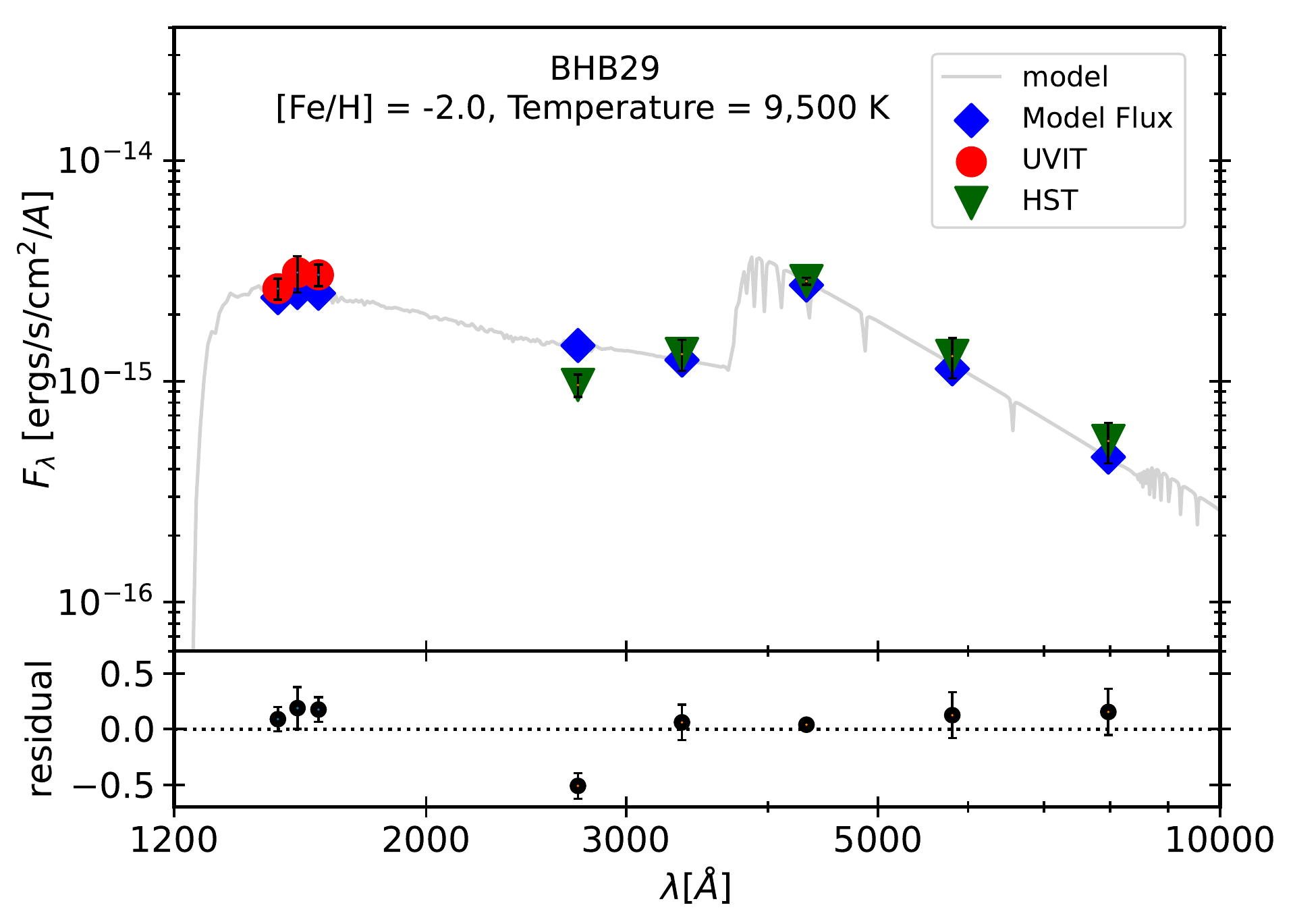}
\includegraphics[width=0.32\columnwidth]{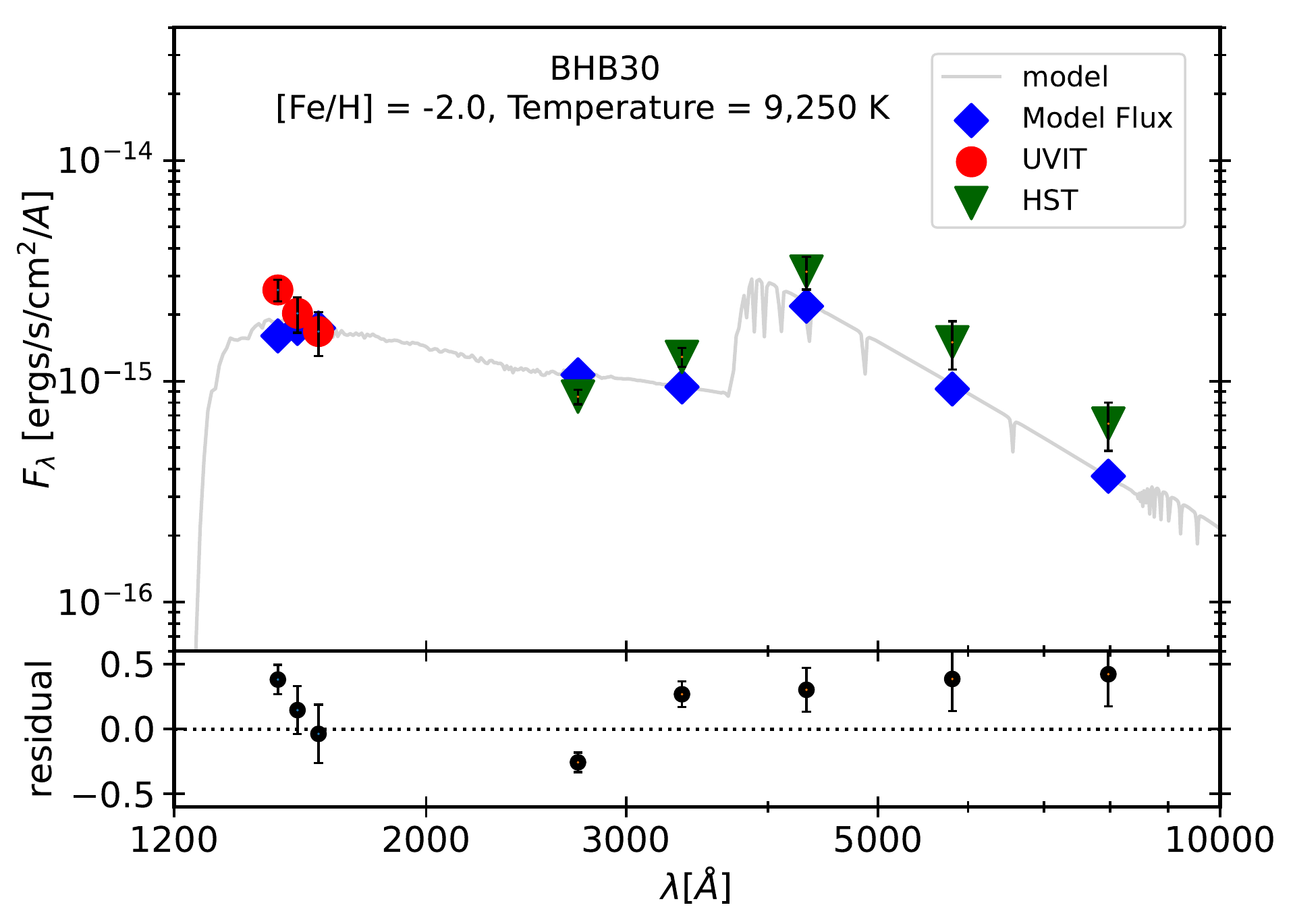}
 \caption{Continued.}
 \label{sed2}
\end{figure*}
\renewcommand{\thefigure}{\arabic{figure}}

\renewcommand{\thefigure}{A\arabic{figure}}
\addtocounter{figure}{-1}
\begin{figure*}[htb]
\centering
\includegraphics[width=0.32\columnwidth]{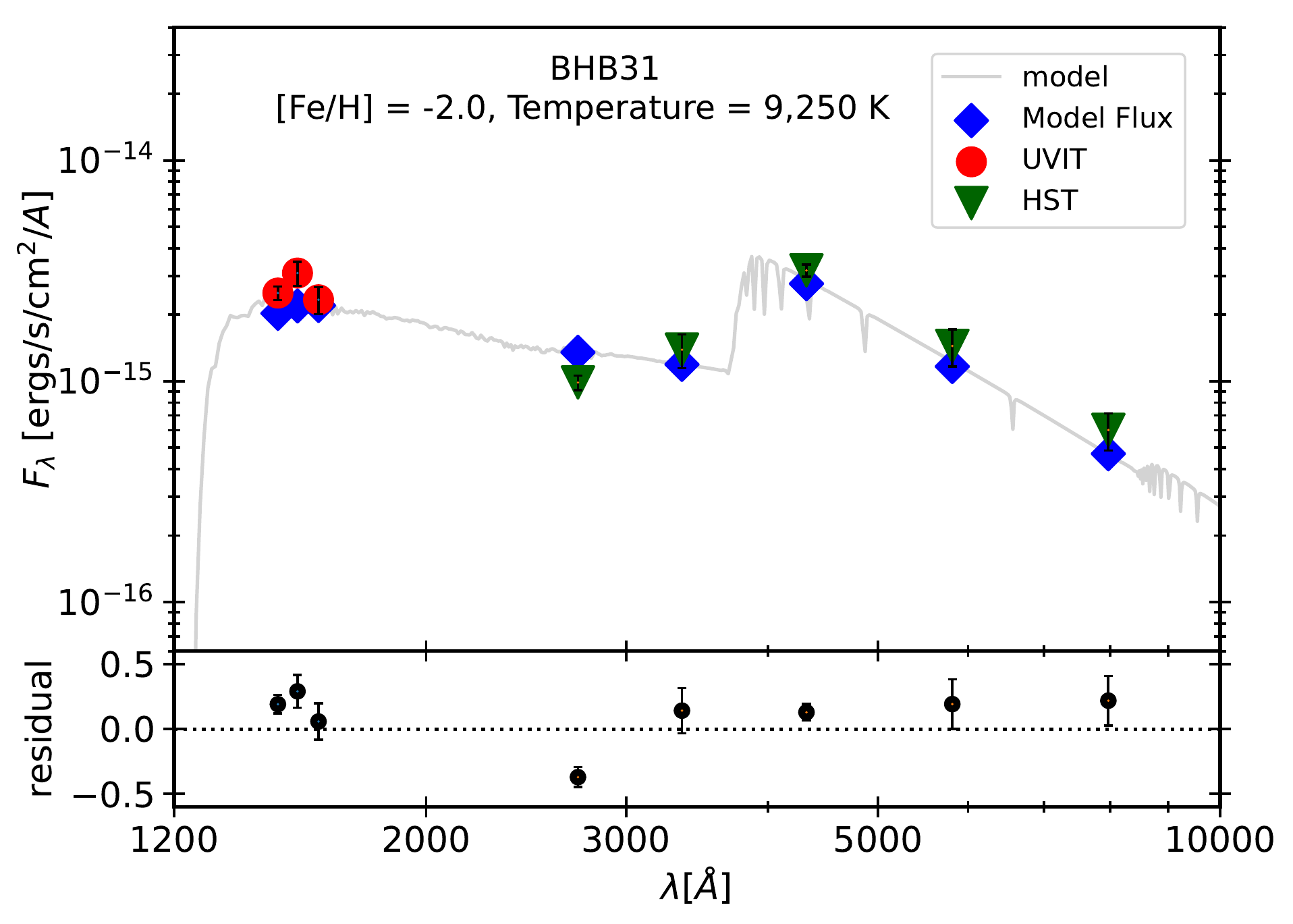}
\includegraphics[width=0.32\columnwidth]{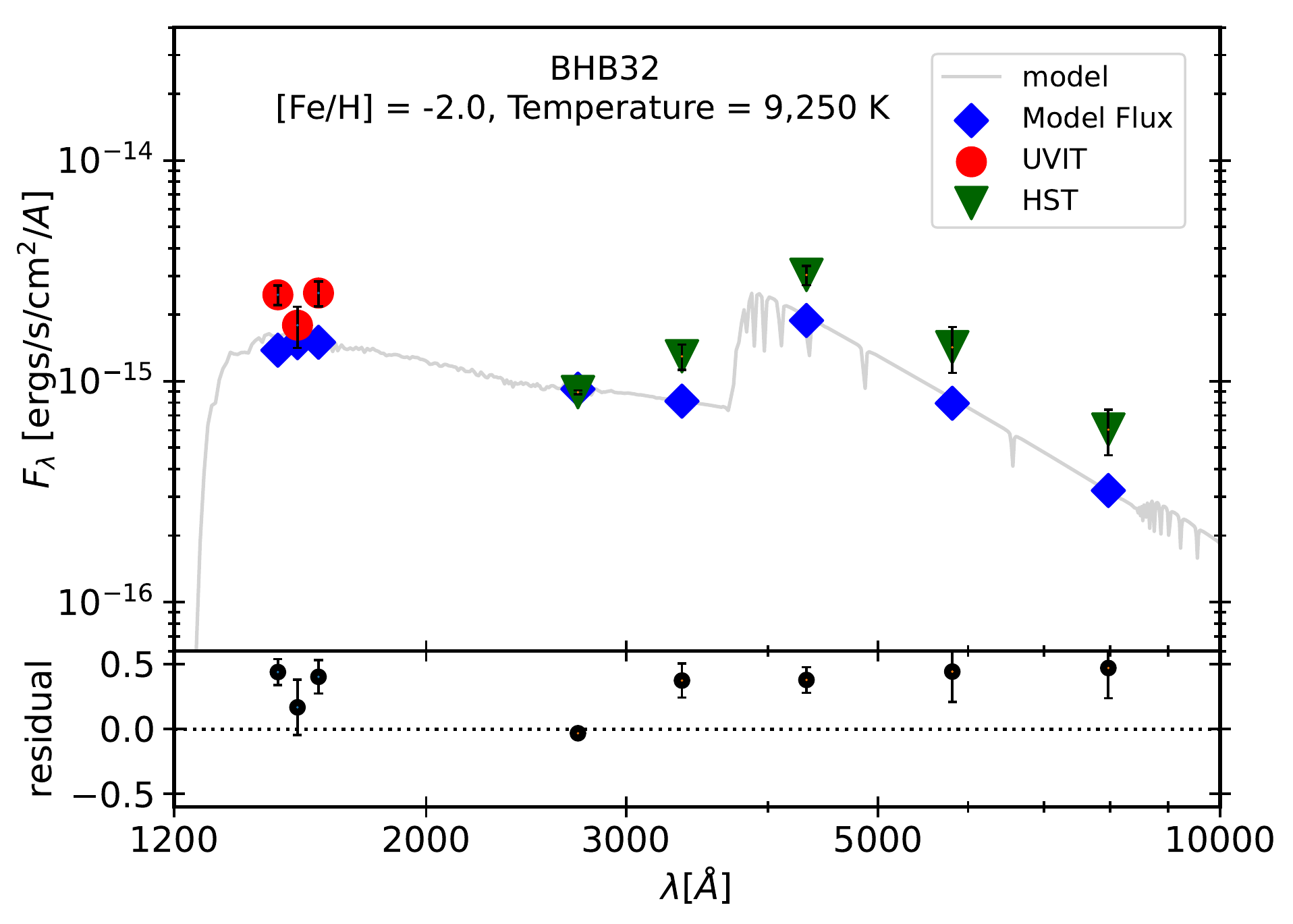}
\includegraphics[width=0.32\columnwidth]{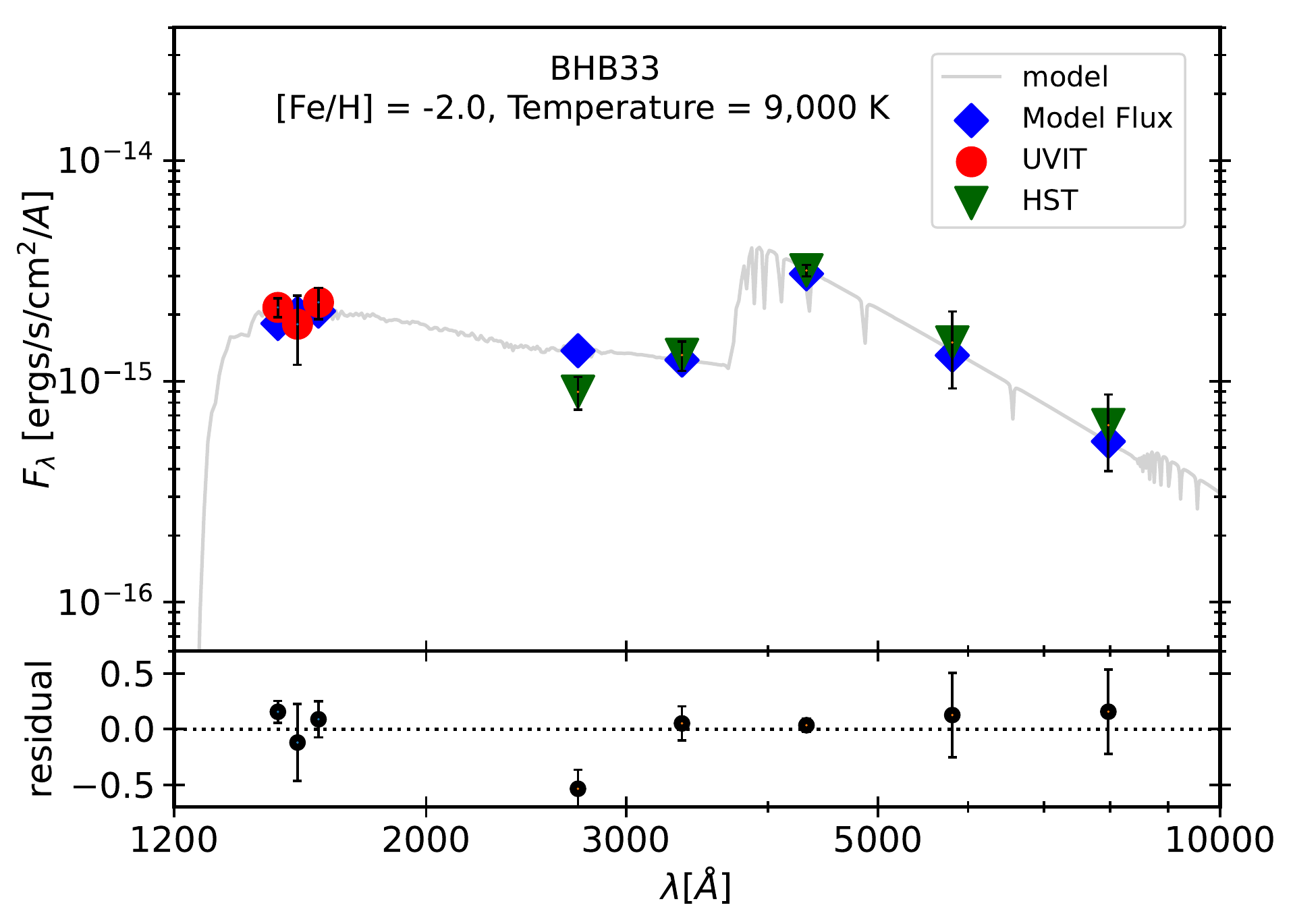}
\includegraphics[width=0.32\columnwidth]{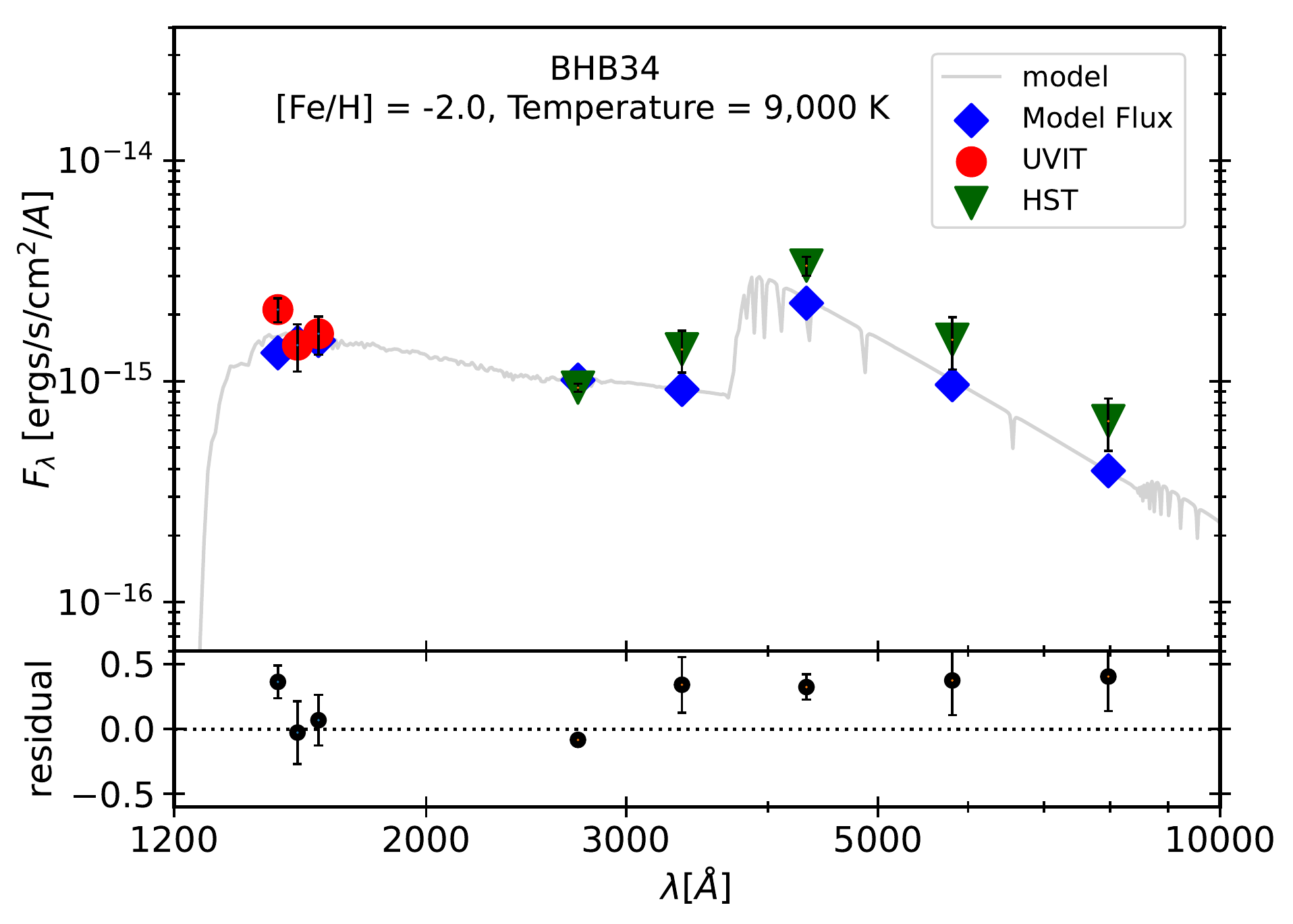}
\includegraphics[width=0.32\columnwidth]{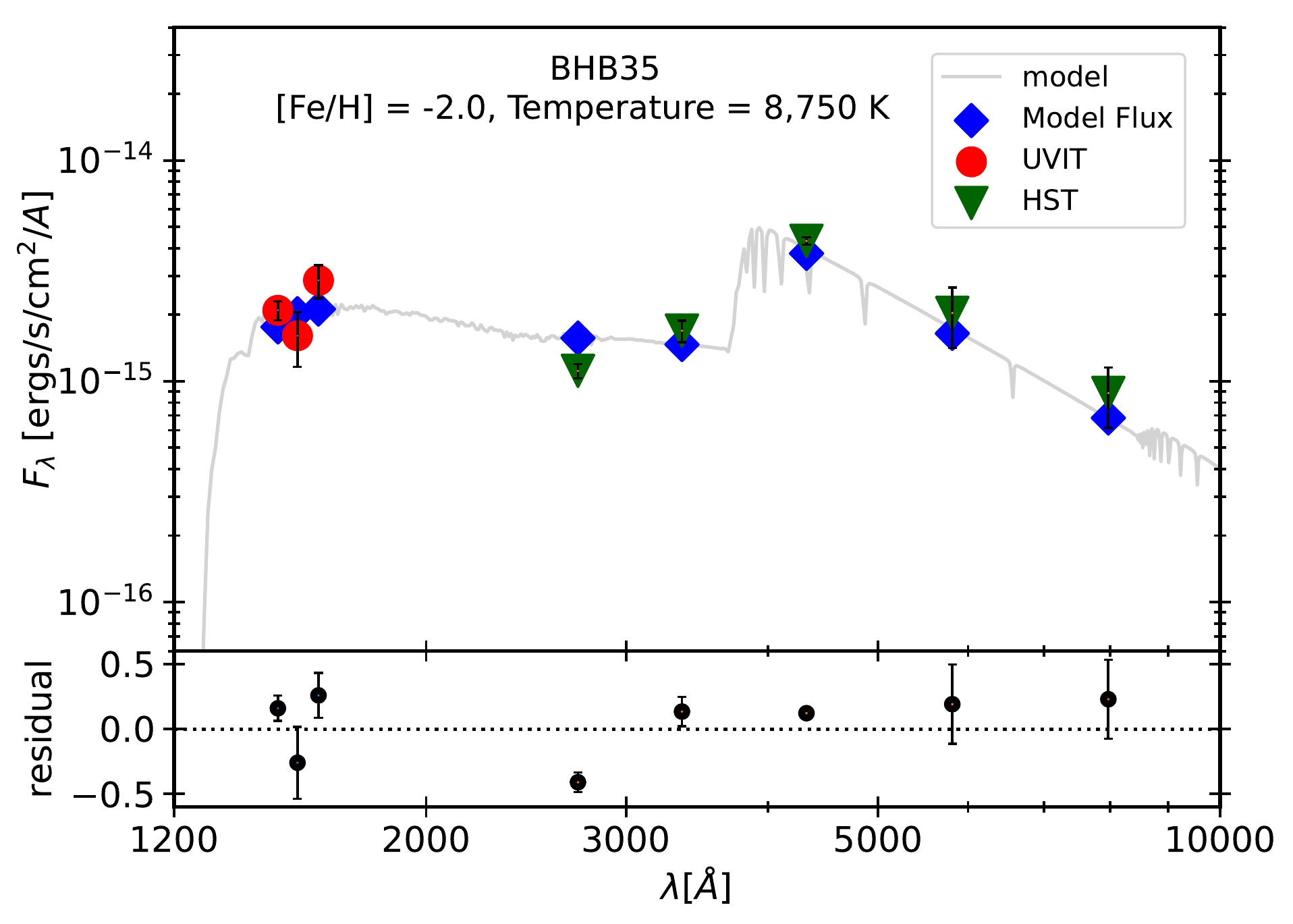}
\includegraphics[width=0.32\columnwidth]{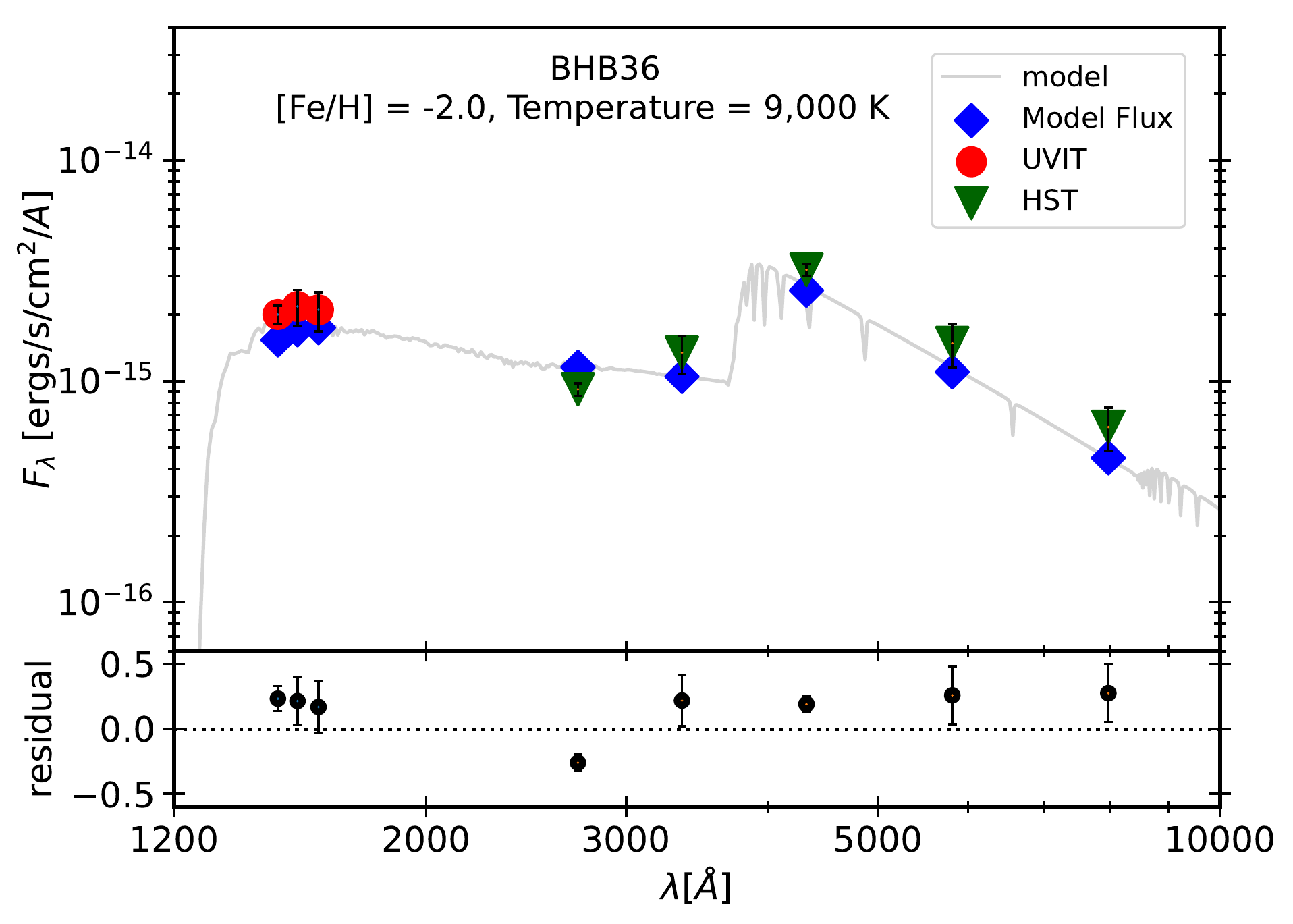}
\includegraphics[width=0.32\columnwidth]{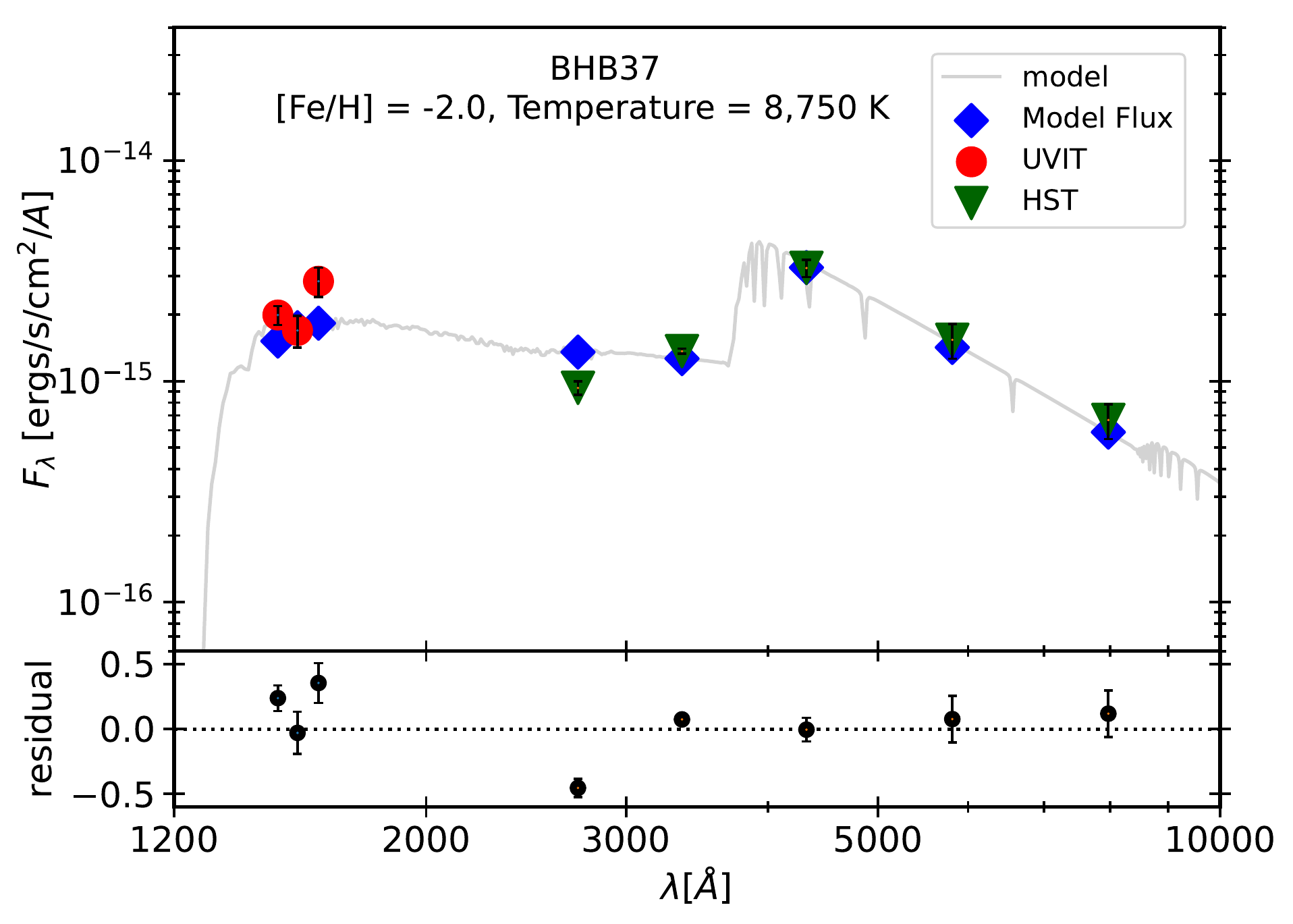}
\includegraphics[width=0.32\columnwidth]{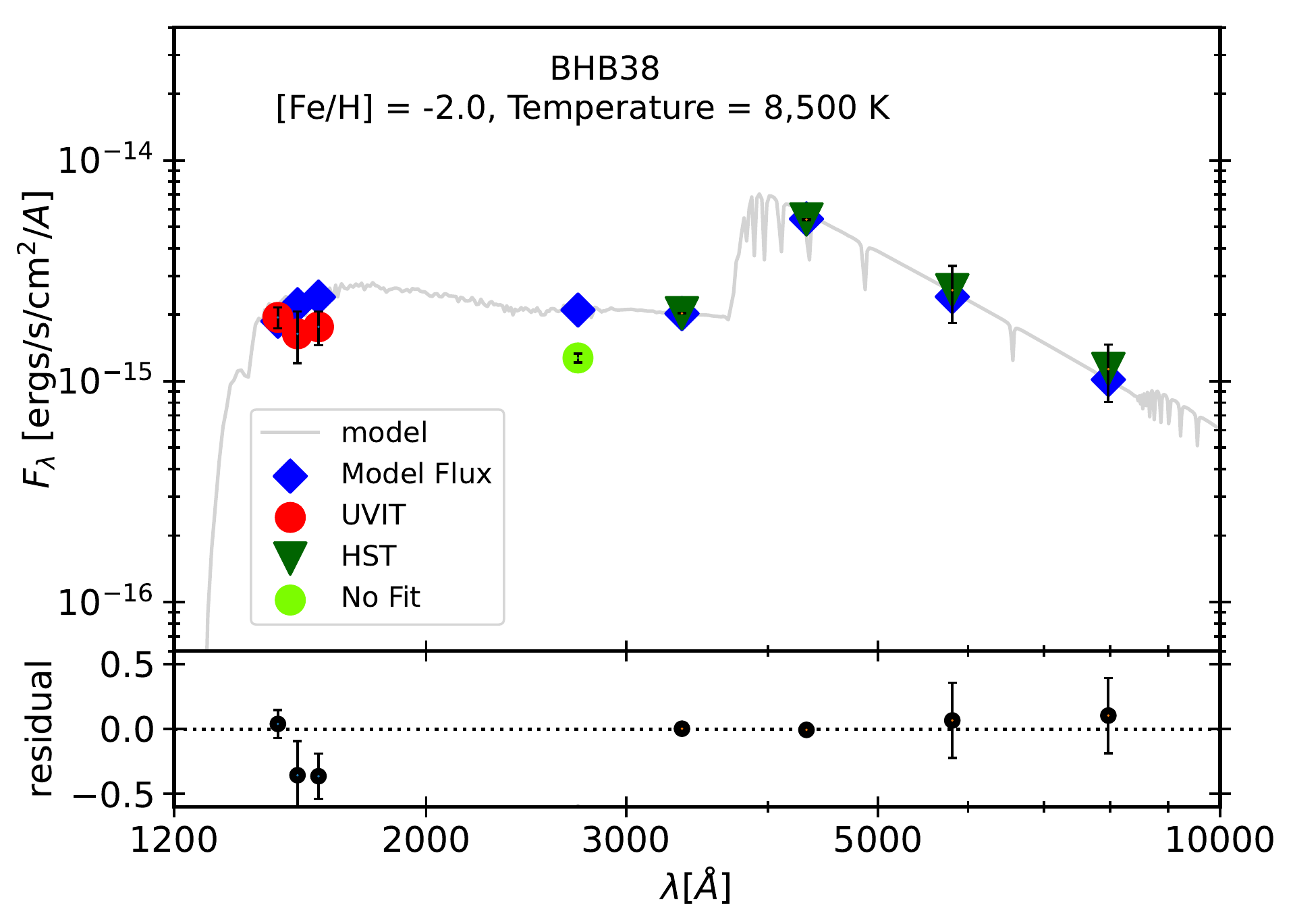}
\includegraphics[width=0.32\columnwidth]{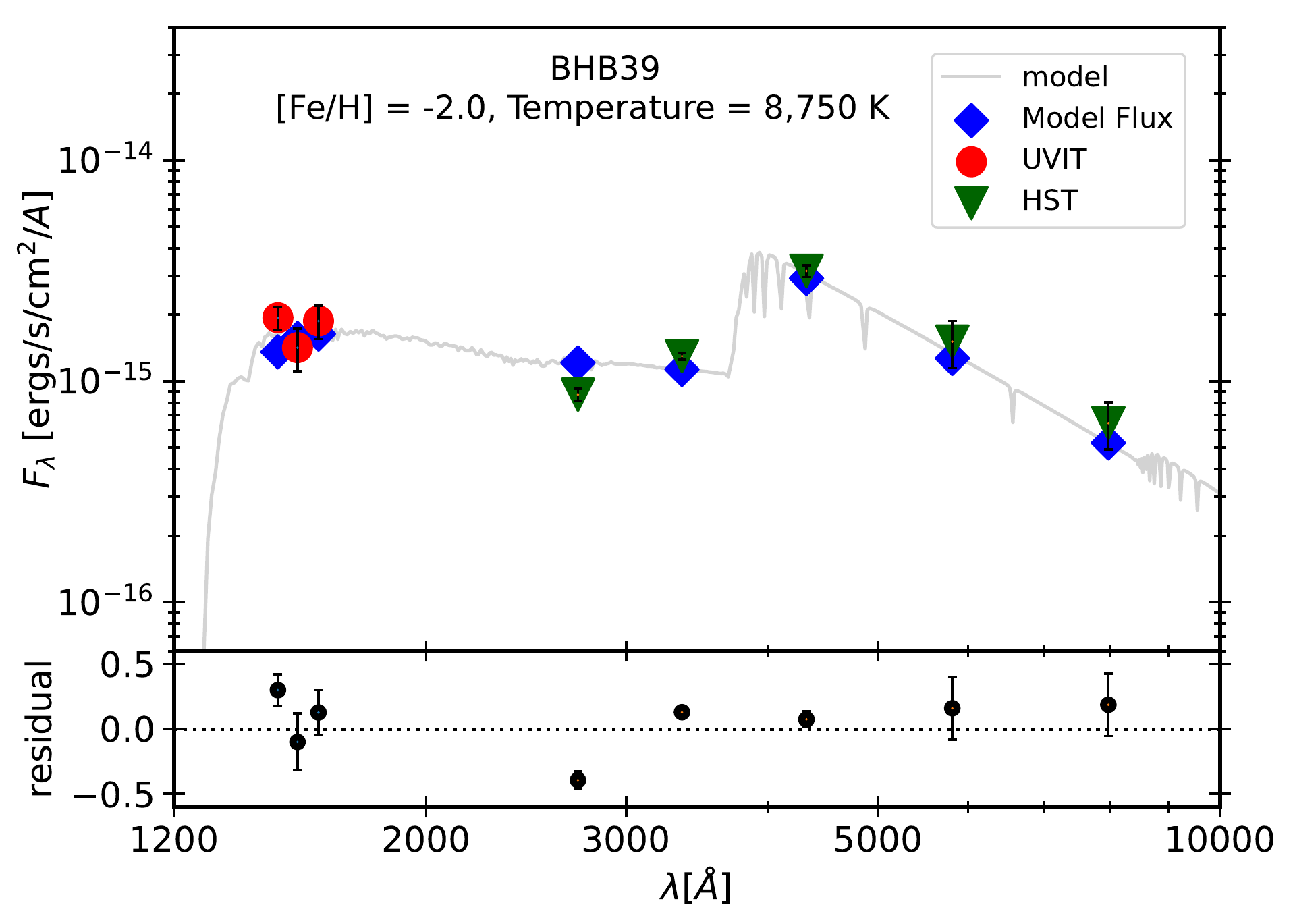}
\includegraphics[width=0.32\columnwidth]{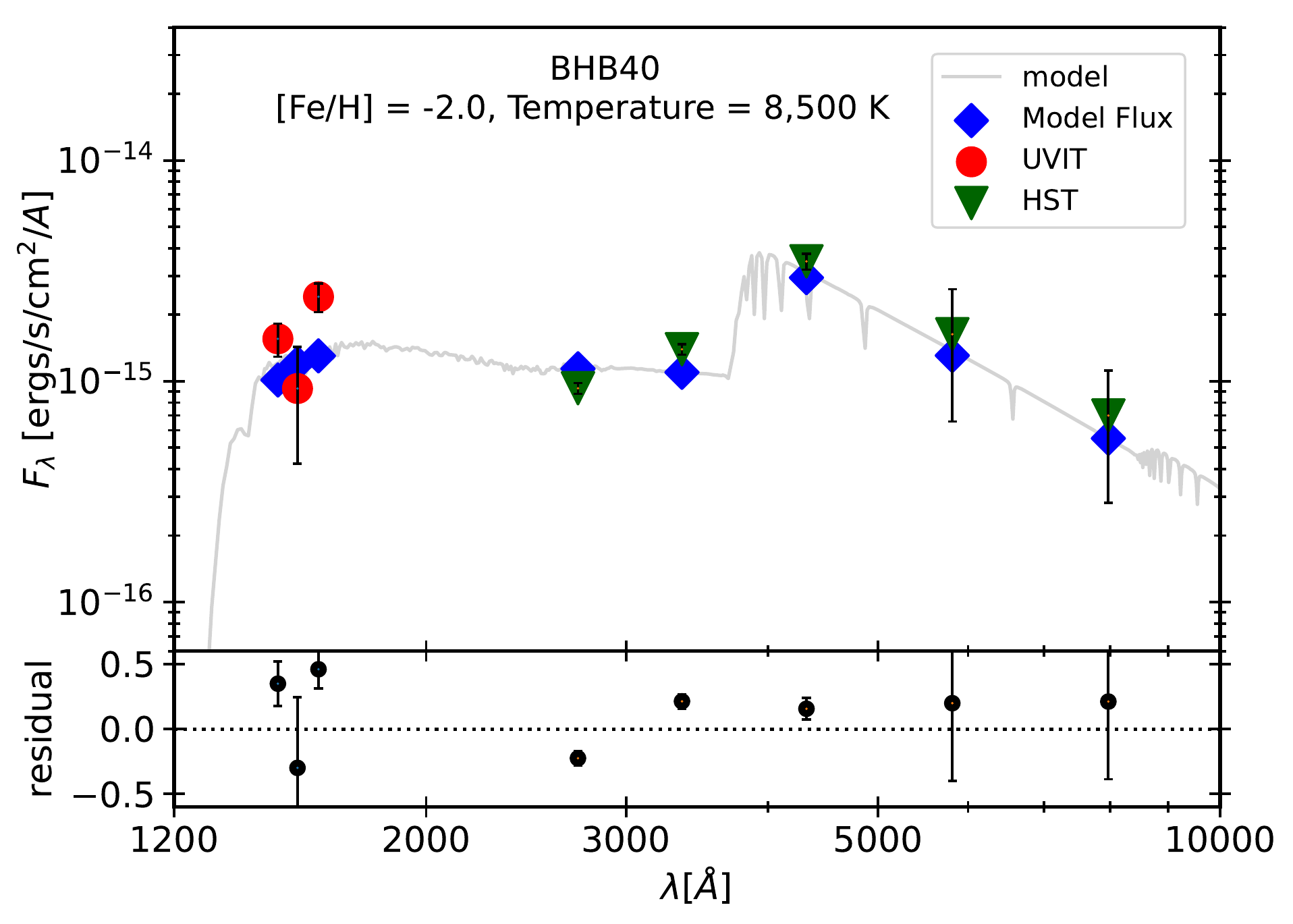}
\includegraphics[width=0.32\columnwidth]{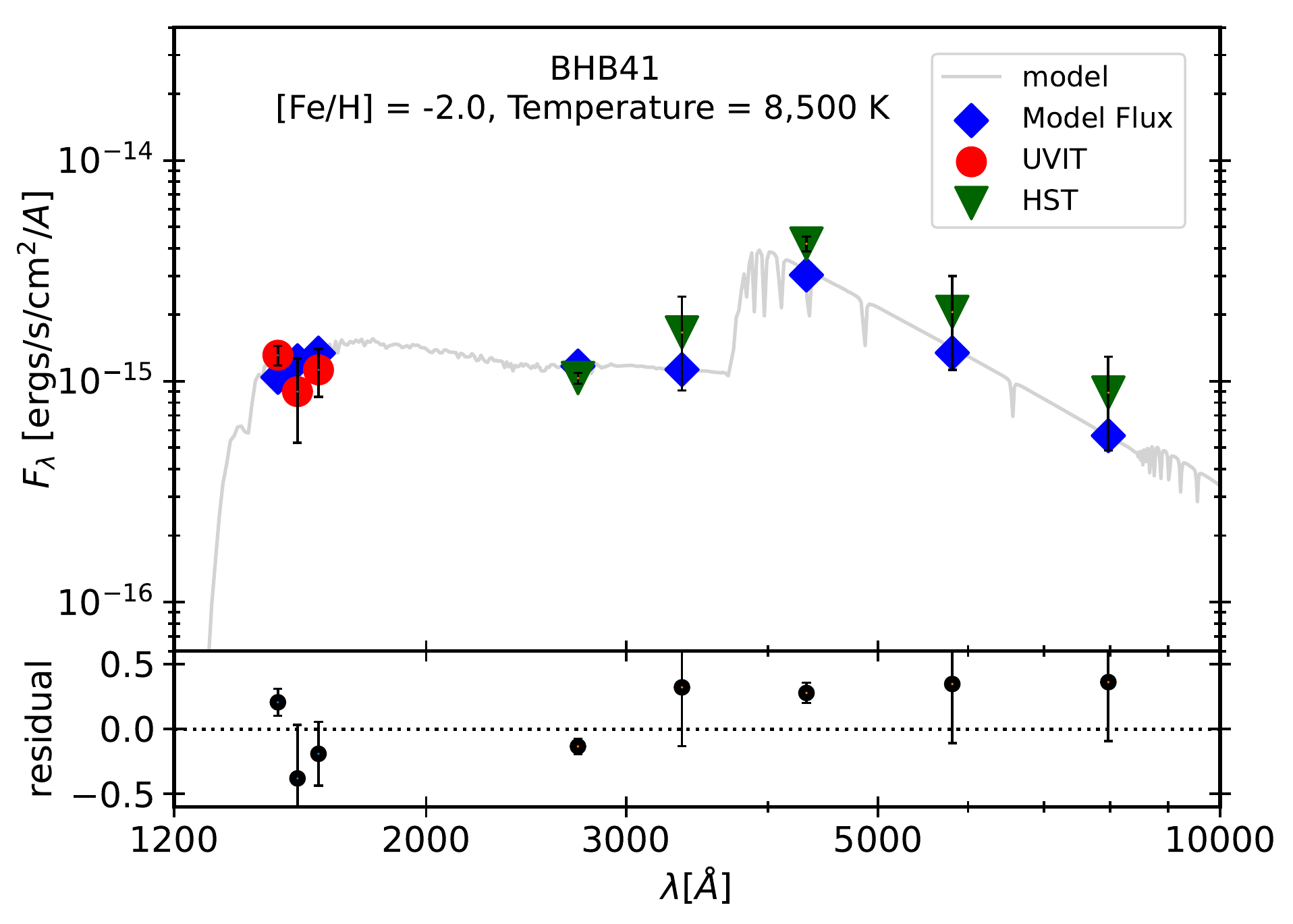}
\includegraphics[width=0.32\columnwidth]{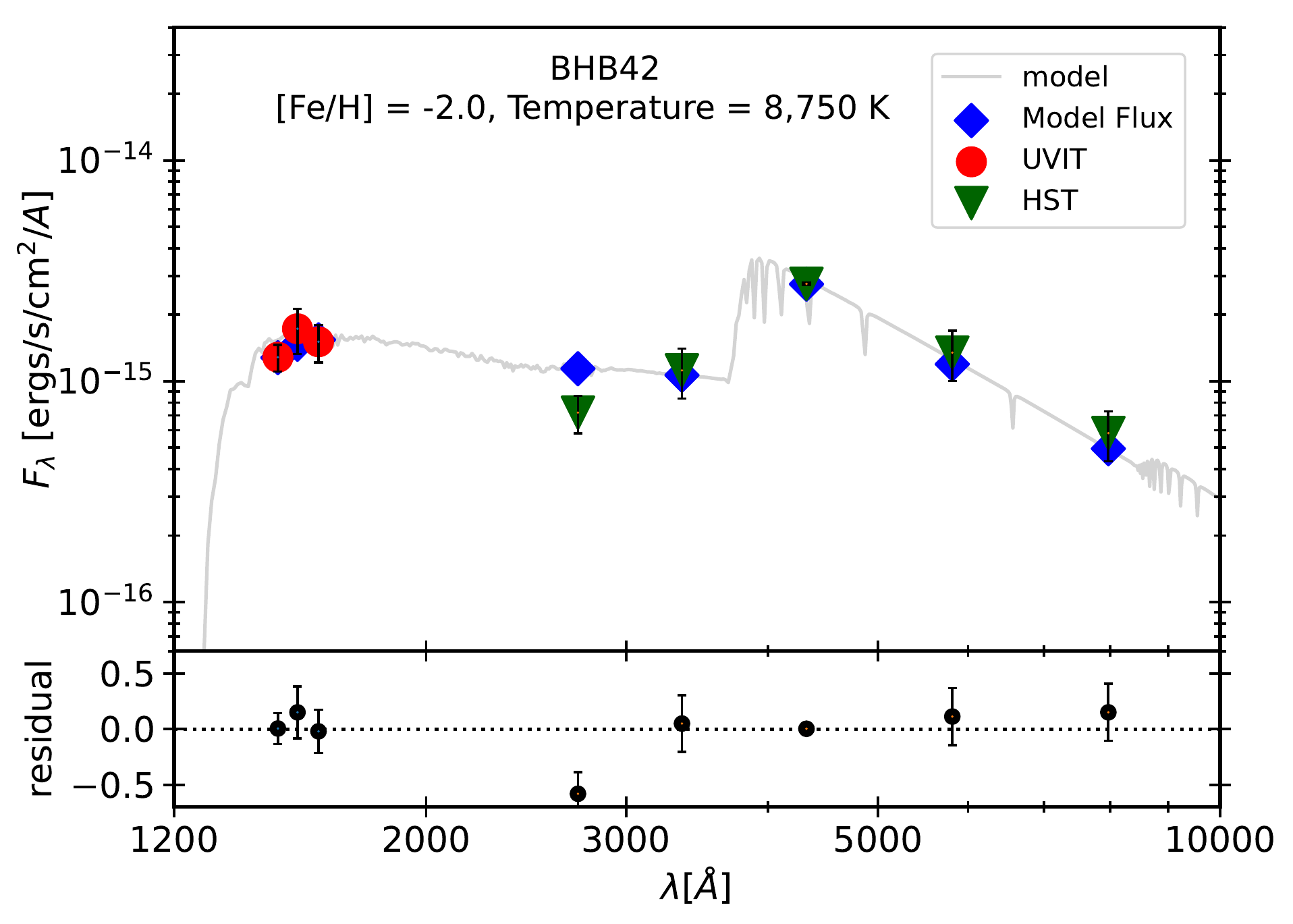}
\includegraphics[width=0.32\columnwidth]{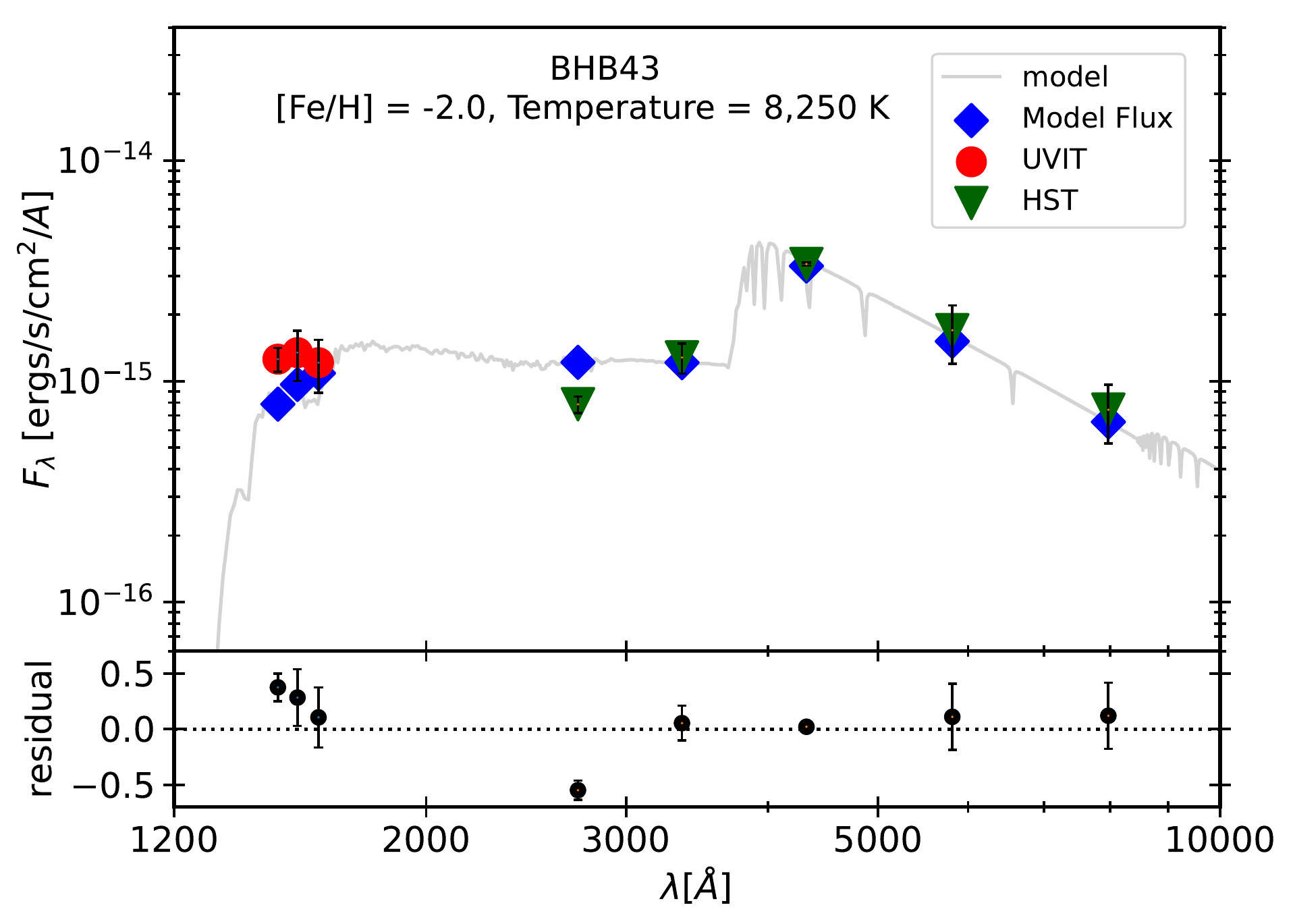}
\includegraphics[width=0.32\columnwidth]{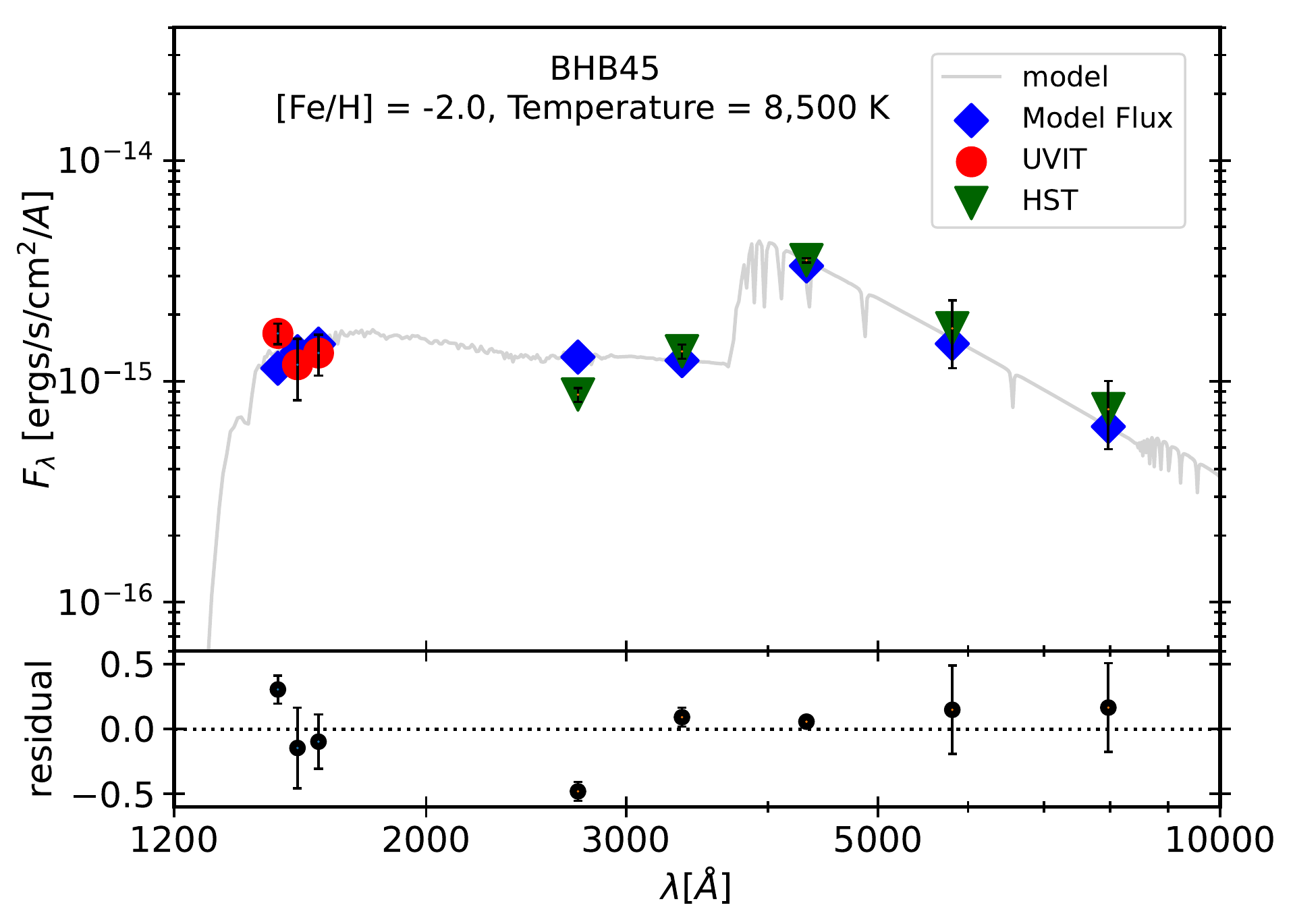}
\includegraphics[width=0.32\columnwidth]{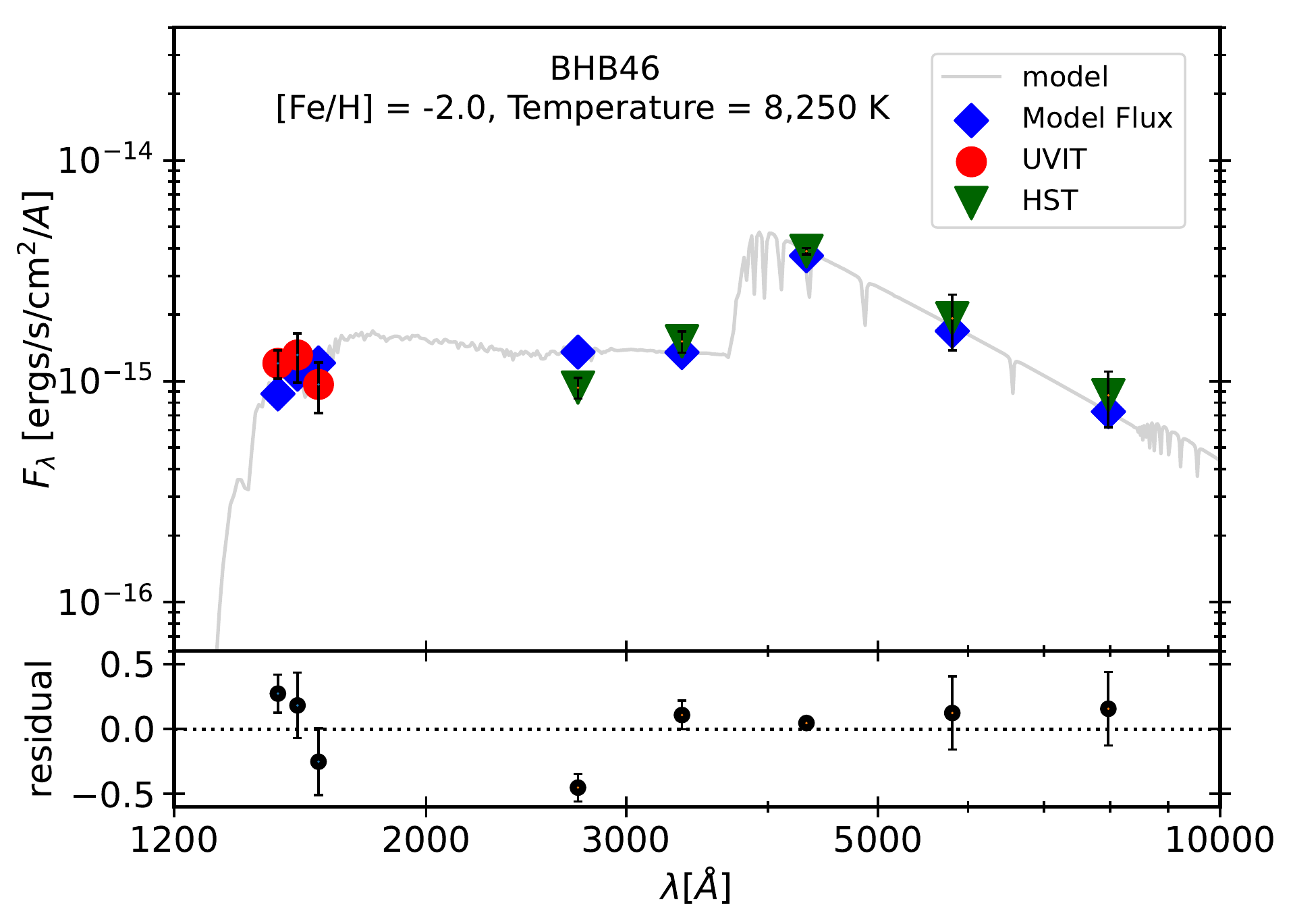}
 \caption{Continued.}
 \label{sed3}
\end{figure*}
\renewcommand{\thefigure}{\arabic{figure}}

\renewcommand{\thefigure}{A\arabic{figure}}
\addtocounter{figure}{-1}
\begin{figure*}[htb!]
    \centering
    \includegraphics[width=0.32\columnwidth]{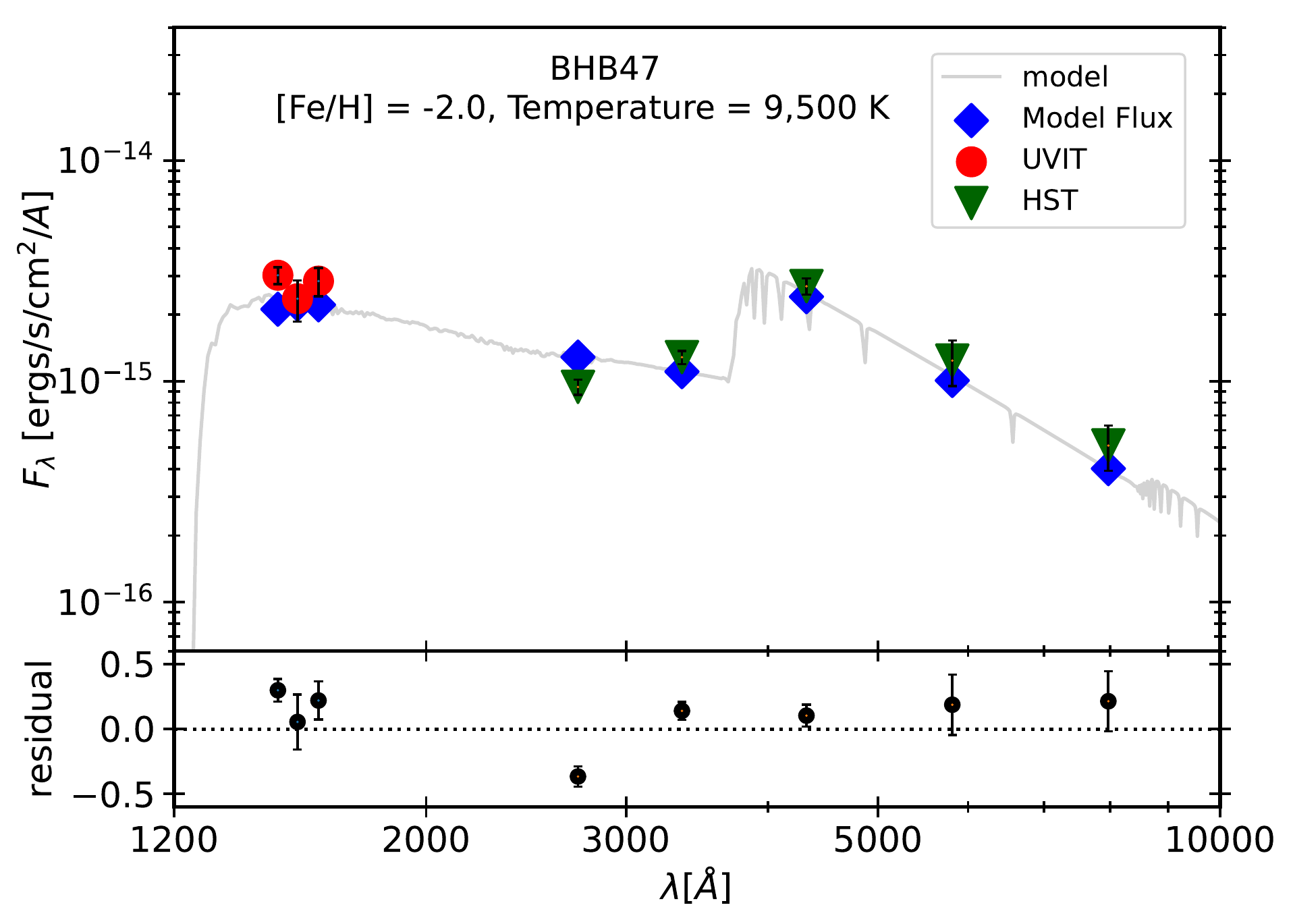}
    \includegraphics[width=0.32\columnwidth]{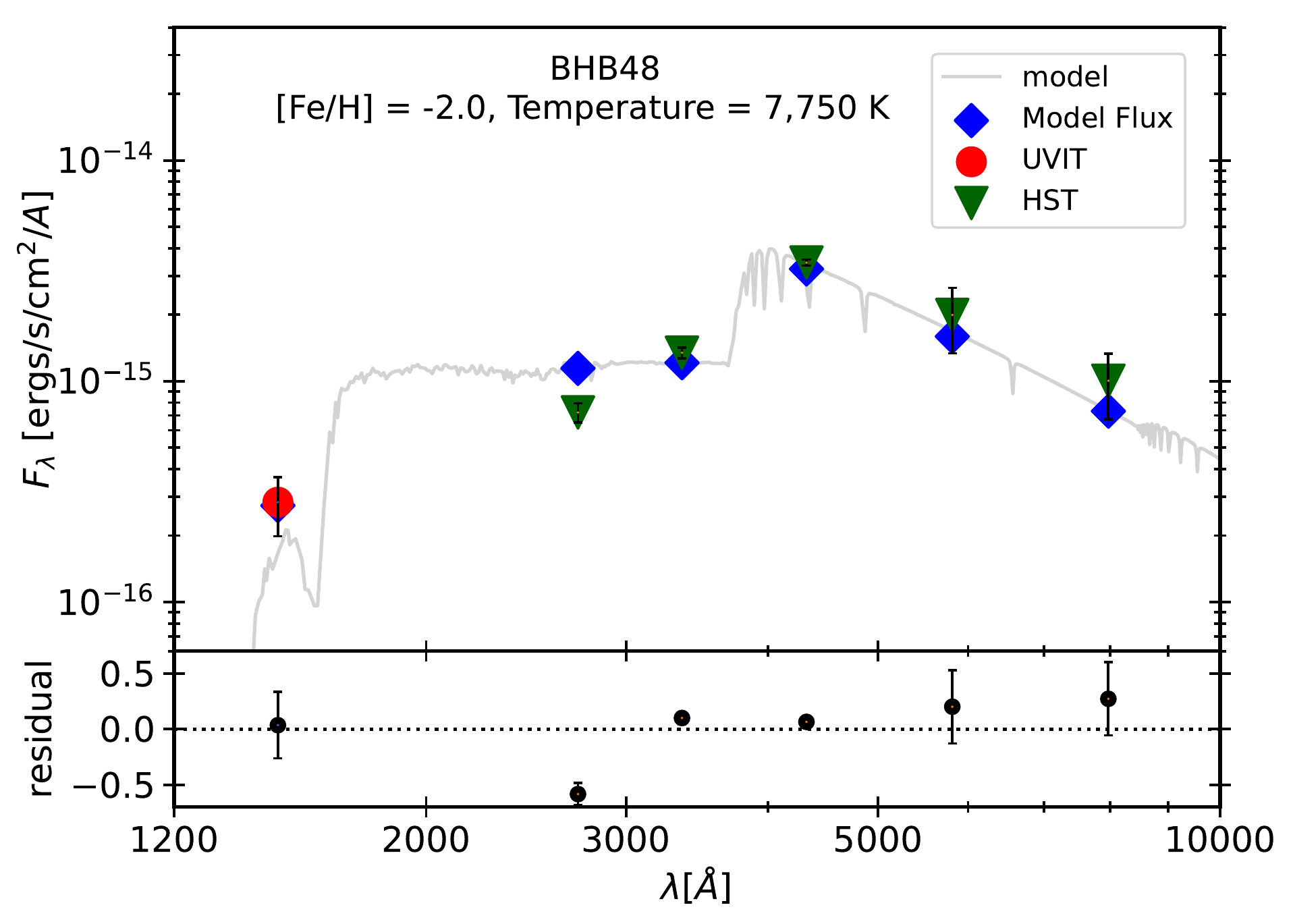}
    \includegraphics[width=0.32\columnwidth]{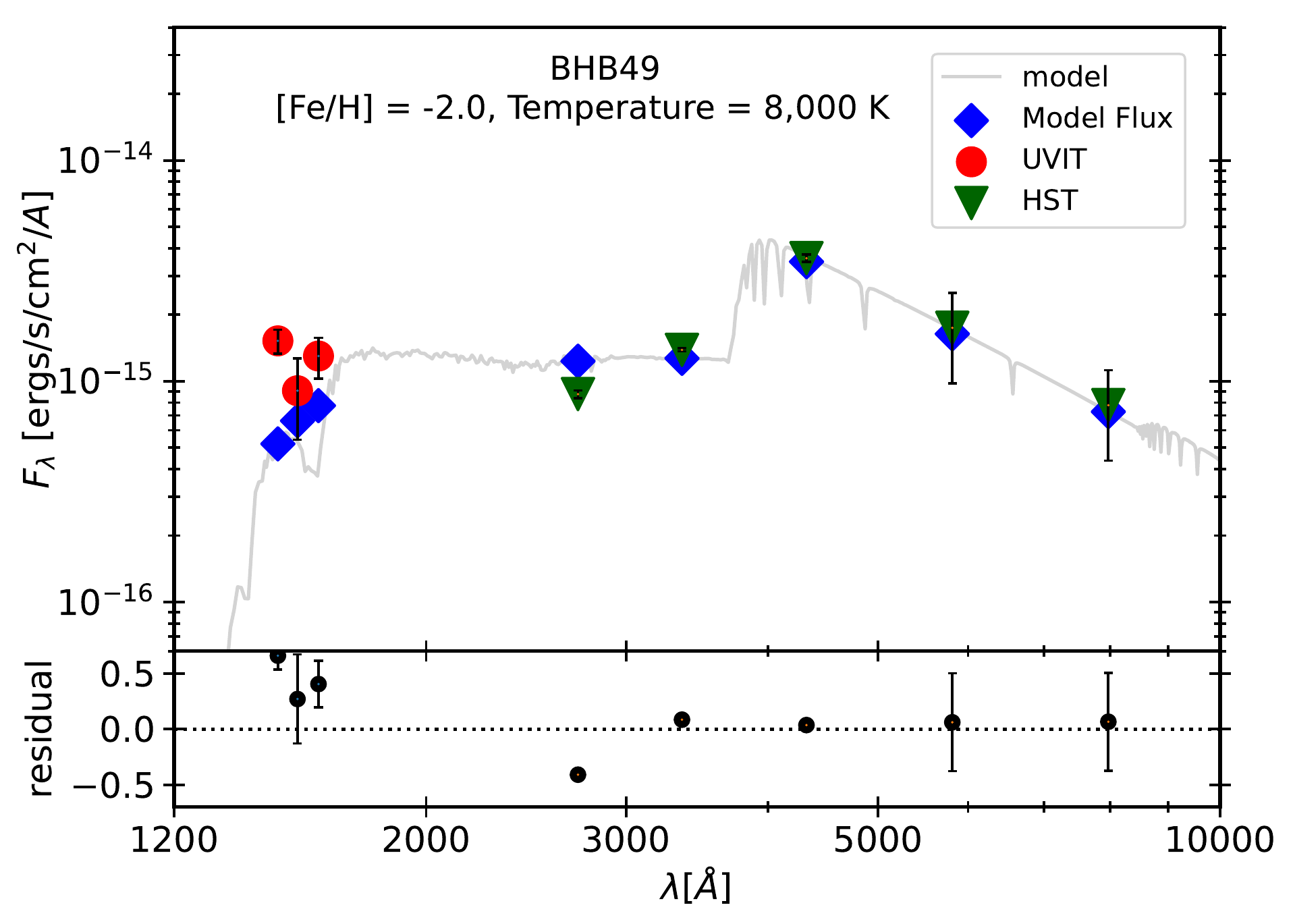}
    \includegraphics[width=0.32\columnwidth]{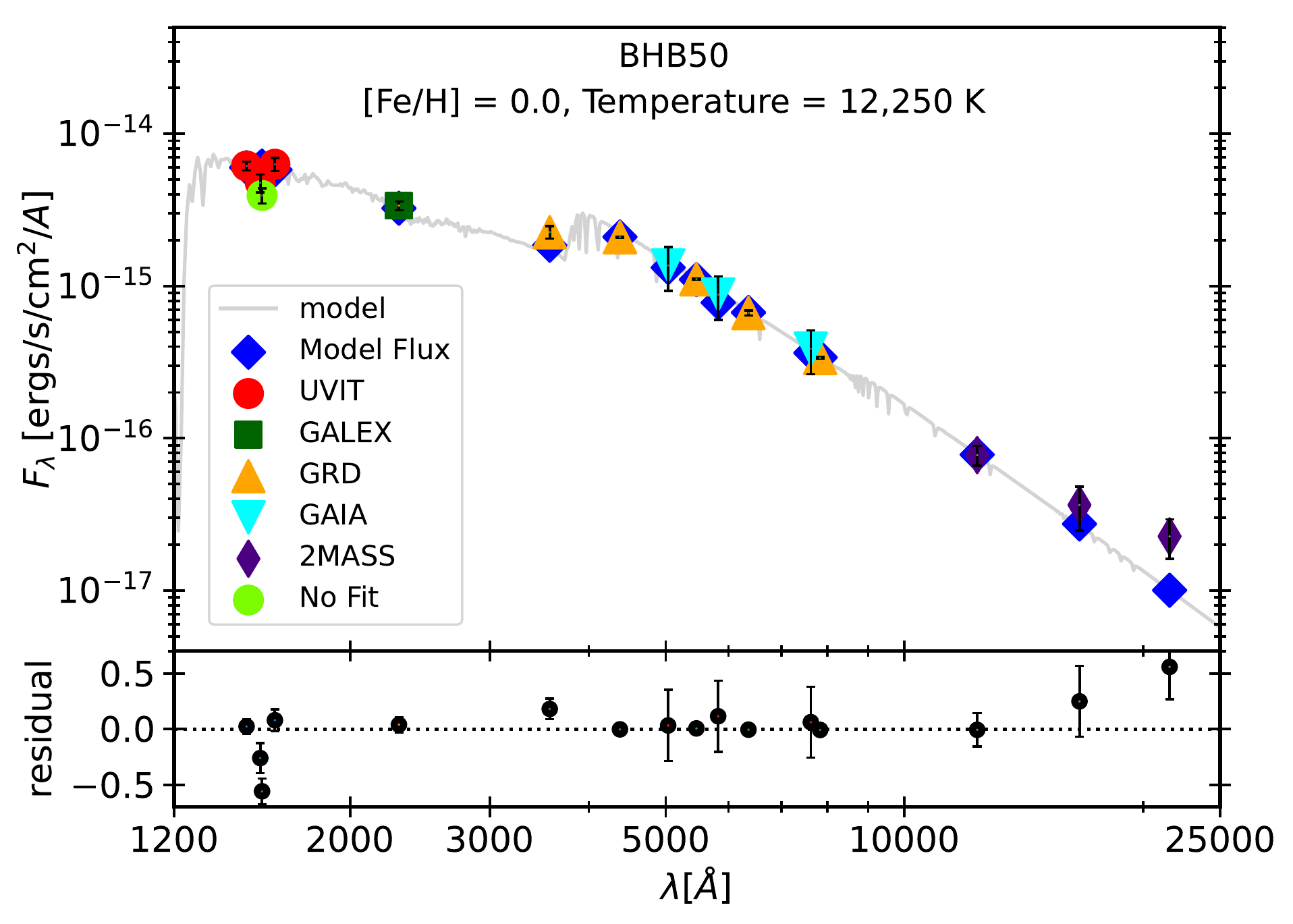}
    \includegraphics[width=0.32\columnwidth]{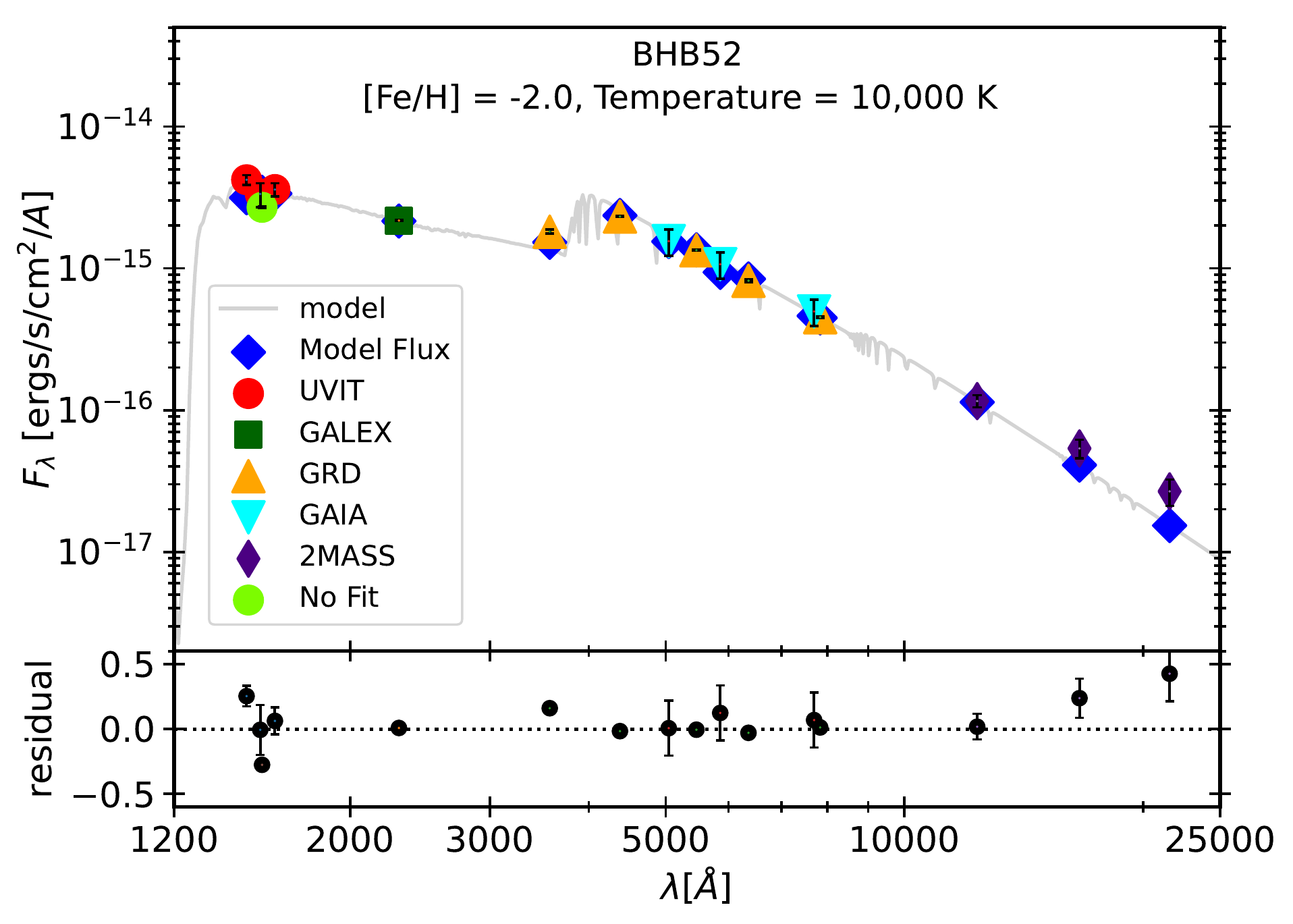}
    \includegraphics[width=0.32\columnwidth]{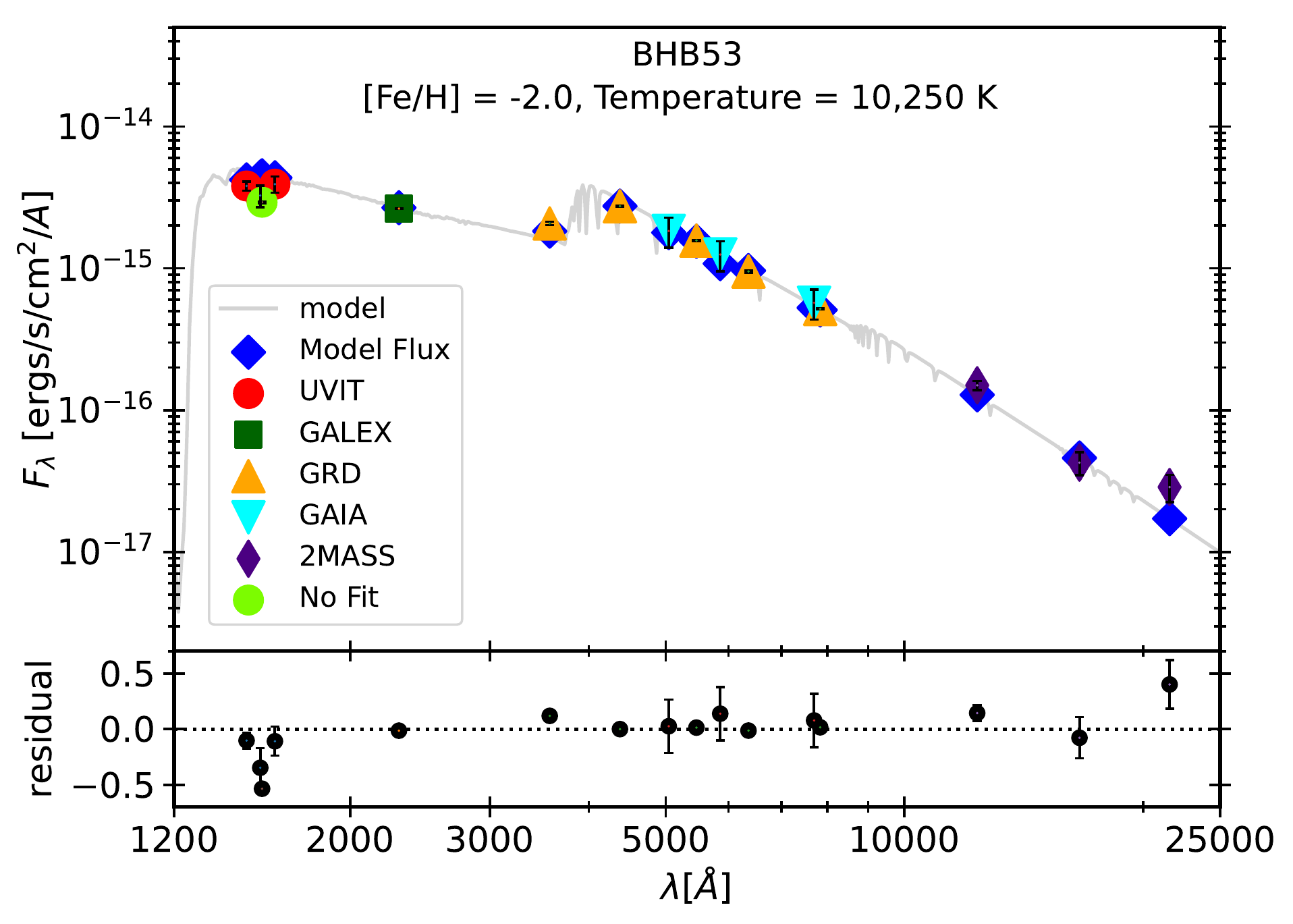}
    \includegraphics[width=0.32\columnwidth]{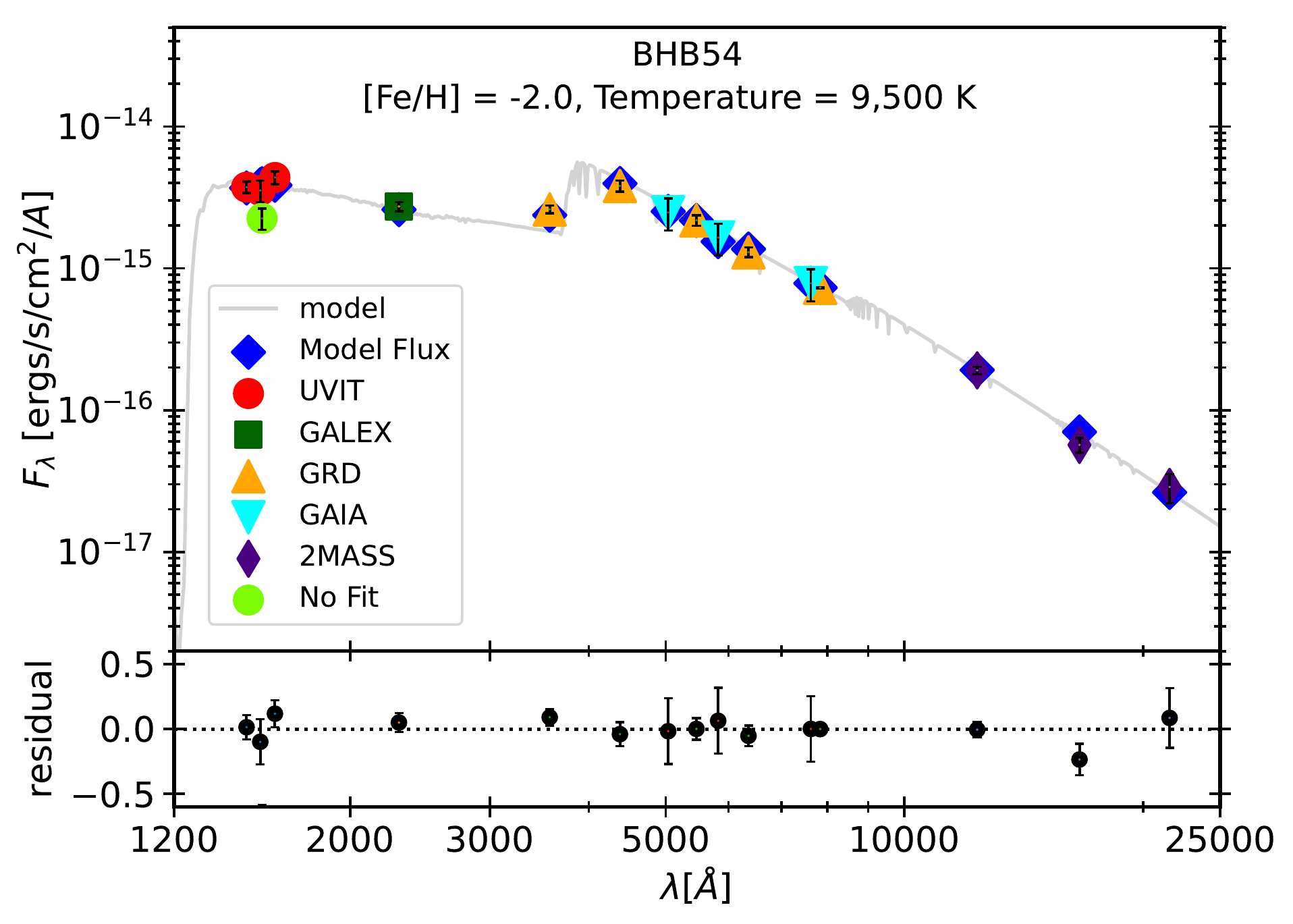}
    \includegraphics[width=0.32\columnwidth]{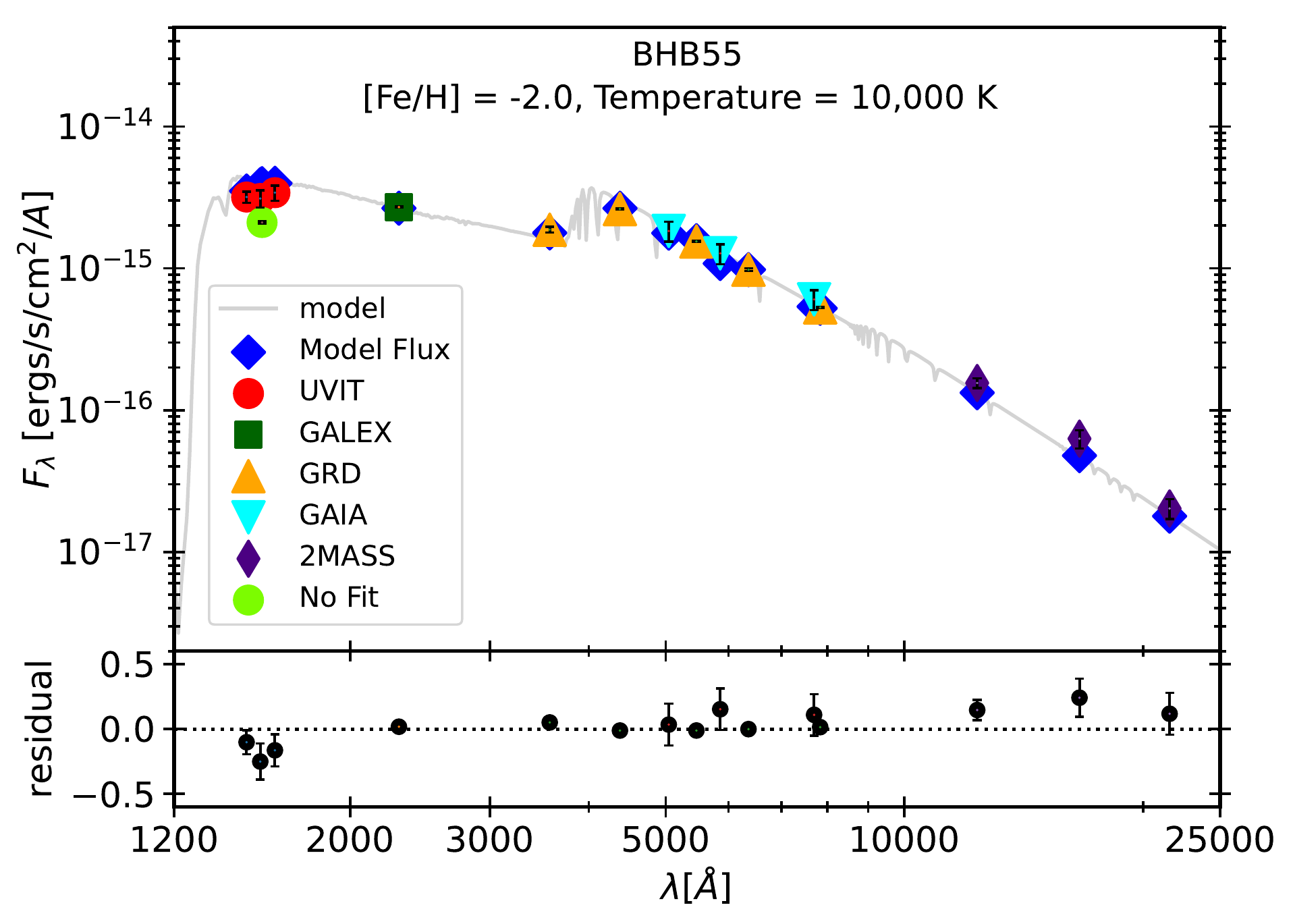}
    \includegraphics[width=0.32\columnwidth]{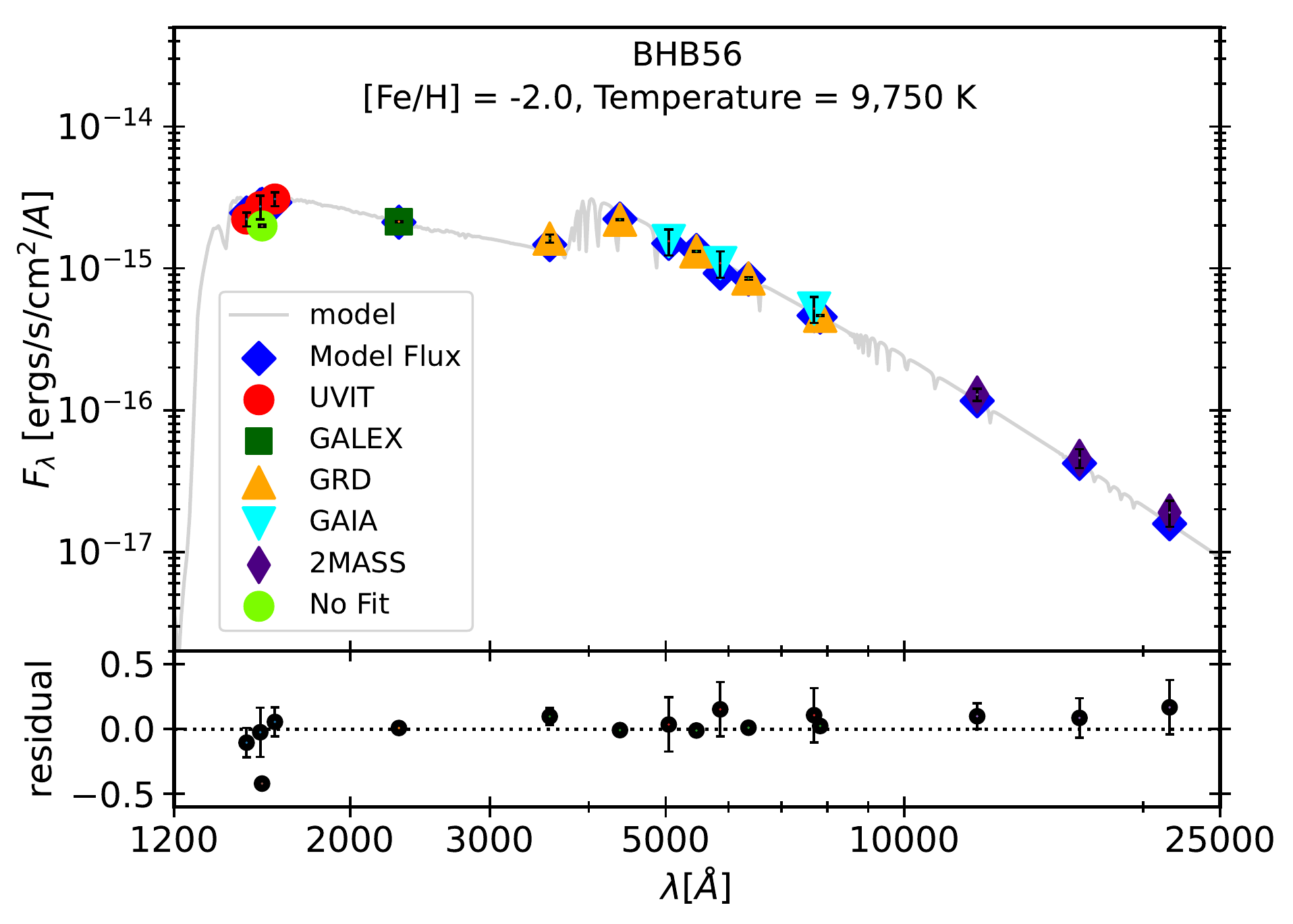}
    \includegraphics[width=0.32\columnwidth]{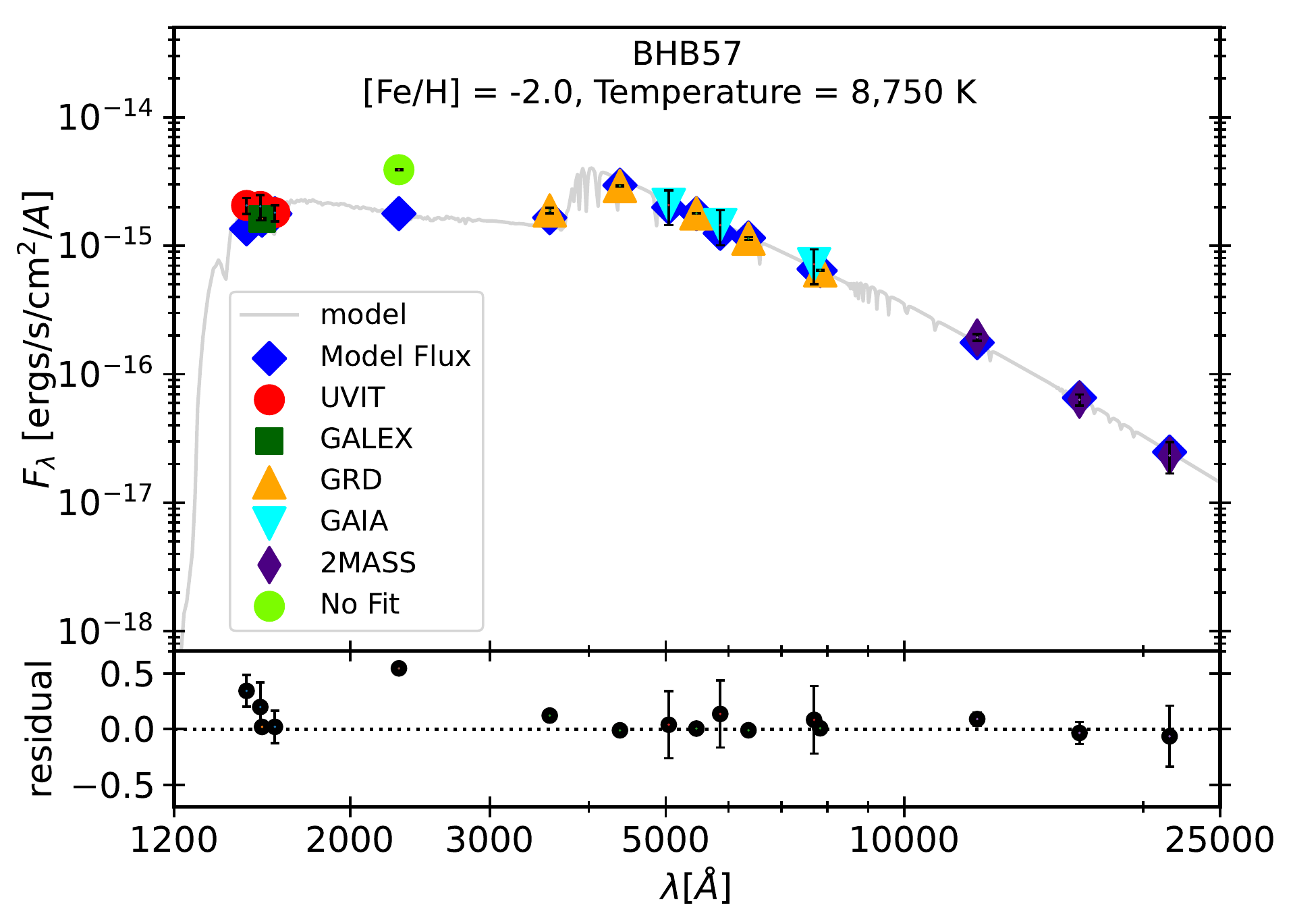}
    \includegraphics[width=0.32\columnwidth]{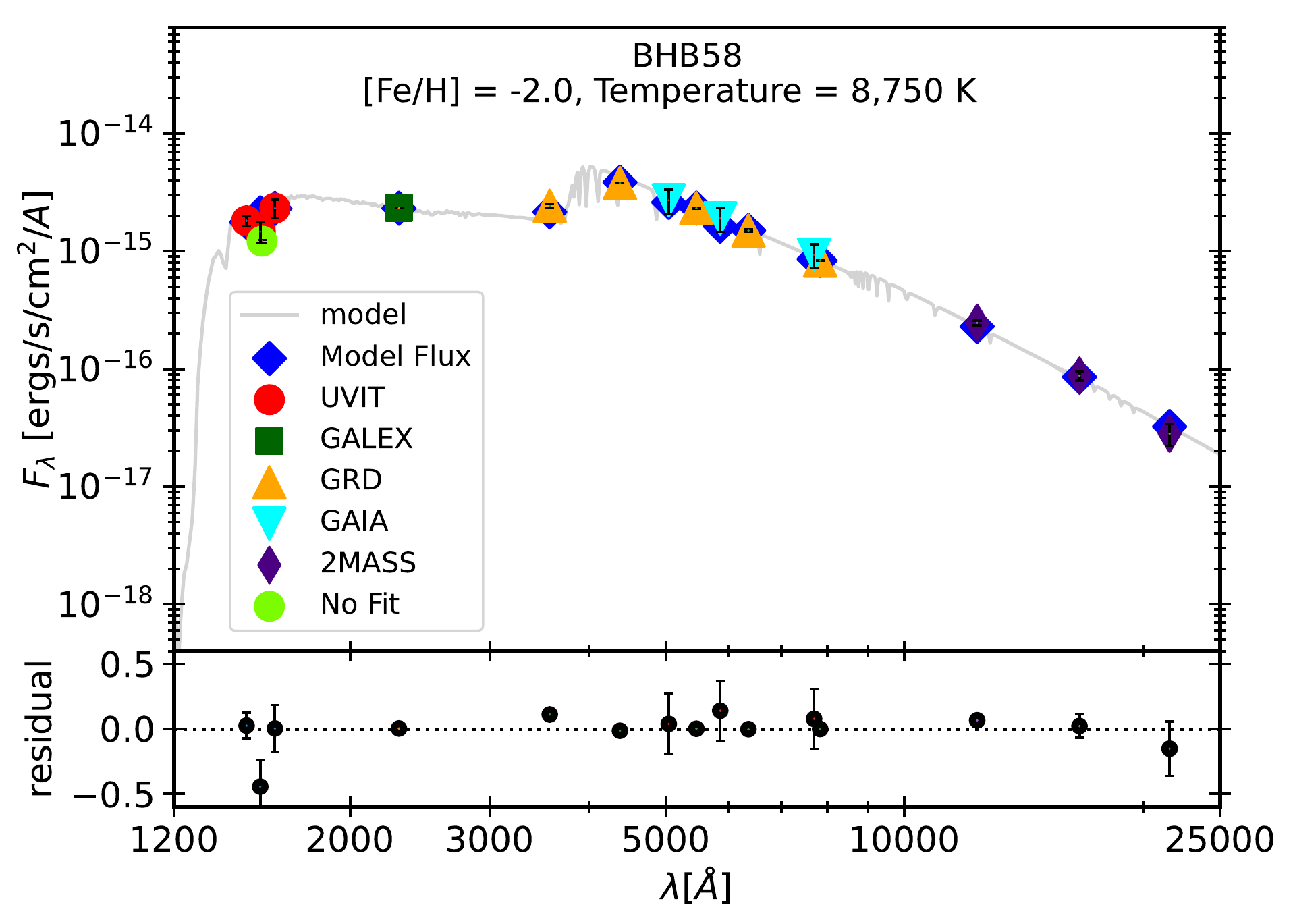}
    \includegraphics[width=0.32\columnwidth]{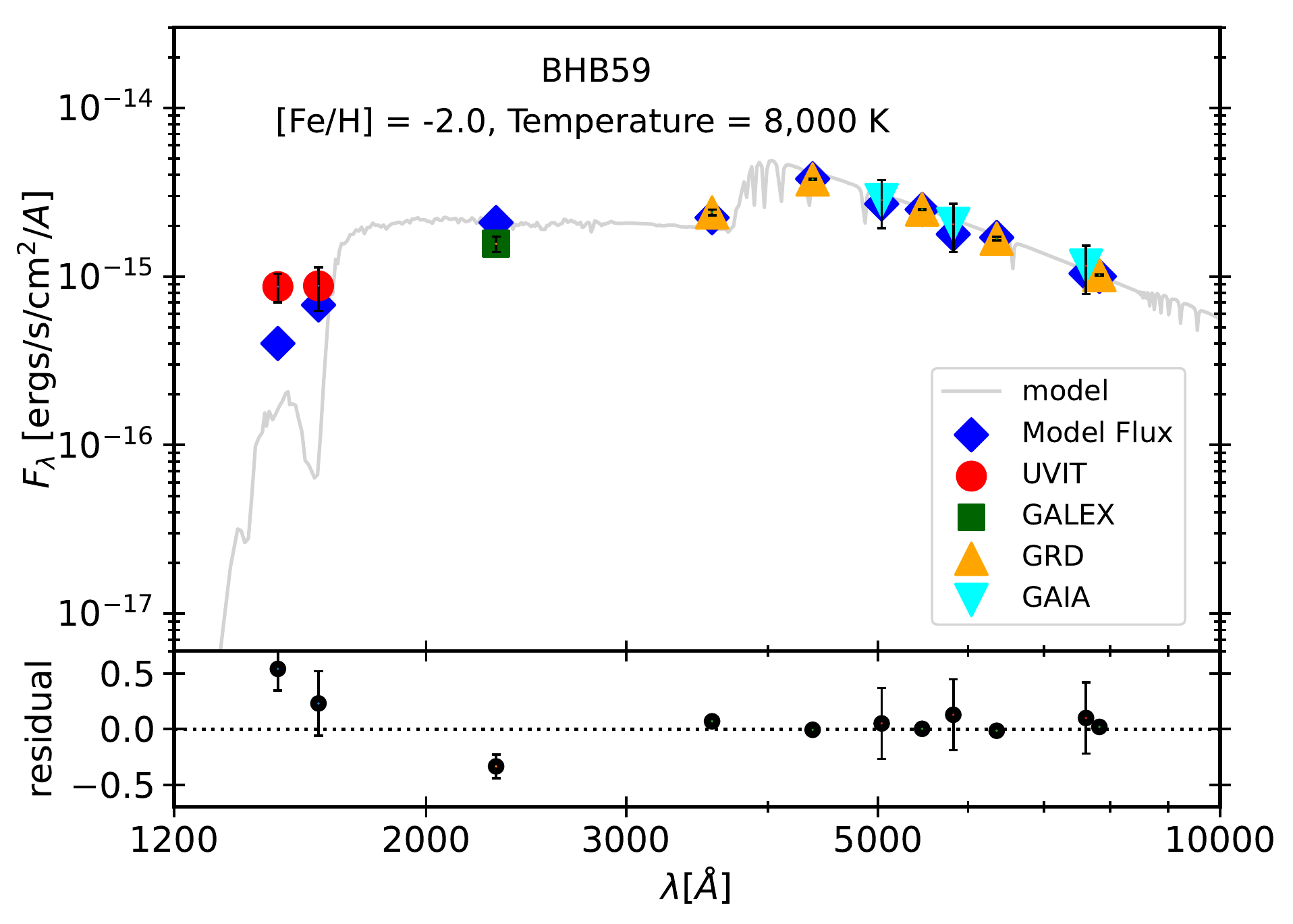}
     \includegraphics[width=0.32\columnwidth]{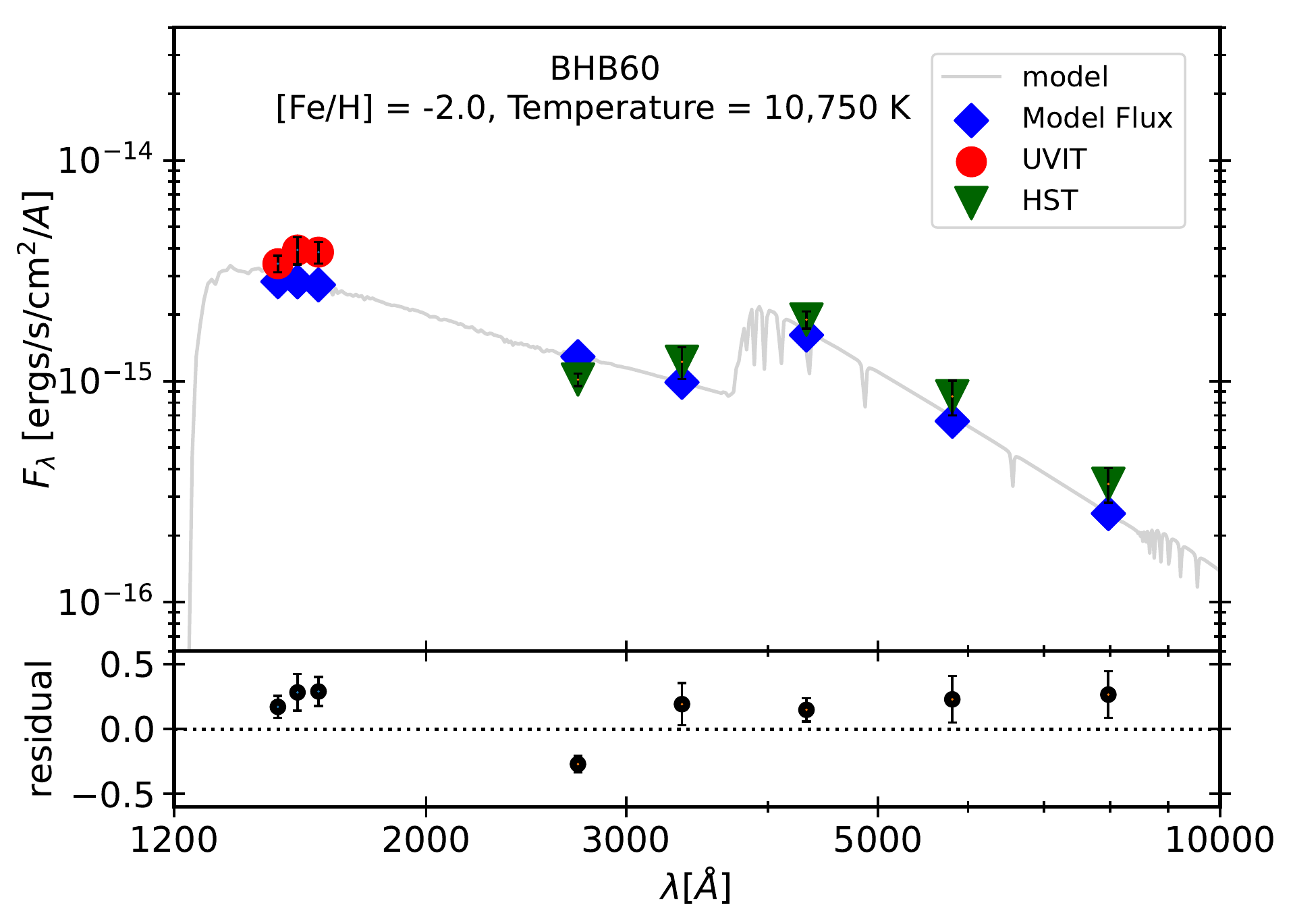}
    \includegraphics[width=0.32\columnwidth]{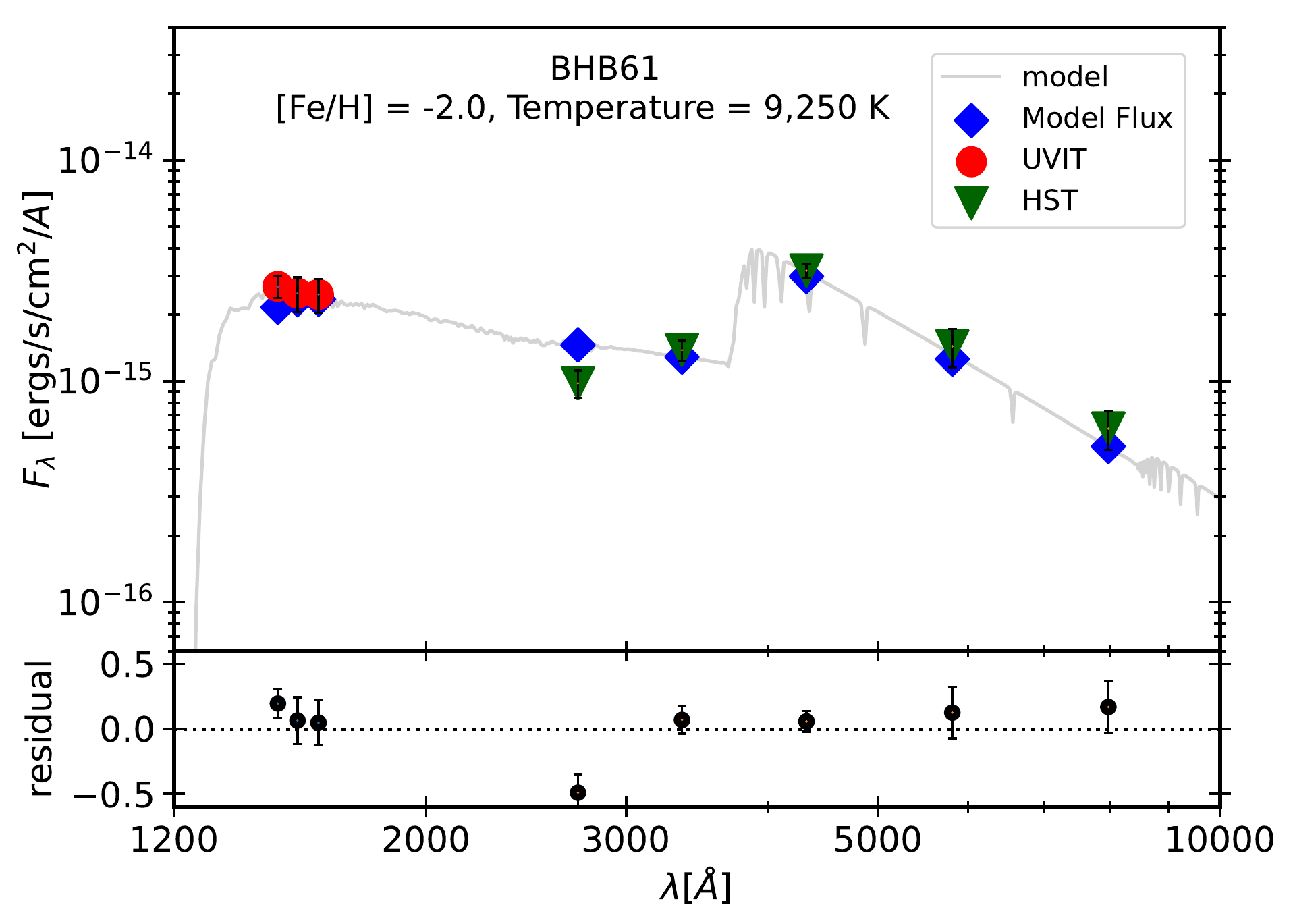}
    \includegraphics[width=0.32\columnwidth]{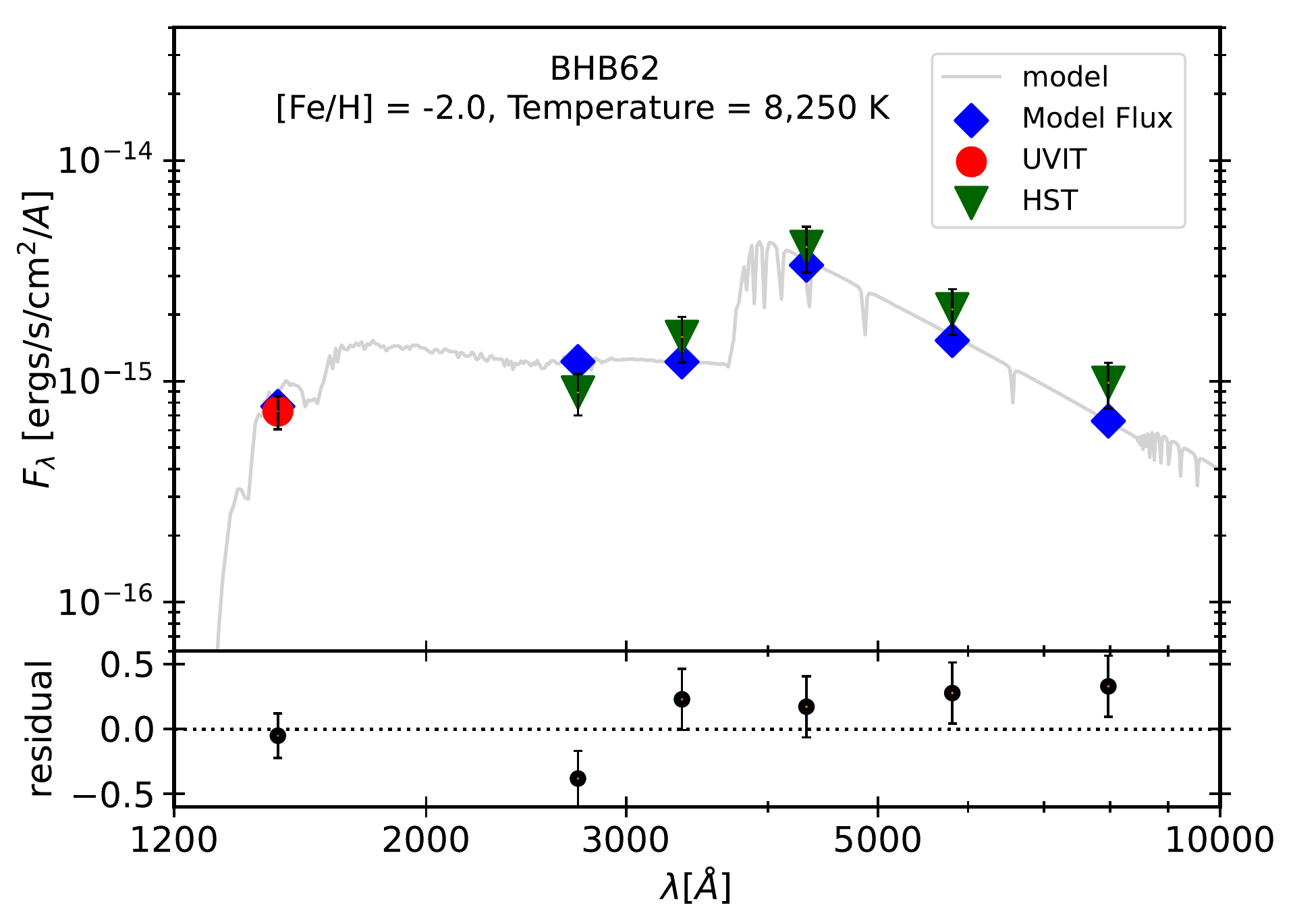}
    \includegraphics[width=0.32\columnwidth]{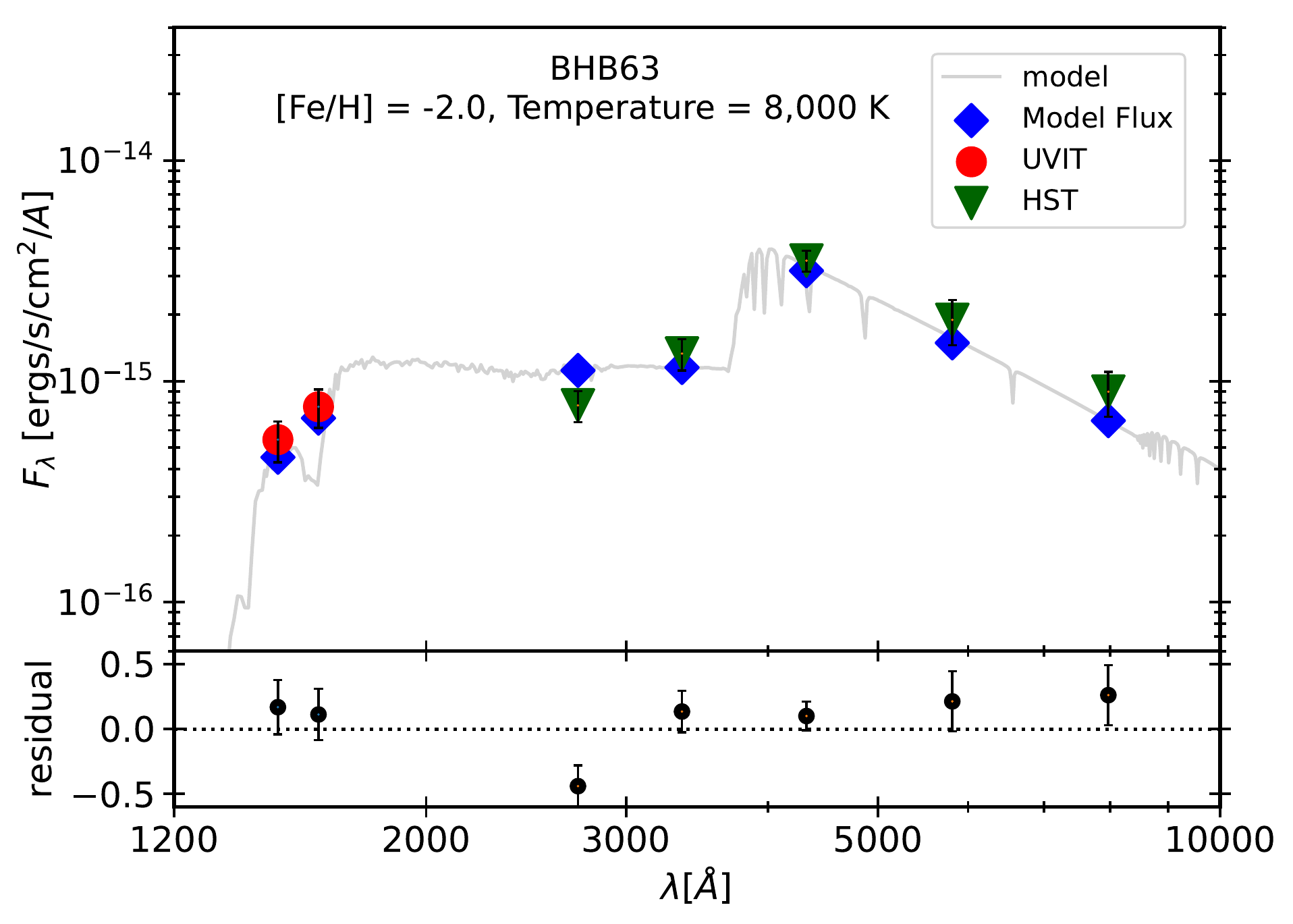}
    \caption{Continued.}
    \label{fig:sed4}
\end{figure*}
\renewcommand{\thefigure}{\arabic{figure}}


\bibliography{references}{}
\bibliographystyle{aasjournal}



\end{document}